\documentclass[11pt]{article}
\pdfoutput=1
\usepackage{soul}
\usepackage{jheppub}
\usepackage{amsfonts}
\usepackage{amsmath}
\usepackage{amsthm}
\allowdisplaybreaks[4]         
\usepackage{amssymb}   
\usepackage{placeins}
\usepackage{euscript}   
\usepackage{cleveref}    
\usepackage[dvipsnames]{xcolor}          
\usepackage{tensor}     
\usepackage{lipsum}  
\usepackage{graphicx}
\usepackage{caption}
\usepackage{afterpage}
\usepackage{subcaption}   
\usepackage{hyperref}
\usepackage{enumerate}
\hypersetup{
	colorlinks,
	linkcolor =NavyBlue,
	citecolor=MidnightBlue,
	urlcolor=NavyBlue,
}

\newcommand{\req}[1]{(\ref{#1})} 

\newcommand{\bea}{\begin{eqnarray}}
\newcommand{\eea}{\end{eqnarray}}
\newcommand{\ba}{\begin{eqnarray}}
\usepackage{braket}
\newcommand{\ea}{\end{eqnarray}}

\newcommand{\beq}{\begin{equation}}
\newcommand{\eeq}{\end{equation} }
\newcommand{\beqa}{\begin{eqnarray}}

\newcommand{\eeqa}{\end{eqnarray}}
\newcommand{\beqar}{\begin{eqnarray*}}
\newcommand{\eeqar}{\end{eqnarray*}}

\newcommand{\nmax}{n_\text{max}}
\newcommand{\mmax}{m_\text{max}}

\newcommand{\be}{\begin{equation}}
\newcommand{\ee}{\end{equation}}
\newcommand{\diff}{\mathrm{d}}

\newcommand{\peff}{p_{\rm eff}}
\renewcommand{\req}[1]{(\ref{#1})}

\newcommand{\eg}{{\it e.g.,}\ }
\newcommand{\ie}{{\it i.e.,}\ }

\newcommand{\Q}{\mathcal{Q}}

\makeatletter
\newcommand{\dal}{\mathop{\mathpalette\dal@\relax}}
\newcommand{\dal@}[2]{%
  \begingroup
  \sbox\z@{$\m@th#1\square$}%
  \dimen0=\fontdimen8
    \ifx#1\displaystyle\textfont\else
    \ifx#1\textstyle\textfont\else
    \ifx#1\scriptstyle\scriptfont\else
    \scriptscriptfont\fi\fi\fi3
  \makebox[\wd\z@]{%
    \hbox to \ht\z@{%
      \vrule width \dimen0
      \kern-\dimen0
      \vbox to \ht\z@{
        \hrule height \dimen0 width \ht\z@
        \vss
        \hrule height 2\dimen0
      }%
      \kern-2.5\dimen0
      \vrule width 2.5\dimen0
    }%
  }%
  \endgroup
}
\makeatother

\definecolor{shadecolor}{rgb}{.25,.25,.25}

\usepackage{caption}
\usepackage{subcaption}

\usepackage{multirow}
\usepackage{tikz}
\usetikzlibrary{decorations.markings,decorations.pathmorphing}
\tikzstyle{singularity}=[carmine,line width=0.5mm,decorate, decoration={zigzag,amplitude=2,segment length=6.17}]

\definecolor{carmine}{rgb}{0.59, 0.0, 0.09}
\definecolor{egyptianblue}{rgb}{0.06, 0.2, 0.65}
\definecolor{frenchlilac}{rgb}{0.53, 0.38, 0.56}
\definecolor{darkspringgreen}{rgb}{0.09, 0.45, 0.27}
\definecolor{ochre}{rgb}{0.8, 0.47, 0.13}

\usepackage{tikz-3dplot}
\tikzstyle{particle1}=[lightseagreen,line width=0.5]
\tikzstyle{particle2}=[ochre,line width=0.5]
\definecolor{lightseagreen}{rgb}{0.13, 0.7, 0.67}
\definecolor{fandango}{rgb}{0.71, 0.2, 0.54}

\title{
\boldmath  On regular charged black holes in three dimensions}

\author[a]{Pablo Bueno,}
\author[b]{Oscar Lasso Andino,}
\author[c,a]{Javier Moreno,}
\author[a]{Guido van der Velde}

\affiliation[a]{Departament de Física Quàntica i Astrofísica, Institut de Ciències del Cosmos\\ Universitat de Barcelona, Martí i Franquès 1, E-08028 Barcelona, Spain
\vspace{0.1cm}}
\affiliation[b]{Escuela de Ciencias Físicas y Matemáticas, Universidad de Las Américas,\\ Redondel del ciclista, Antigua vía a Nayón, C.P. 170504, Quito, Ecuador
\vspace{0.1cm}}
\affiliation[c]{Departamento de Física, Universidad de Concepción,\\ Casilla, 160-C, Concepción, Chile \vspace{0.1cm}}

\emailAdd{pablobueno@ub.edu}
\emailAdd{oscar.lasso@udla.edu.ec}
\emailAdd{fjaviermoreno@udec.cl}
\emailAdd{guidovandervelde@ub.edu}

\date{\today}
\abstract{As argued in \href{http://www.arXiv.org/abs/2104.10172}{{\tt arXiv:2104.10172}}, introducing a non-minimally coupled scalar field, three-dimensional Einstein gravity can be extended by infinite families of theories which admit simple analytic generalizations of the charged BTZ black hole. Depending on the gravitational couplings, the solutions may describe black holes with one or several horizons and  with curvature or BTZ-like singularities. In other cases, the metric function behaves as $f(r)\overset{(r\rightarrow 0)}{\sim} \mathcal{O}(r^{2s})$ with $s\geq 1$, and the black holes are completely regular---a feature unique to three dimensions. Regularity arises generically \emph{i.e.,}
without requiring any fine-tuning of parameters. In this paper we show that all these theories satisfy Birkhoff theorems, so that the most general spherically-symmetric solutions are given by the corresponding static black holes. We perform a thorough characterization of the Penrose diagrams of the solutions, finding a rich structure which includes, in particular, cases which tessellate the plane and others in which the diagrams cannot be drawn in a single plane. We also study the motion of probe particles on the black holes, finding that observers falling to regular black holes reach the center after a finite proper time. Contrary to the singular cases, the particles are not torn apart by tidal forces, so they oscillate between antipodal points describing many-universe orbits.  We argue that in those cases the region $r=0$ can be interpreted as a horizon with vanishing surface gravity, giving rise to generic inner-extremal regular black hole solutions. We also analyze the deep interior region of the solutions identifying the presence of Kasner eons and the conditions under which they take place. Finally, we construct new black hole solutions in the case in which infinite towers of terms are included in the action.}

\begin{document} 
\vspace*{1\fill}
\maketitle
\flushbottom


\section{Introduction}
The three-dimensional version of general relativity is much simpler than its four-dimensional counterpart. Indeed, in $D=3$ all curvature is captured by the Ricci tensor, which means that Einstein spacetimes are locally maximally symmetric. In addition, metric perturbations propagate no degrees of freedom \cite{Deser:1983tn}. Despite these features, Ba\~nados, Teitelboim, and Zanelli (BTZ) showed that the theory is rich enough to admit black hole solutions in the presence of a cosmological constant \cite{Banados:1992wn,Banados:1992gq}. The BTZ black holes share many of the features of their higher-dimensional cousins, such as the presence of event (and Cauchy) horizons, their interpretation as thermodynamic objects or their role as thermal states in holographic systems.
They also present some notable differences, \eg the metric is equivalent to pure AdS$_3$ almost everywhere and curvature invariants remain constant except at the very singularity locus.

Additional three-dimensional black holes can be constructed when other fields are coupled to gravity and in the presence of higher-curvature corrections. In the case of the latter, a prototypical example is that of New Massive Gravity \cite{Bergshoeff:2009hq} and its extensions \cite{Gullu:2010pc, Sinha:2010ai,Paulos:2010ke,Bueno:2022lhf}, which allow for black holes which differ from the BTZ one 
for special values of the gravitational couplings. These solutions feature new asymptotic behaviors, they are not locally maximally symmetric and they generically possess curvature singularities \cite{Bergshoeff:2009aq,Oliva:2009ip,Clement:2009gq,Alkac:2016xlr,Barnich:2015dvt,AyonBeato:2009nh,Gabadadze:2012xv,Ayon-Beato:2014wla,Fareghbal:2014kfa,Nam:2010dd,Gurses:2019wpb,Bravo-Gaete:2020ftn,Chernicoff:2024dll, Chernicoff:2024lkj}. 
On the other hand, including electromagnetic and scalar fields also leads to new solutions, such as those found for the Einstein-Maxwell \cite{Clement:1993kc,Kamata:1995zu,Martinez:1999qi,  Hirschmann:1995he, Cataldo:1996yr, Dias:2002ps, Cataldo:2004uw,  Cataldo:2002fh}, Einstein-Maxwell-dilaton  \cite{Chan:1994qa, Fernando:1999bh,    Chen:1998sa,Koikawa:1997am,Edery:2020kof,Edery:2022crs,Karakasis:2022fep,Priyadarshinee:2023cmi}  and  Maxwell-Brans-Dicke \cite{Sa:1995vs,Dias:2001xt} systems, as well as for minimally and non-minimally coupled scalars \cite{Martinez:1996gn,Henneaux:2002wm,Correa:2011dt,Zhao:2013isa,Tang:2019jkn,Karakasis:2021lnq,Baake:2020tgk,Nashed:2021ldz,Karakasis:2021ttn,Arias:2022jax,Desa:2022gtw,Cardenas:2022jtz,Karakasis:2023hni,Maluf:2024svb}, including theories obtained from certain three-dimensional limits of Lovelock theories \cite{Hennigar:2020drx,Hennigar:2020fkv,Ma:2020ufk,Konoplya:2020ibi,Lu:2020iav}. Again, curvature singularities are a prototypical feature of the solutions and so is the presence of logarithmic profiles for some of the fields, although globally regular scalars also arise in some cases. Additional solutions are found for Born-Infeld-like and non-linear electrodynamics models  \cite{Cataldo:1999wr,Myung:2008kd,Mazharimousavi:2011nd,Mazharimousavi:2014vza,Hendi:2017mgb,Guerrero:2021avm,Gonzalez:2021vwp,Maluf:2022jjc,Sardeshpande:2024bnk} which, in some cases, correspond to regular black holes when certain action parameters and charges of the solutions are adjusted  \cite{Cataldo:2000ns,He:2017ujy,HabibMazharimousavi:2011gh}. Yet another context in which new three-dimensional black holes arise is the one of braneworld holography, where modified versions of the BTZ black hole including the full backreaction of strongly coupled quantum fields have been constructed \cite{Emparan:2002px,Emparan:2020znc,Emparan:2022ijy,Bueno:2023dpl,Panella:2024sor,Panella:2023lsi}.

New families of analytic generalizations of the BTZ black hole were constructed in  \cite{Bueno:2021krl}. These solve the equations of a class of extensions of three-dimensional Einstein gravity known as \emph{Electromagnetic Quasi-topological gravities} (EQT). The theories involve two towers of terms weighted by arbitrary couplings and built from contractions of derivatives of a real scalar field and the Ricci tensor, which appears at most linearly in all densities---see \req{IEQTG} below. 
The solutions are characterized by a single metric function, $f(r)$, and the scalar $\phi$ takes a universal ``magnetic'' form---being proportional to the angular coordinate---which automatically solves its equation of motion, namely,
\begin{equation}\label{eq:SSSM}
\diff s^2=-f(r)\diff t^2+\frac{\diff r^2}{f(r)}+r^2\diff\varphi \, ,\quad \diff \phi=p \diff \varphi\, , 
\end{equation}
where $r\geq 0$, $\varphi=[0,2\pi)$, and where  $p$  is an arbitrary constant proportional to  the charge of the solution. Observe that in this configuration the scalar field is proportional to the polar coordinate so, analogously to a four-dimensional monopole, one needs two coordinate charts to describe the solution. Remarkably, the solution for  the metric function $f(r)$ can be obtained fully analytically for general values of the gravitational couplings---see \req{fEQT2}. 
Depending on such values, the solutions describe various kinds of spacetimes \cite{Bueno:2021krl}. These include naked singularities, globally regular horizonless solutions and black holes with one or multiple horizons which may hide a (BTZ-like, conical or curvature) singularity. In some cases, corresponding to the metric function behaving as
\begin{equation}\label{regui}
    f(r)\overset{(r\rightarrow 0)}{\longrightarrow} \mathcal{O}(r^{2s})\, , \quad \text{with}\quad s\geq 1\, ,
\end{equation}
 the solutions turn out to be regular. Just like in similar higher-dimensional constructions of regular black holes for theories involving  higher-derivative corrections to Einstein gravity \cite{Cano:2020ezi,Bueno:2024dgm}, regularity takes place in these solutions without any sort of fine tuning between the gravitational couplings and/or the physical parameters of the solutions. On the other hand, the type of regularity achieved through \req{regui} is exclusive to $D=3$ and relies on the vanishing intrinsic curvature of the $(D-2)$-sphere in that number of dimensions.    

In this paper we perform a thorough study of various properties of EQT theories and their black hole solutions.  We start in Section~\ref{sec:black holes in EQTG} by reviewing the basic definition of EQT gravity in three dimensions as introduced in \cite{Bueno:2021krl}. We make a detailed account of its black hole solutions, which we classify according to their horizon and singularity structures.  In Section~\ref{sec:non-local} we comment on the possibility of reinterpreting black holes as solutions to a dual theory with an electric field and explain how various solutions in the context of non-linear electrodynamics can be more naturally understood as solutions to certain EQT theories. We also consider the effects of including infinite towers of terms in the original EQT frame, giving rise to nonlocal models, and construct new black holes to those. In Section~\ref{birk} we prove that our theories satisfy a Birkhoff theorem. Namely, we show that imposing spherical symmetry plus the azimuthal ansatz for the scalar necessarily implies staticity,
\begin{equation}
\diff s^2= -N(t,r) f(t,r) \diff t^2 + \frac{\diff r^2}{f(t,r) }+r^2 \diff \varphi^2 \quad \underset{\diff \phi=p \diff \varphi}{\overset{\rm EQT} {\Longrightarrow}}\quad 
\diff s^2= -f(r) \diff t^2 + \frac{\diff r^2}{f(r) }+r^2 \diff \varphi^2\, , 
\end{equation}
where $f(r)$ is given, for a given theory, by the corresponding black hole metric. This opens the door to future studies of dynamical collapse within this framework. In Section~\ref{penro} we perform a very detailed study of the possible Penrose diagrams of the black holes depending on the number of horizons and regularity properties. We introduce a systematic approach which allows us to characterize all possible diagrams, finding a very rich structure. In particular, some of the black holes present diagrams which tessellate the plane---see \eg Figure~\ref{fig:tes}. For others, it is not possible to draw the diagram in a single plane, in other words, they possess non-planar Penrose diagrams---see Figure~\ref{fig:nonplanar}. In Section~\ref{sec:geodesics} we study the motion of neutral probe particles on the black hole backgrounds. We find different behaviors depending on the regularity of the solutions. In the case of regular black holes, we find the non-radial trajectories of infalling particles to consist of spirals accumulating around the $r = 0$ locus, which is reached in a finite proper time. On the other hand, radial trajectories involve never-ending multiple-universe oscillations between antipodal points which cross $r = 0$ repeatedly. The tidal forces remain finite at all times for both types of trajectories. In Section~\ref{eons} we analyze the deep interior region of single-horizon black holes. In particular, we study the conditions under which  the metric can be approximated by Kasner spacetimes. We characterize the various ``Kasner eons'' controlled by the different higher-derivative densities, and the transitions between them. We conclude in Section~\ref{conc}. Besides outlining plans for potential future research, we make further comments on the meaning of the $r=0$ region for regular black holes of the (\ref{regui}) type. We argue that this should be interpreted as an extremal horizon and that, as a consequence, solutions of this kind are the first instances of generic inner-extremal regular black holes. We connect this feature to previous considerations in higher-dimensions, which suggest that they may be dynamically stable.

\section{Black holes in Electromagnetic Quasi-topological gravity}\label{sec:black holes in EQTG}
The static charged BTZ black hole is a solution of the three-dimensional Einstein-Maxwell system \cite{Clement:1993kc,Hirschmann:1995he,Kamata:1996vg,Cataldo:1996yr,Martinez:1999qi}. Dualizing the Maxwell field---see Section~\ref{dualf}---a solution with the same metric can be obtained for the Einstein-dilaton action\footnote{A canonically normalized scalar has length dimension $[\phi]=-1/2$ in three dimensions. However, here we find it more convenient to set $[\phi]=0$. On the other hand, the Newton constant has length dimension $[G]=1$.} 
\begin{equation}\label{eq:Edil}
I=\frac{1}{16\pi G}\int {\mathrm d}^3x \sqrt{|g|} \mathcal{L}_0 \, , \quad \text{where}\quad \mathcal{L}_0= R+ \frac{2}{L^2}-\alpha_1 (\partial \phi)^2 \, ,
\end{equation}
where $(\partial \phi)^2\equiv g^{ab}\partial_a \phi \partial_b\phi$ and $\alpha_1$ is a dimensionless coupling. 
Indeed, the above theory admits a static and spherically symmetric solution of the form \req{eq:SSSM},\footnote{A more standard set of conventions both for the action  and for this solution in the absence of higher-curvature corrections would involve setting $\alpha_1\rightarrow 8\pi G$ and $p\rightarrow Q/\sqrt{16 \pi G}$. This would restore $[\phi]=-1/2$ as well as a slightly more familiar expression for the charged BTZ solution \cite{Martinez:1999qi}.} and
where the metric function takes the usual charged BTZ form
\begin{equation}\label{fEQT}
f(r)= \displaystyle \frac{r^2}{L^2}-\lambda- \alpha_1 p^2 \log\left( \frac{r}{r_0}\right)  \, ,
\end{equation}
where $r_0$ is some length scale and $\lambda$ is an integration constant related to the mass of the solution.

As anticipated in the introduction, in this paper we will be interested in a set of modifications of this solution recently constructed in \cite{Bueno:2021krl,Bueno:2022ewf}.\footnote{See also \cite{Alkac:2022tvc,Ditta:2022fjz,deOliveira:2024pab,Beissen:2024hpy} for recent related works.} These are in turn solutions of the EQT theories described in those papers. The full action of the theory of interest is given 
\begin{equation}\label{IEQTG}
I_{\rm EQT}=\frac{1}{16\pi G}\int {\mathrm d}^3x \sqrt{|g|} \left[\mathcal{L}_0 - \Q_\alpha+\Q_\beta\right]\, ,
\end{equation}
where the Einstein-dilaton action receives the corrections
\begin{align}\label{qa}
\Q_\alpha&=\sum_{n=2} L^{2(n-1)}\alpha_n (\partial \phi)^{2n}\, ,\\ \label{qb}
\Q_\beta&=  \sum_{m=0} L^{2(m+1)}\beta_m (\partial \phi)^{2m} \left[(3+2m)R^{bc}\partial_b\phi \partial_c \phi -(\partial \phi)^2 R \right]\,,
\end{align}
where  the $\alpha_n$ and $\beta_m$ are dimensionless constants. 
Remarkably, the above theory admits an analytic generalization of the charged BTZ solution of the form  \cite{Bueno:2021krl}
\begin{equation}\label{fEQT2}
f(r)= \frac{\displaystyle \left[ \frac{r^2}{L^2}-\lambda- \alpha_1 p^2 \log\left( \frac{r}{r_0}\right)+\sum_{n=2}\frac{\alpha_n L^{2(n-1)}p^{2n}}{2(n-1)r^{2(n-1)}} \right]}{\displaystyle \left[ 1+ \sum_{m=0} \beta_m(2m+1)\left(\frac{pL}{r}\right)^{2(m+1)} \right]} \, , \quad \diff \phi = p \diff \varphi \, . 
\end{equation}
In the following subsections we perform a detailed study of the different types of black holes described by the above solution according to their horizon and singularity structure. Before doing so, let us make a few comments regarding the connection between the three-dimensional EQT theories considered here and similar higher-curvature generalizations of Einstein gravity constructed in higher dimensions.  

Such higher-dimensional models admit multiparametric generalizations of the Schwarzschild black hole, prototypically characterized by a single metric function whose equation of motion can be integrated once, leading either to an algebraic equation---in which case we refer to the theories as \emph{Quasi-topological gravities} (QT)---or to a second-order differential equation---in which case we refer to them as  \emph{Generalized Quasi-topological gravities} (GQT) \cite{Oliva:2010eb,Myers:2010ru,Dehghani:2011vu,Bueno:2016xff,Hennigar:2016gkm,Bueno:2016lrh,Hennigar:2017ego,Hennigar:2017umz,Bueno:2017sui,Ahmed:2017jod,Feng:2017tev,Bueno:2017qce,Bueno:2019ycr,Cisterna:2017umf,Arciniega:2018fxj,Cisterna:2018tgx,Bueno:2019ltp,Frassino:2020zuv,KordZangeneh:2020qeg,Bueno:2022res,Moreno:2023rfl}. As argued in \cite{Cano:2020ezi,Cano:2020qhy,Cano:2022ord}, matter fields can be easily introduced in this framework. The idea boils down to non-minimally coupling gravity to a $(D-2)$-form field-strength $F$. Considering a ``magnetic'' configuration of $F $ in which it is proportional to the volume form of the space transverse to $(t,r)$,\footnote{For instance, in four dimensions, $F= p \sin \theta \diff \theta \wedge \diff \phi$.} the equation of motion of $F$ is automatically solved.
On the other hand, when $F$ is included, it modifies the relevant higher-curvature corrections, which now involve contractions of $F$ and the Riemann tensor, and one is led to the notion of \emph{Electromagnetic Quasi-topological gravities} (EQT) and \emph{Electromagnetic Generalized Quasi-topological gravities} (EGQT) respectively, depending again on the algebraic/differential character of the equation of $f(r)$. 
 In three dimensions, the $1$-form field strength is nothing but  $\diff \phi$, and the magnetic ansatz is just $\diff \phi = p\, \diff \varphi$, as   shown  in \req{eq:SSSM}.  
 
 An important difference with respect to higher-dimensional models is that neither QT nor GQT gravities exist in three dimensions \cite{Bueno:2022lhf}.  Three-dimensional EQT gravities in turn present some important differences with respect to their higher-dimensional counterparts. First of all they only involve terms in the action which are linear in the curvature and including higher-curvature terms necessarily leads to EGQT theories \cite{Bueno:2022ewf}. On the other hand, in $D=4$ there exist EQTs involving terms of arbitrarily high orders in curvature, and the same applies in $D\geq 5$, both in the QT and EQT cases. In all such higher-dimensional models, the algebraic equation which determines $f(r)$ includes powers of that function up to the same order as the highest-order curvature density appearing in the action. This in turn prevents one from solving the equation of $f(r)$ explicitly, except for the lowest-order cases. This contrasts with the three-dimensional EQT theories considered here, for which $f(r)$ can be obtained explicitly in the most general case, which makes them, in this respect, rather unique theories.

\subsection{Black hole types according to their horizon structure}\label{sec:number_roots}

Every positive real root of the numerator of $f(r)$ corresponds to a horizon and, by playing with the values of the couplings, it is possible to achieve solutions which include several of them. In this subsection we present some results regarding the existence of horizons for the metric defined by \eqref{eq:SSSM} and \eqref{fEQT2}. Throughout this subsection we define $x \equiv (r/L)^2$, which simplifies the analysis.    We are interested in the real and positive roots of $f(x)$, which correspond to one horizon each. Indeed, for each positive root, $f(x_i)=0$, $x_i>0$, there are two possible values of $r_i=\pm  L \sqrt{x_i}$, but only the positive one corresponds to a horizon.\footnote{For now we just ignore the additional root at $r=0$ which takes place for regular black holes of the (\ref{regui}) type. We comment on this later, particularly in Section \ref{conc}. } 

Due to the fact that the known results about the existence of roots are stated for polynomials of a given degree, let us consider the case $\alpha_1 =0$ first. Taking the numerator of $f(r)$ as in \eqref{fEQT2} and multiplying by $x^{n_\text{max}-1}$ we find an expression of the form
\begin{equation}
    g(x)\equiv x^{n_\text{max}}-\lambda  x^{n_\text{max}-1}+\sum_{n=2}^{n_\text{max}} \alpha_n h(n)x^{(n_\text{max}-n)}\, .
\end{equation}
where we have defined $h(n)\equiv p^{2n}/(2(n-1))\geq 0$. It is clear that for a given $n_\text{max}$, the polynomial $g(x)$ has a degree $ n_\text{max}$ and therefore there can be $ n_\text{max}$ roots  at most. According to Descartes' rule of signs, the number of positive roots of a given polynomial  with real coefficients which have been arranged by descending variable exponents---such as $g(x)$ with $\alpha_1=0$ above---either equals the number of sign changes between consecutive nonvanishing coefficients, or it is less than it by an even number.\footnote{A root of multiplicity $k$ is counted as $k$ roots.} Hence, if $\lambda>0$ and we take all the couplings to be positive, $\alpha_{n\geq 2} \geq 0$, $g(x)$ has either 2 positive roots, or none. On the other hand, if $\alpha_{n\geq 2} \geq 0$, and $\lambda<0$, then the polynomial $g(x)$ has no changes of sign and therefore no positive real roots exist. Naturally, for a fixed sign of $\lambda$, the number of roots of the polynomial $g(x)$ depends on all the signs of the $\alpha_{n\geq 2}$. We can carry out a few cases explicitly.
 Assume first that $\lambda>0$ and $\alpha_{n\geq2}>0$. Then, the number of roots of the polynomial $g(x)$ is either $2$ or zero. If we modify one of the $\alpha_{j>2}$ so that it becomes negative, the number of roots will be $4$, $2$ or $0$ if $j<n_\text{max}$ and $3$ or $1$ if $j=n_\text{max}$. If another $\alpha_k$ becomes negative, there are two possibilities. If $k=j+1$ or $k=j-1$, the maximum number of roots remains the same. On the other hand, if $\alpha_k$ is not adjacent to $\alpha_j$, then the maximum number of roots will increase by $2$, except if $k=k_\text{nmax}$, in which case it will increase by $1$. The procedure can be repeated every time an $\alpha_i$ flips sign. The discussion is analogous for $\lambda<0$.



\begin{figure}[!t]
\begin{center}
\includegraphics[width=\textwidth]{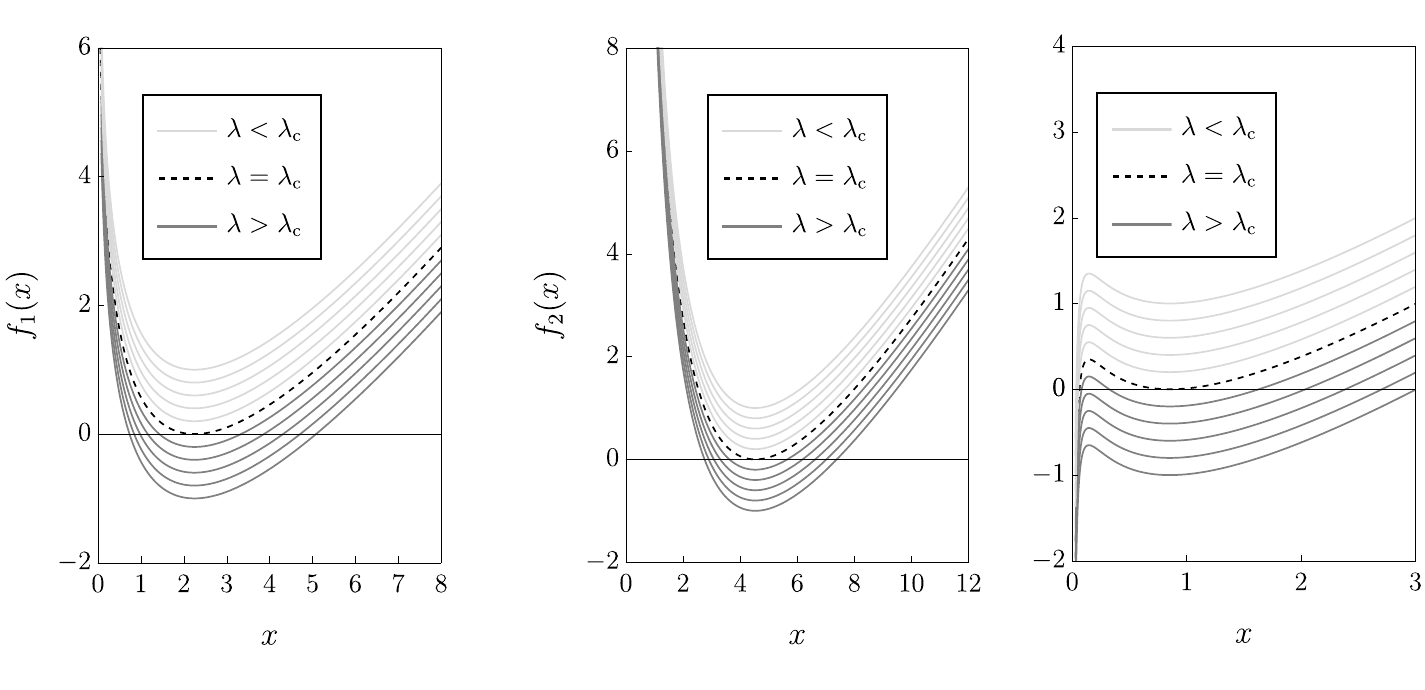}
\caption{We plot the functions $f_1(x)$ and $f_2(x)$ for different values of $p$ and $\lambda$. The gray lines represent curves with $\lambda=\lambda_{\rm c}+n/5$ while the light gray ones correspond to $\lambda=\lambda_c-n/5$, with $n=1,\ldots,5$. (Left) We plot $f_1(x)$ setting $p=3$ and $\lambda_{\rm c}=p^2/4\left[1-\log\left(p^2/4 \right)\,  \right]$. (Center and Right) We plot $f_2(x)$ using $p=2$ and $\lambda_{\rm c}=p^2/4\left[\sqrt{1+32\alpha_2}-\log\left(p^2/8\left[1+\sqrt{1+32\alpha_2}\right] \right)\,  \right]$ with $\alpha_2=2$ and $\alpha_2=-1/64$ in each plot, respectively.}
\label{fig:f1roots}
\end{center}
\end{figure}

Because of the presence of the logarithmic term, things change when $\alpha_1 \neq 0$. Consider first the charged BTZ case, for which $\alpha_1=1/2$ and $\alpha_{n\geq 2}=0$.  Then we have
\begin{equation}
    f_1(x)=x-\lambda-\frac{p^2}{4}\log x\,.
\end{equation}
The function $f_1(x)$ reaches a minimum at $x_\text{min}=p^2/4$. 
Now, it is easy to see that  $f_1(x_\text{min})$ vanishes for \cite{Martinez:1999qi}
\begin{equation}
    \lambda=\lambda_{\rm c}\, \quad \text{where}\quad \lambda_{\rm c}\equiv \frac{p^2}{4}\left[1-\log\left(\frac{p^2}{4} \right)\,  \right]\, ,
\end{equation}
which delimits the separation, in parameter space, between solutions with 2 and 0 horizons. Namely, whenever $\lambda > \lambda_{\rm c}$ the solution is a black hole with two horizons. These merge into a single one for  $\lambda = \lambda_{\rm c}$, which corresponds to an extremal black hole. Finally, no horizons exist for $\lambda < \lambda_{\rm c}$. In Figure~\ref{fig:f1roots} we have plotted $f_{1}(x)$ for different values of $\lambda$ and  fixed $p=3$. The value $\lambda=\lambda_c\approx 0.425407$ defines a curve that has one horizon and divides the graph in two regions: the one where the solutions have no horizon ($\lambda<\lambda_c$) and the other where the solutions have two horizons ($\lambda>\lambda_c)$.



Consider next turning on $\alpha_2$. The metric function reads
\begin{equation}\label{f2}
 f_2(x)=x- \lambda  -\frac{p^2}{4} \log x +\frac{\alpha_2 p^4}{2x}\,.
\end{equation}
This function has extrema for 
\begin{equation}
    x_{\pm}= \frac{p^2}{8} \left[1\pm\sqrt{1+32\alpha_2} \right] \, .
\end{equation}
Then, $x_+$ is positive whenever $\alpha_2\geq -1/32$ and $x_-$ is positive whenever $0>\alpha_2\geq -1/32$. For $\alpha_2< -1/32$,  $f_2(x)$ is therefore monotonously increasing with $x$ which, along with the fact that $\lim_{x\rightarrow0} f_2 =-\infty$, means that the solution possesses a single horizon. For $\alpha_2\geq 0$, only $x_+$ is positive, and it always corresponds to a minimum of $f_2(x)$, since $\lim_{x\rightarrow0} f_2 =+\infty$ in that case. Now, $f_2(x_+)$ vanishes for
\begin{equation}
    \lambda=\lambda_{\rm c}\, \quad \text{where}\quad \lambda_{\rm c}\equiv \frac{p^2}{4}\left[\sqrt{1+32\alpha_2}-\log\left(\frac{p^2}{8}\left[1+\sqrt{1+32\alpha_2}\right] \right)\,  \right]\, ,
\end{equation}
which again delimits the separation between solutions with two and zero horizons. The former occurs whenever  $\lambda>\lambda_{\rm c}$ and the latter whenever $\lambda<\lambda_{\rm c}$. The extremal case, with a single horizon, corresponds to $\lambda=\lambda_{\rm c}$. In the left panel of Figure~\ref{fig:f1roots} we have plotted $f_2(x)$ with $\alpha_2=2$ and $p=3$ fixed. It is clear that the dashed curve is the boundary that divides both zones, the one where $f_2(x)$ has no horizons and the zone where it has $2$ horizons.
Finally, when $0>\alpha_2\geq -1/32$, both $x_+$ and $x_-$ are positive and  $\lim_{x\rightarrow0} f_2 =-\infty$. Then there must be at least a horizon.  We have several options. If $f_2(x_{+})$ and $f_2(x_{-})$ have different signs then we have three horizons. If $f_2(x_{+})=0$ then $f_2(x_{-})$ can be positive, in which case we have two horizons, or negative, in which case we have a single horizon. Finally, when both $f_2(x_{+})$ and $f_2(x_{-})$ are negative we have  a single horizon. In the center and right panels of Figure~\ref{fig:f1roots}  we have plotted $f_{2}(x)$ with fixed $\alpha_2=-1/64$ and $p=2$. Again, the case $\lambda=\lambda_c$ (the dashed curve) is the boundary of both zones. The light gray curves have only one horizon and the darker gray ones have one, two, or three horizons, depending on the values of $\lambda$. 

\begin{figure}[!t]
\begin{center}
\includegraphics[width=\textwidth]{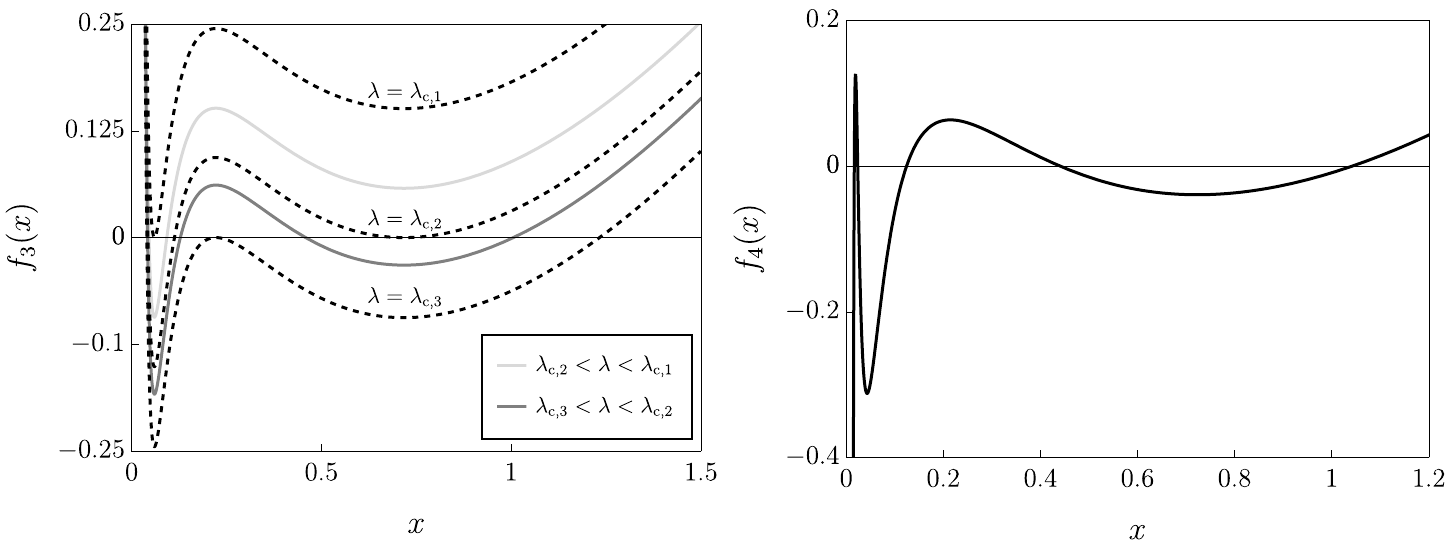}
\end{center}
\caption{We plot the metric functions $f_3(x)$ and $f_4(x)$ using $p=2$. (Left) We show $f_3(r)$ for $\alpha_2=-0.027$ and $\alpha_3=0.0003$ using different values of $\lambda$. The dashed lines correspond to critical values of $\lambda$ that degenerate two of the roots in the polynomial into one, namely, $\lambda=\lambda_{\text c,1}=0.606732$, $\lambda=\lambda_{\text c, 2}=0.757761$, and $\lambda=\lambda_{\text c, 3}=0.8515$. In addition, we include two curves with $\lambda_{\text c, 2}<\lambda=0.7<\lambda_{\text c, 1}$ and $\lambda_{\text c, 3}<\lambda=0.79<\lambda_{\text c, 2}$ corresponding to two and four horizons, respectively. (Right) We plot $f_4(x)$ for $\alpha_2=-0.0268$, $\alpha_3=0.000326$ and $\alpha_4=-0.000001$ as well as $\lambda=0.8$. For these values we have a five-horizon black hole.}
\label{fig:f3f4}
\end{figure}

By turning on more $\alpha$'s a richer horizon structure arises. This can be readily verified by introducing, for instance, $\alpha_3$ and adjusting the parameter values such that the corresponding polynomial $f_3(x)$ has real positive roots. The explicit expression is given by 
\begin{equation}
    f_3(x)=x-\lambda-\frac{p^2}{4}\log x+\frac{\alpha_2p^4}{2x}+\frac{\alpha_3p^6}{4x^2}\, .
\end{equation}
When $\alpha_2<0$ and $\alpha_3>0$, the derivative $f'_3(x)$ has either three or one real root, making it possible to have up to four horizons. For instance, taking $p=2$, $\alpha_2=-0.027$, and $\alpha_3=0.0003$, the equation $f'_3(x)=0$ has three real solutions, leading to three critical values of $\lambda$: $\lambda_{\text{c}, 1} \simeq 0.6067$, $\lambda_{\text{c}, 2} \simeq 0.7578$, and $\lambda_{\text{c}, 3} \simeq 0.8515$. For $\lambda > \lambda_{\text{c}, 1}$, there are no horizons. A single-horizon extremal case occurs when $\lambda = \lambda_{\text{c}, 1}$. In the range $\lambda_{\text{c}, 1} < \lambda < \lambda_{\text{c}, 2}$, the function has two horizons. Another extremal case, with three horizons, arises when $\lambda = \lambda_{\text{c}, 2}$. When $\lambda_{\text{c}, 2} < \lambda < \lambda_{\text{c}, 3}$, the function develops four horizons. Finally, for $\lambda > \lambda_{\text{c}, 3}$, the solution has a single horizon. This discussion is illustrated in Figure~\ref{fig:f3f4} (Left).  

From these examples it seems clear that by turning on couplings up to $n_{\rm max}$, it is always possible to find solutions with up to $n_{\rm max} + 1$ horizons. As a final example, in Figure~\ref{fig:f3f4} (Right) we plot a case with 5 horizons corresponding to $n_{\rm max}=4$.

\subsection{Black hole types according to their singularity structure}
\begin{figure}[!t]
 \includegraphics[width=0.357\linewidth]{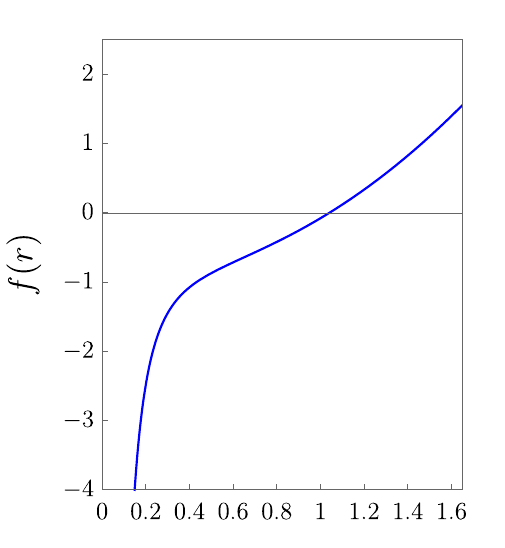}  
    \includegraphics[width=0.313\linewidth]{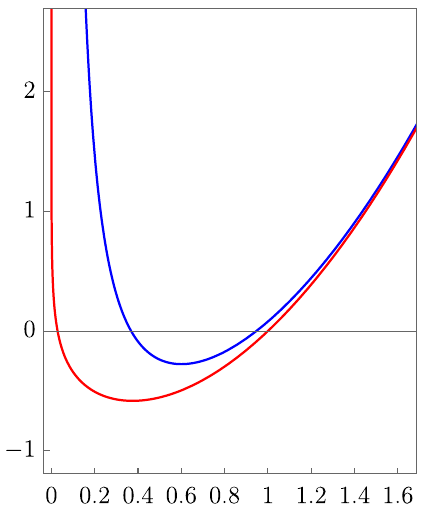}  
    \includegraphics[width=0.315\linewidth]{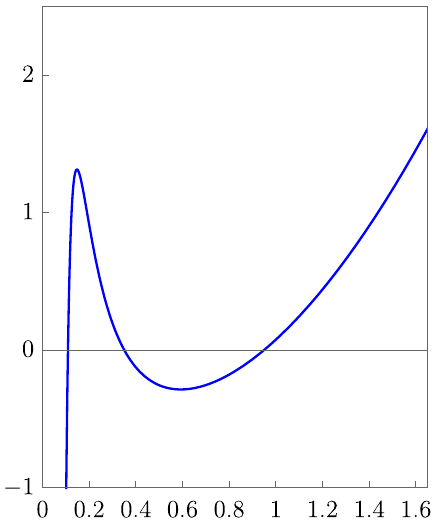}  
    \caption{We show the metric function $f(r)$ for black holes with a curvature singularity at $r=0$ and one, two---the red curve corresponds to the charged BTZ---and three horizons respectively.}
    \label{fig:curvsing}
\end{figure}

\begin{figure}[!t]
 \includegraphics[width=0.352\linewidth]{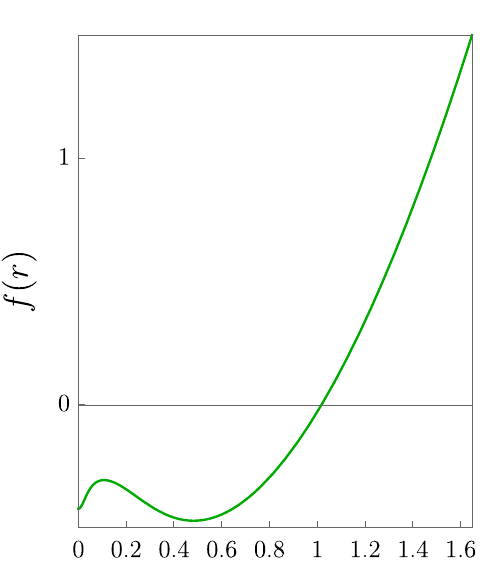}  
    \includegraphics[width=0.315\linewidth]{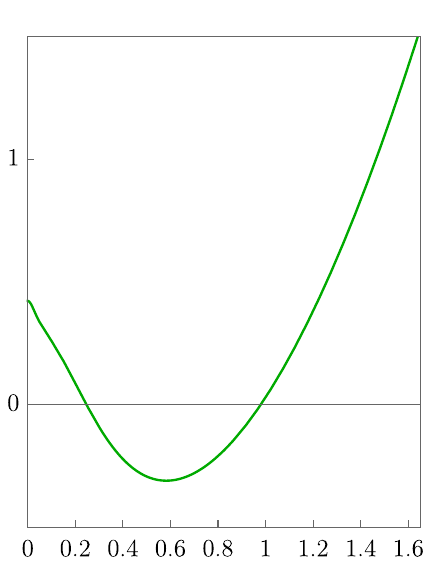}  
    \includegraphics[width=0.312\linewidth]{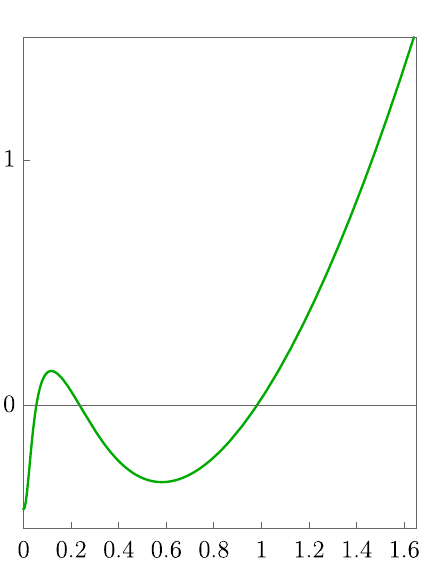}  
       \caption{We show the metric function $f(r)$ for black holes with a conical or pseudo-conical (BTZ-like) singularity at $r=0$ and one, two and three horizons.}
    \label{fig:BTZlike}
\end{figure}

The solutions described by \req{eq:SSSM} and \eqref{fEQT2}  may or may not exhibit a singularity at $r=0$. In $D=3$ there are only three independent curvature invariants which can be built from contractions of the Riemann tensor and the metric \cite{Paulos:2010ke}. For spacetimes of the form \req{eq:SSSM} only two of them remain independent \cite{Bueno:2022lhf}. They can be chosen to be $R$ and $S_a^bS_b^a$, where $S_{ab}\equiv R_{ab}-\frac{1}{3} g_{ab} R$ is the traceless Ricci tensor. In the present case, they read
\begin{align}
\left. R\right|_f=-\frac{(2f'+r f'')}{r}\,,\quad 
\left.S_a^bS_b^a\right|_f= \frac{(f'-r f'')^2}{6r^2}\, .
\end{align}
Hence, regular black holes are defined by the absence of divergences in both invariants at any point. In Figures~\ref{fig:curvsing}, \ref{fig:BTZlike} and \ref{fig:reg} we illustrate the three main classes of black holes---with three subcases each---which will be the main focus of this paper. Let us say a few things about each of them.

\subsubsection*{Black holes with curvature singularities at $r=0$}\label{sbhs}
The first class of black holes we are interested in are those which include a curvature singularity at $r=0$. 
This is the case of black holes  for which $n_{\rm max}> m_{\rm max}+2$ or $n_{\rm max}=1$, $\beta_m=0$ $\forall m$ (namely, the charged BTZ) \cite{Bueno:2021krl}.  The curvature invariants diverge at  $r=0$ as
\begin{align}
&\left.R\right|_f\overset{(r\rightarrow 0)}{=}-\frac{(5-2\nmax-2\mmax)(2-\nmax+\mmax) \alpha_{ n_{\rm max}}}{(1+2\mmax)(\nmax-1)L^2\beta_{\mmax}}\left(\frac{p L}{r}\right)^{2(n_{\rm max}-m_{\rm max}-1)}\,,\\ \notag 
&\left.S_a^bS_b^a\right|_f\overset{(r\rightarrow 0)}{=}\frac{(1-\nmax+\mmax)^2\left(1+2[\mmax+(2-\nmax)]^2\right)\alpha_{\nmax}^2}{(\nmax-1)^2(2\mmax+1)^2L^4\beta_{\mmax}^2 }\left(\frac{p L}{r}\right)^{4(n_{\rm max}-m_{\rm max}-1)}\, .
\end{align}
There are a couple of special cases corresponding to $n_{\rm max}=\{1,2\}$, $m_{\rm max}=0$. For these, the metric function does not diverge at $r=0$, but the curvature invariants diverge logarithmically at that point \cite{Bueno:2021krl}. As far as the singularity is concerned, this type of solutions is the one which resembles the most the situation encountered in the usual higher-dimensional cases---although note that here the curvature is Ricci curvature, whereas \eg for the Schwarzschild black hole in $D\geq 4$, the above invariants simply vanish as it is a Ricci-flat spacetime.

In Figure~\,\ref{fig:curvsing} we present instances of this type of black holes corresponding to $\alpha_1 =1/2$, $\beta_m=0$ $\forall m$, $\alpha_n =0$ $\forall n \geq 4$, and: $\alpha_2=-1/2$, $\alpha_3=0$ (one horizon); $\alpha_2=1/2$, $\alpha_3=0$ (two horizons); $\alpha_2=1/2$, $\alpha_3=-1/50$ (three horizons). In all cases, we set $L=1,\lambda=1,p=3/4$. Solutions with an arbitrary number of horizons can also be found, as explained earlier, as long as we turn on a sufficiently large number of couplings.

\subsubsection*{Black holes with conical/BTZ-like singularities}\label{conicalbhs}
Another class of solutions described by \req{fEQT2} corresponds to black holes for which $f(r)$ tends to a constant value at $r=0$ and  the $(r,\varphi)$ components of the metric behave as
\begin{equation}\label{conic}
    \mathrm{d}s_{(r,\varphi)}^2\overset{(r\rightarrow 0)}{=}\pm \frac{\mathrm{d}r^2}{\left[1-\frac{\vartheta}{2\pi} \right]^2}+r^2 \mathrm{d}\varphi^2\, .
\end{equation}
This describes a conical singularity with deficit angle $\vartheta$ for the plus-sign case, and a causal-structure singularity similar to the Taub-NUT one \cite{Banados:1992gq} for the minus-sign one. Both situations occur in the case of the neutral BTZ solution, which corresponds to setting $\alpha_1=\alpha_n=\beta_m=0$ $\forall \,n,m$ in \req{fEQT2}. In that case, the mass parameter is related to the deficit angle by $\vartheta =2\pi (1- \sqrt{|\lambda|})$, and the solution describes black holes for $\lambda >0$ and naked singularities for $\lambda<0$ except for $\lambda=-1$, which corresponds to pure AdS$_3$. For more general values of the gravitational couplings, singularities of this kind arise whenever $n_{\rm max}=m_{\rm max} +2$ \cite{Bueno:2021krl}. In that case, when $\alpha_{n_{\rm max}}$ and $\beta_{m_{\rm max}}$ have the same sign, the solution has a conical singularity, whereas when they have opposite signs, the solution has a causal-structure singularity instead and \req{conic} holds with the minus sign. 

In Figure~\ref{fig:BTZlike} we present instances of this type of solutions corresponding to $\beta_0=1/5$, $\beta_1=1/4500$, $\beta_m=0$  $\forall m\geq 2$, $\alpha_1=1/2$, $\alpha_n =0$ $\forall n \geq 4$, and: $\alpha_2=-1/5$, $\alpha_3=-1/500$ (one horizon); $\alpha_2=1/5$, $\alpha_3=1/500$ (two horizons); $\alpha_2=1/5$, $\alpha_3=-1/500$ (three horizons).  
In all cases, we set $L=1,\lambda=1,p=3/4$.

\begin{figure}[!t]
  \includegraphics[width=0.35\linewidth]{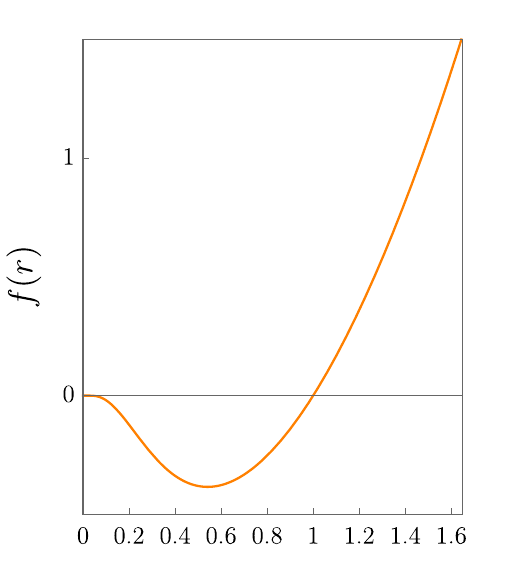}  
    \includegraphics[width=0.31\linewidth]{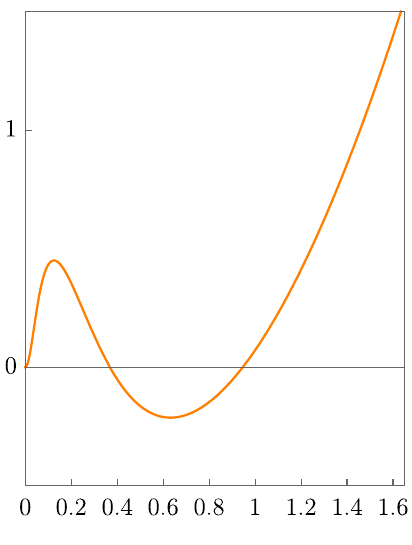}  
    \includegraphics[width=0.315\linewidth]{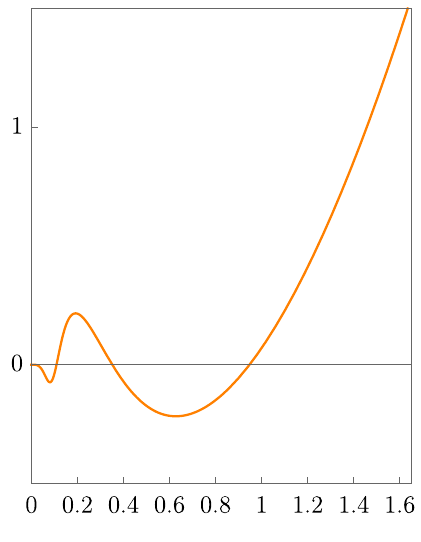}  
   \caption{We show the metric function $f(r)$ for regular black holes such that $f(r)\sim \mathcal{O}(r^{2s})$, $s\geq 1$ at $r=0$ and one, two and three horizons.}
    \label{fig:reg}
\end{figure}

\subsubsection*{Regular black holes with $f(r)\overset{r\sim 0}{=} \mathcal{O}(r^{2s})$, $s\geq 1$}\label{rbhs}
Regular black holes are black holes for which all curvature invariants remain finite everywhere. In general $D\geq 4$ spacetime dimensions, static and spherically symmetric regular black holes are characterized by metrics such that $f(r)=1+\mathcal{O}(r^2)$ at $r=0$. These kinds of solutions exist in $D=3$ as particular cases of black holes of the class presented in the previous subsection, corresponding to situations in which $\beta_{n_{\rm max}-2}$ is fixed in terms of $\alpha_{n_{\rm max}}$ in a way such that the condition holds---see Section~\ref{special} below. Therefore, regular black holes of this kind are not generic in the present context, but they require some degree of fine-tuning.\footnote{As shown in \cite{Bueno:2024dgm}, regular black holes of this type can be in fact generic in $D\geq 5$ scenarios in which the Einstein action is supplemented by infinite towers of higher-curvature corrections---see also Section~\ref{sec:non-local}.} However, as argued in \cite{Bueno:2021krl}, a different and previously unnoticed class of regular black hole spacetimes exists in three dimensions. This corresponds to metric functions for which
\begin{equation}\label{freg}
    f(r)\overset{(r\rightarrow 0)}{=} \mathcal{O}\left(r^{2s}\right)\, , \quad s \geq 1\, .
\end{equation}
For those, the curvature invariants behave as
\begin{equation}
R\overset{(r\rightarrow 0)}{=} \mathcal{O}\left(r^{2(s-1)}\right)\,,\quad 
S_a^bS_b^a\overset{(r\rightarrow 0)}{=} \mathcal{O}\left(r^{4(s-1)}\right)\, ,
\end{equation}
which therefore either vanish or take a constant value at $r=0$. Observe that curvature invariants in $D\geq 4$ for spherically symmetric spacetimes with a metric function behaving as \req{freg} always involve contributions which diverge at least as $1/r^2$ at $r=0$. The key difference is that a $(D-2)$-sphere of radius $r$ has an intrinsic curvature which grows as $r\rightarrow 0$ in $D\geq 4$---and diverges for $r=0$---whereas it just vanishes in the case of a circle, namely, for $D=3$. Note also that, as opposed to the $f(r)\rightarrow 1+\mathcal{O}(r^2)$ case, regularity can be achieved in this case for spacetimes with an odd number of horizons---including the case of a single horizon---which may avoid issues related to the instability of inner horizons \cite{Carballo-Rubio:2018pmi,Carballo-Rubio:2024dca}.\footnote{We thank Pablo A. Cano for this observation.}
Regarding this, it is important to clarify what exactly is the nature of the $r=0$ locus for this class of solutions. In the following sections we study this issue through the analysis of the Penrose diagrams and geodesic structure of the solutions, and additional comments are made in the conclusions. In the present context, regular black holes of this class occur whenever $m_{\rm max}> n_{\rm max}-2$ if $n_{\rm max} \geq 2$ or whenever $m_{\rm max} \geq 1$ if $n_{\rm max}=1$.\footnote{To be more precise, the leading behavior of the regular solutions with $n_{\text{max}}=1$ and $m_{\text{max}}\geq 1$ has an extra logarithmic factor, such that $$f(r)\overset{(r\rightarrow 0)}{=} \mathcal{O}\left(r^{2s}\log r\right)\, , \quad s \geq 1\, .$$ This produces a similar logarithmic correction to the curvature invariants, as compared to the general regular case with $m_{\text{max}}> n_{\text{max}} -2$ and $n_{\text{max}}\geq 2$.} Observe that these conditions are generic and do not require any kind of fine tuning between different couplings. 

Once again, here we focus on black holes of this class including one, two and three horizons---see Figure~\ref{fig:reg}. The instances shown there correspond to $\beta_0=1/5$, $\beta_1=1/1800$, $\beta_m=0$  $\forall m\geq 4$, $\alpha_1 =1/2$, $\alpha_n =0$ $\forall n \geq 4$, and: $\beta_2=0$, $\beta_3=0$, $\alpha_2=0$, $\alpha_3=0$ (one horizon); 
$\beta_2=0$, $\beta_3=0$, $\alpha_2=1/2$, $\alpha_3=0$ (two horizons); 
$\beta_2=1/250000$, $\beta_3=1/4900000$, $\alpha_2=1/2$, $\alpha_3=-1/50$ (three horizons).  
In all cases, we set $L=1,\lambda=1,p=3/4$.

\subsubsection*{Black holes with finite-volume singularities}\label{fvs}
There is in fact an additional generic case which was not considered explicitly in \cite{Bueno:2021krl}. This corresponds to black holes which contain a curvature singularity which occurs at finite volume, \ie at some  finite value of $r$. Solutions of this kind feature in higher-dimensions \eg for Lovelock gravities---see \cite{Bueno:2024fzg,Bueno:2024qhh}. In that case, such singularities are associated to $f(r)$ becoming complex when trying to cross beyond the singularity locus. $f(r)$ takes a finite value there, but the curvature invariants diverge. In the present case, on the other hand, they occur at the largest positive zero of the denominator of $f(r)$, which we denote by $r_\star$. Provided the putative horizon radius is greater than $r_\star$, the solution describes a black hole. Otherwise, it is a naked singularity. Observe that this situation typically takes place whenever some of the $\beta_m$ have negative values. The most general case is the one in which $f(r)$  has a degenerate pole of order $\gamma$, at $r_\star$, $f(r)\overset{(r\rightarrow r_\star)}{\sim} 1/ (r-r_{\star})^\gamma$,  which is also the largest of its positive  poles. Then, the curvature invariants diverge as
\begin{equation}
    \left. R\right|_f \overset{(r\rightarrow r_\star)}{\sim} \frac{1}{(r-r_\star)^{\gamma+2}}\, ,\quad \left.S_a^bS^a_b\right|_f \overset{(r\rightarrow r_\star)}{\sim} \frac{1}{(r-r_\star)^{2(\gamma+2)}}\, .
\end{equation}
The generic case corresponds to $\gamma=1$, which occurs, for instance, at $r_\star=p L \sqrt{|\beta_0|}$ in the simplest possible situation, corresponding to $\beta_0<0$, $\beta_m=0$ $\forall m\geq 1$.  It would be interesting to study these kinds of solutions in more detail, but in this paper we will focus on the three previous types. 

\subsection{Special cases}\label{special}
The cases presented in the previous subsections correspond to generic classes of solutions which occur for certain values of the couplings, as long as they remain within certain ranges. On the other hand, by imposing constrains which relate numerically the values of pairs of couplings with one another---\ie by fine tuning them---it is possible to achieve ``special'' types of solutions. We will pay some attention to these, particularly in Section~\ref{penro}, so  let us say a few things about them.


\subsubsection*{Extremal black holes}
An obvious non-generic case corresponds to extremal black holes, namely, solutions for which some of the horizons are degenerate, \ie they are such that there exist certain $r_+>0$ such that  $f(r_+)=f'(r_+)=0$. The simplest instance of this situation in the present framework takes place for $\alpha_1=\alpha_{n\geq 3}=0$ in the special point in parameter space for which $\alpha_2=\lambda^2/(2p^4)$.

\subsubsection*{Regular black holes with $f(r)\overset{r\sim 0}{=} 1+\mathcal{O}(r^2)$}
Another relevant non-generic case occurs for solutions behaving as $f(r)\rightarrow 1+\mathcal{O}(r^2)$ at $r=0$. This corresponds to regular black holes of a more familiar type than the ones described in subsection~\ref{rbhs}. In the present context, black holes of this type occur as particular instances of the family which includes conical/BTZ-like singularities. For instance, if $\alpha_j$ and $\beta_{j-2}$ are the only active couplings, the black hole belongs to this class if the condition
\begin{equation}
  \beta_{j-2} = \frac{ \alpha_j p^2}{2(j-1)(2j-3)}  \, 
\end{equation}
    holds. Previous regular black holes constructed in $D=3$ not belonging to the new class presented in \cite{Bueno:2021krl}---\eg those in \cite{Cataldo:2000ns,He:2017ujy,HabibMazharimousavi:2011gh}---would be of this type. In all known cases, in order for such solutions to exist, a certain degree of fine-tuning between action parameters and integration constants of the solutions is required in order to achieve regularity.

\section{Black holes in non-local Electromagnetic Quasi-topological gravity}\label{sec:non-local}

As we saw in Section~\ref{rbhs}, some of the EQT theories posses regular black holes as their only vacuum spherically symmetric solutions---see also Section~\ref{birk}. The fact that it is possible to tame the charged BTZ black hole singularity with local effective theories, consisting of a finite number of densities and linear in the curvature, is remarkable. In higher dimensions, a similar feature takes place within the corresponding EQT framework (although in that case the corrections involve higher-curvature terms), as shown in \cite{Cano:2020ezi,Cano:2020qhy}. On the other hand, most of the proposed ultraviolet completions of Einstein gravity, such as the one attained by string theory, involve a non-local gravity sector. Following a bottom-up approach, theories involving infinite towers of QT densities have been constructed which generically lead to regular black hole solutions in \cite{Bueno:2024dgm}. Previous instances of regular black holes constructed in $D=3$ at particular points in parameter space also take place for theories which have a non-local nature \cite{Cataldo:2000ns,He:2017ujy,HabibMazharimousavi:2011gh}.
In this section we consider the case in which our EQT theories involve infinitely many terms, giving rise to non-local Lagrangians. We will see that the analysis of the solutions is essentially identical to the case studied in the previous section. We will construct the corresponding black holes in a few explicit cases, including new instances of regular black holes and some solutions already studied in the context of non-linear electrodynamics.


\subsection{Dual frame}\label{dualf}
As argued in \cite{Bueno:2021krl}, we can consider a dual frame in which the dynamics is described in terms of an ``electric'' two-form field strength $F_{ab}$ instead of our ``magnetic'' scalar and in which the metric function takes the same form---see also \cite{Alkac:2021seh}.
This field strength is defined in terms of the scalar field as
\begin{equation}\label{F_phi}
F_{a b}=-\frac{1}{2}\epsilon_{a b c} \frac{\partial \mathcal{L}}{\partial (\partial_c \phi)} \, .  
\end{equation}
The dual description is thus obtained through the Legendre transform
\begin{equation}\label{Ldual}
\mathcal{L}_{\rm dual}=\mathcal{L}-F_{a b}\partial_c \phi \epsilon^{a b c}\, .
\end{equation}
However, in order to get the dual Lagrangian as a function of the field strength $\mathcal{L}_{\rm dual}(F)$ one needs to invert the relation (\ref{F_phi}) to find $\partial\phi(F)$. When any coupling constant besides $\alpha_1$ is turned on, this cannot be achieved exaclty, but only perturbatively in powers of the length scale $L$. Consequently, $\mathcal{L}_{\rm dual}(F)$ involves an infinite sum of terms of arbitrary high order in derivatives, coupling the field strength to the Ricci tensor. The first two orders read \cite{Bueno:2021krl}
\begin{equation}
\mathcal{L}_{\rm dual}=R+\frac{2}{L^2}-\frac{1}{2\alpha_1}F^2 + L^2\left[ -\frac{\alpha_2}{4\alpha_1^4}(F^2)^2+3\frac{\beta_0}{\alpha_1^2}F_a^{~c} F^{a b}R_{\langle c b\rangle}\right] +\mathcal{O}(L^4) \, , 
\end{equation}
where $R_{\langle c b\rangle}$ is the traceless part of the Ricci tensor. The relation (\ref{F_phi}) also shows that although the scalar acquires a very simple, vortex-like configuration (\ref{fEQT}), the expression for the electrostatic potential is way more involved
\begin{equation}\label{potential}
A_t(r)=-\alpha_1 p \log{\left(\frac{r}{r_0}\right)}+\sum_{n=2}\frac{n\alpha_n p}{2(n-1)}\left( \frac{L p}{r}\right)^{2(n-1)}+f'(r)L\sum_{m=0}\beta_m(m+1)\left( \frac{L p}{r}\right)^{2m+1}\, .
\end{equation}
Note that these are higher-order corrections on top of the first term, which is the Maxwell potential for an electric charge $p$ in three dimensions.

\subsection{Non-local EQT}

Having discussed the connection with the dual frame, in this section we explore EQT theories with infinitely many non-vanishing coupling constants $\alpha_n$, $\beta_m$. That is, we go a bit further and study non-local theories already in the scalar frame. To do so, we first note that the sums in (\ref{qa}) and (\ref{qb}) might be packed into a pair of analytic functions on the scalar kinetic term. More precisely, the infinite series could be resumed into the non-local action   
\begin{equation}\label{IEQTnon-local}
I_{\rm {EQT}}=\frac{1}{16 \pi G}\int {\mathrm d}^3x\sqrt{g}\left[ R- \mathcal{Q} \right]\, ,
\end{equation}
with 
\begin{equation}
\mathcal{Q}\equiv-\frac{2}{L^2}K(X)+ X F(X)R-\left[ 3 F(X)+2 X F'(X)\right]L^2 R^{a b}\partial_a\phi\partial_b\phi\, , 
\end{equation}
and $X\equiv L^2(\partial\phi)^2$ representing the kinetic term. Of course, by writing the functions $K(X)$ and $F(X)$ in power series, i.e.,
\begin{equation}
K(X)=\sum_{n=0} \alpha_n X^n\,,\quad F(X)=\sum_{m=0} \beta_m X^m\, .
\end{equation}
we recover \eqref{qa} and \eqref{qb} respectively, provided that $\alpha_0=1$ as this term is identified with the cosmological constant. With this simple parametrization, the metric function of the SSS solution is given by the compact expression
\begin{equation}\label{fEQTnon-local}
f(X)=\frac{-\lambda -p^2\int \text{d} X\frac{K(X)}{ X^2}}{1+X F(X)+2 X^2 F'(X)}\, , \quad X=\frac{L^2 p^2}{r^2}\, .
\end{equation}
From this approach, we may recover the particular solutions analyzed in Section~\ref{sec:black holes in EQTG} by taking $K(X)$ and $F(X)$ to be polynomial functions. For instance, the choice $K(X)=1$ and $F(X)=0$ trivially reduces \eqref{IEQTnon-local} to the Einstein action supplemented with a cosmological constant and hence $f(X)$ in \eqref{fEQTnon-local} reproduces the BTZ black hole \eqref{fEQT}. Likewise, $K(X)=1-\frac{1}{2}\alpha_1 X$ and $F(X)=0$ reduces to the Einstein-dilaton action presented in \eqref{eq:Edil}, with $f(X)$ matching the charged BTZ black hole described in \eqref{fEQT}. Nevertheless, if $K(X)$ or $F(X)$ are analytic functions with a power series expansion involving infinitely many terms, (\ref{fEQTnon-local}) gives the most general spherically symmetric solution to non-local  EQT theories in the scalar frame, sourced by an electric charge. 

\subsection{Examples of black holes in non-local EQT}

Interestingly, turning $F(X)$ off leads to dilaton theories minimally coupled to Einstein gravity, which in the dual frame correspond to non-linear electrodynamics  models. In that case, it is easy to obtain $f(r)$ for a given Lagrangian function $K(X)$ or, alternatively, to cook up a Lagrangian function such that a given metric function solves the corresponding equations of motion. Indeed, from \req{fEQTnon-local} with $F(X)\equiv 0$, we have
\begin{equation}\label{maqui}
    K(X)=-\frac{f'(X)X^2}{p^2}\, ,
\end{equation}
namely, given a metric function expressed in terms of $X$, we can straightforwardly obtain a non-local Lagrangian for the scalar field coupled to Einstein gravity such that $f(r)$ is automatically a solution. For example, we can recover the black hole presented in \cite{Cataldo:2000ns} from a theory with a resummed, non-perturbative expression. More concretely, using
\begin{equation}\label{cataldo1}
     K(X)=1-\frac{X}{1+\alpha  X}=\sum_{n=0}^\infty\alpha_{n} X^n\,,\quad \text{with }\quad \alpha_0=1\, ,\,  \alpha_{n\geq 1}=(-1)^n\alpha^{n-1} \,, 
\end{equation}
we arrive at the metric function
\begin{equation}\label{fCat}
f(r)=\frac{r^2}{L^2}-\lambda- p^2 \log{\left(\frac{r^2}{p^2 L^2}+ \alpha\right)}\, ,
\end{equation}
where we absorbed an integration constant in $\lambda$. Alternatively, starting from \req{fCat} and inserting it in \req{maqui}, we can derive $K(X)$ as in \req{cataldo1}.  The solution is identical to the one presented in \cite{Cataldo:2000ns} up to some trivial parameter redefinitions.
Note that this solution is not strictly regular at $r=0$, although a naive computation of the curvature invariants yields finite results. Following the terminology of Section~\ref{sec:black holes in EQTG}, for $\lambda>0$ this has a BTZ-like singularity. Moreover, according to (\ref{potential}), the radial component of the electric field $E_r=\partial_r A_t$ reads
\begin{equation}
E_r=\frac{- p~ K'(X)}{r}= \frac{ p ~r^3}{( r^2+\alpha p^2 L^2)^2}\, .
\end{equation}
This reflects a charge screening due to the self interaction of the electromagnetic field. The length scale of these interactions is $\sim \sqrt{\alpha } p L$, which plays the role of the length ``$a$'' in \cite{Cataldo:2000ns}. 
We again stress the convenience of working with the scalar description, since it is not possible to write a closed expression for the action in terms of the field strength at all distance scales. The first orders in powers of $L^2$ would be
\begin{equation}
\mathcal{L}_{\rm dual}=\frac{1}{8\pi}\left[F^2+(F^2)^2 \alpha L^2+9(F^2)^3 \alpha^2L^4\right]+\mathcal{O}(L^6)\, .
\end{equation}

Another interesting example is the one of Born-Infeld electrodynamics \cite{Cataldo:1999wr}. This corresponds to
\begin{equation}
    K(X)=(1+2 b^2)-2b^2\sqrt{1+\frac{X}{b^2}}=\sum_{n=0}^\infty\alpha_{n} X^n\, ,\quad \text{with} \quad \alpha_0=1\, , \alpha_{n\geq 1}= -2 b^{2-2n}\binom{1/2}{n} 
\end{equation}
giving
\begin{eqnarray}
f(r)&=&\frac{r^2}{L^2}(1+2 b^2)-\lambda-2 \frac{b^2}{L^2}r\sqrt{r^2+\frac{p^2 L^2}{b^2}}-2 p^2 \log{\left(r+\sqrt{r^2+ \frac{p^2 L^2}{b^2}}\right)}\, ,\\
E_r&=& \frac{p}{\sqrt{r^2+\frac{p^2L^2}{b^2}}}\, .
\end{eqnarray}
These reduce to the metric function and the electric field in \cite{Cataldo:1999wr} provided that $b/L$ is identified with the Born-Infeld parameter. This is a singular solution because the curvature invariants diverge at $r=0$ \cite{Cataldo:1999wr}.

Using \req{maqui} we can make further experiments. For instance, a version of the three-dimensional Hayward black hole \cite{Hayward:2005gi}, with metric function
\begin{equation}
    f(r)=\frac{r^2}{L^2}-\lambda- \frac{r^2 p^2}{r^2 + \alpha p^2 L^2}\, ,
\end{equation}
can be obtained for
\begin{equation}
    K(X)=1- \frac{\alpha X^2}{(1+\alpha X)^2}=\sum_{n=0}^{\infty}\alpha_n X^n\, , \quad \text{with} \quad \alpha_0=1\, , \alpha_{n\geq 1}=(n-1)(-\alpha)^{n-1}\, .
\end{equation}
This is a regular solution of the $f(r)\overset{(r\rightarrow 0)}{=}1+\mathcal{O}(r^2)$ type whenever $\lambda=-1$, which is obviously a non-generic situation. In the analogy with the higher-dimensional Hayward metric, the role of the mass parameter is played here by the electric charge $p$. Observe also that $K(X)$ does not contain a linear term, so the kinetic term for the scalar field is absent.

In order to find regular black hole solutions with $f(r)\overset{r\sim 0}{=} \mathcal{O}(r^{2s})$ we need to turn $F(X)$ on, that is, the dilaton must couple non-minimally to the curvature. We require that $f(X)\rightarrow 0$ as $X\rightarrow \infty$, which occurs whenever the denominator in (\ref{fEQTnon-local}) grows faster at infinity than the numerator. A simple example entails considering $K(X)$ as in \req{cataldo1}  and $F(X)=\beta$, leading to
\begin{equation}
f(r)=\frac{\frac{r^2}{L^2}-\lambda- p^2 \log{\left(\frac{r^2}{p^2 L^2}+ \alpha\right)}}{1+\frac{\beta L^2p^2}{r^2}}\, ,
\end{equation}
which behaves as $f(r)\sim r^2$ at the origin. We could also find a regular solution with a metric function going to zero faster than any polynomial by taking, for example,  the non-local density
\begin{equation}
 F(X)= \frac{\sinh{\beta X}}{\beta X}=\sum_{m=0}^{\infty}\beta_mX^m\, , \quad \text{with} \quad \beta_m= \frac{\beta^m(1+(-1)^m)}{2(m+1)!}\, ,  
\end{equation}
together with \req{cataldo1}. This gives
\begin{equation}
f(r)=\frac{\frac{r^2}{L^2}-\lambda- p^2 \log{\left(\frac{r^2}{p^2 L^2}+ \alpha\right)}}{1+\frac{2 L^2p^2}{r^2} \cosh{(\beta L^2 p^2/r^2)}-\frac{1}{\beta}\sinh{(\beta L^2 p^2/r^2)}}\, ,    
\end{equation}
which behaves as $f(r)\sim r^2 ~\text{e}^{-\beta p^2 L^2/r^2}$ near $r=0$.
\section{Birkhoff theorem}\label{birk}
The aim of this section is to prove that the single function static metrics (\ref{eq:SSSM})  are in fact the most general spherically symmetric solutions of the Electromagnetic Quasi-topological gravities (\ref{IEQTnon-local}). Namely, we  will show that our gravitational theories, sourced by an electric charge, satisfy a Birkhoff theorem. This is a property that holds for very few higher-derivative gravity theories. In fact, no $D=4$ higher-derivative theory beyond GR has been shown to satisfy a Birkhoff theorem. In $D\geq 5$, only Lovelock theories and a certain subset of QT gravities have been shown to have this property \cite{Oliva:2010eb,Oliva:2011xu,Bueno:2024eig,Bueno:2024zsx}.

With this purpose in mind, let us first review how the equations of motion for an arbitrary higher-order gravitational theory look like. In the presence of matter sources with energy-momentum tensor
\begin{equation}
T_{a b}\equiv -\frac{2}{\sqrt{\vert g\vert}}\frac{\delta S_{\text{matter}}}{\delta g^{a b}}\, ,
\end{equation}
the full equations of motion read
\begin{equation}\label{fulleom}
\mathcal{E}_{a b}=8 \pi G T_{a b}\, ,
\end{equation}
where $\mathcal{E}_{a b}$ is the generalized Einstein tensor
\begin{equation}
\mathcal{E}_{a b}\equiv  \frac{16 \pi G}{\sqrt{\vert g\vert}}\frac{\delta I}{\delta g^{a b}}\, .
\end{equation}
Moreover, if the action is constructed from contractions of the Ricci tensor and the metric, that is 
\begin{equation}
I=\int \diff x \sqrt{\vert g\vert}\mathcal{L}(R_{a b},g^{a b})\, ,   
\end{equation}
then it follows that
\begin{eqnarray}
\mathcal{E}_{a b}&=&\left.\frac{\partial \mathcal{L}}{\partial g^{a b}}\right\vert_{R_{a b}}-\frac{1}{2}g_{a b}\mathcal{L}-\nabla_c\nabla_{(a}Q_{b)}^{~c}+\frac{1}{2}\dal Q_{a b}+\frac{1}{2}g_{a b} \nabla_c\nabla_d Q^{c d}\, ,\\ 
Q^{a b}&\equiv& \left.\frac{\partial \mathcal{L}}{\partial R_{a b}}\right\vert_{g^{a b}}\, .
\end{eqnarray}
Now, for the general EQT theory (\ref{IEQTnon-local}), we get 
\begin{eqnarray}
\left.\frac{\partial \mathcal{L}}{\partial g^{a b}}\right\vert_{R_{a b}}&=&(1- X F)R_{a b}+2(3 F+2 X F')L^2 R_{(a c}\partial_{b)}\phi \partial^c\phi-\partial_X \mathcal{Q}\partial_a\phi\partial_b\phi\, ,\\
Q^{a b}&=& g^{a b}(1-X F)+(3 X+2 X F') \partial^a\phi\partial^b\phi\, ,
\end{eqnarray}
which leads to
\begin{eqnarray}\nonumber
\mathcal{E}_{a b}&=&G_{a b}-g_{a b} \frac{K}{L^2}-X F R_{a b} + 2 (3 F+2 X F')L^2 \partial_{(a}\phi \partial^c \phi R_{b) c}\\ \nonumber
&& -\frac{1}{2}g_{a b} R^{c d}Q_{c d}- \nabla^c \nabla_{(a} Q_{b) c}+\frac{1}{2}\dal Q_{a b}+\frac{1}{2}g_{a b}\nabla_c\nabla_d Q^{c d}\\
&& -\left[-2\frac{K'}{L^2}+(F+X F')R -(5 F'+2 X F'')L^2 R^{c d}\partial_c\phi\partial_d\phi\right]\partial_a\phi\partial_b\phi L^2\, ,\\\label{magnetic_ansatz}
X&=& \frac{L^2 p^2}{r^2}\, ,
\end{eqnarray}
where $G_{a b}=R_{ab}-g_{ab}R/2$ is the Einstein tensor. In this expression we already imposed the magnetic ansatz for the scalar field, $\diff \phi = p \,\diff \varphi$, which automatically solves the scalar field equation \cite{Bueno:2021krl}.

Let us consider now a general spherically symmetric configuration for the metric, namely,
\begin{equation}
\diff s^2= -N(t,r) f(t,r) \diff t^2 + \frac{\diff r^2}{f(t,r) }+r^2 \diff \varphi^2\, .
\end{equation}
In this background, only the following three components of the Einstein tensor remain independent,
\begin{eqnarray}
\mathcal{E}_{t t}&=&-\frac{f N}{2 r^2}\left[-2\frac{r^2}{L^2}K+r\partial_r\left(f H\right)\right]\, ,\\
\mathcal{E}_{t r}&=&-\frac{1}{2 r f}H\partial_t f\, ,\\
\mathcal{E}_{r r}&=&\frac{1}{2 r N}H\partial_r N-\frac{\mathcal{E}_{t t}}{f^2 N} \, ,
\end{eqnarray}
where $H\equiv 1+X F+2 X^2 F'$.  Consequently, the equations of motion (\ref{fulleom}) reduce to
\begin{eqnarray}\label{eomM1}
p^2 K+ X^2 \partial_X \left(f H\right)&=&\frac{8 \pi G p^2 L^2}{f N} T_{t t}\, ,\\ \label{eomM2}
\frac{\partial_t f}{r f}&=&- \frac{16 \pi G}{H} T_{t r}\, ,\\ \label{eomM3}
\frac{\partial_r N}{r N}&=&\frac{16 \pi G}{H}\left(T_{t t}+T_{r r} f^2 N\right) \, ,
\end{eqnarray}
together with the conservation equation $\nabla^aT_{a b}=0$. In particular, this means that the vacuum solutions satisfy
\begin{eqnarray}
& X^2\partial_X \left[f(t,r) H\right]+p^2 K&=0\, ,\\
&\partial_t f(t,r)&=0\, ,\\
&\partial_r N(t,r)&=0\, .
\end{eqnarray}
The third equation implies that $N=N(t)$ so, without loss of generality, this function can be absorbed into a redefinition of the time coordinate.  On the other hand, the second imposes $f(t,r)=f(r)$ and therefore the metric is static.
Finally, the first equation is nothing but the familiar equation for $f(r)$ previously obtained.
This completes the proof of the Birkhoff theorem. For each value of $p$, $\lambda$ and action functions $F(X),K(X)$---or, alternatively, gravitational couplings $\alpha_n,\beta_m$---the most general spherically symmetric solution is static and given by \req{fEQTnon-local}---or, alternatively, by \req{fEQT2}. 

Observe that \req{eomM1}-\req{eomM3} could be used to study the collapse of spherical matter and the dynamical formation of the black holes studied here, along the lines of \cite{Bueno:2024eig,Bueno:2024zsx}.

\section{Penrose diagrams}\label{penro}

In this section, we adopt a more systematic approach to explore the space of possible black holes described by the metric function \eqref{fEQT2}. We focus on characterizing their global properties by examining three key aspects: the number of zeros of the function $f(r)$, the sign of this function between the zeros, and the behavior of $f(r)$ in the near-origin region ($ r \to 0 $).\footnote{More generally, in order to study the global structure of a solution we have also to examine the behavior in the asymptotic region $r\rightarrow\infty$. However, for our solutions, the asymptotic behavior is fixed to be asymptotically AdS and thus we omit this variable from our analysis.} To facilitate this analysis, we assign block diagrams to represent each region between the roots of $f(r)$. With them, we construct the corresponding Penrose diagram of the solution, making manifest properties of the global structure.\footnote{For a related alternative notion of two-dimensional conformal diagram see \cite{walker1970block,Chrusciel:2012gz}} In the case of regular black holes of the generic type, we do not count $r=0$ as a horizon---we relegate this to the conclusions---but we do make some comments about solutions of that kind which do not possess any additional horizons in subsection \ref{zeroh}.

\subsection{Black holes with one horizon}
As we have seen, 
the zeros of $f(r)$ are determined solely by the roots of its numerator, which depend on the values of the constants $\alpha_i$. Let us start with the following question. Suppose that there is a single zero, \ie there exists a unique $r = r_+$ such that $f(r_+) = 0$. In that situation, how many distinct global spacetimes exist?

\begin{figure}
\begin{minipage}{0.215\textwidth}
\includegraphics[width=1\textwidth]{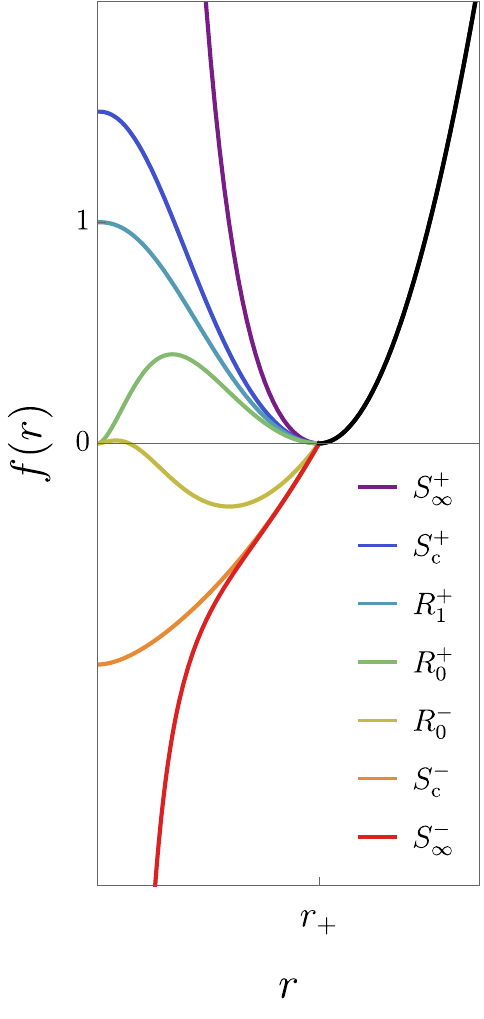}       
\end{minipage}
\begin{minipage}{0.3395\textwidth}
   \includegraphics[width=1\textwidth]{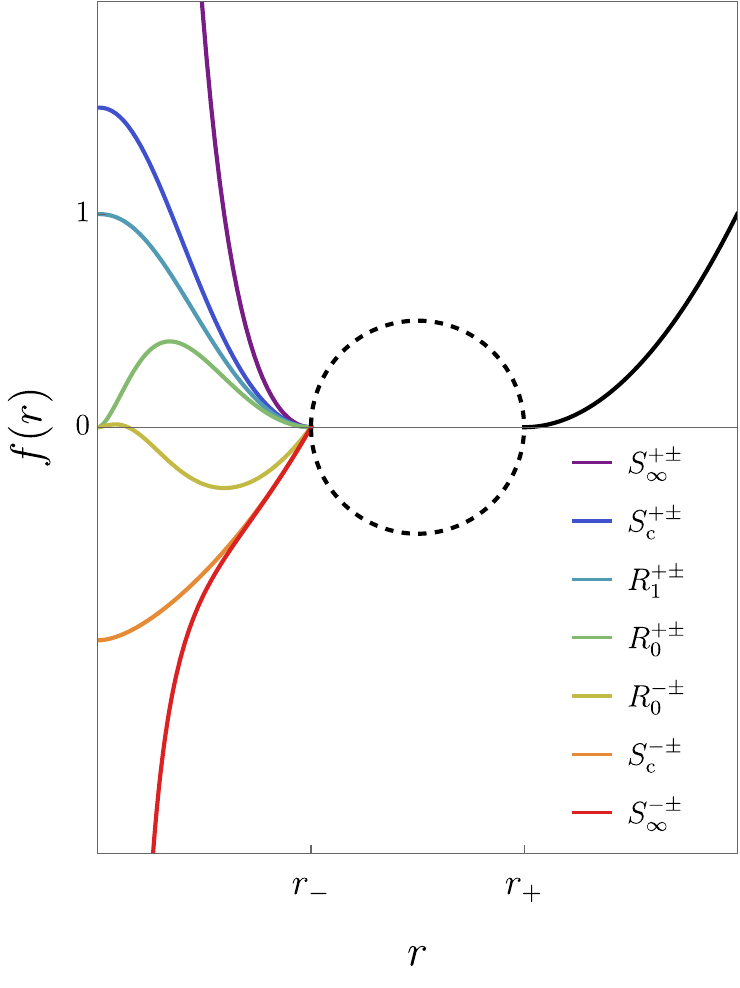}      
\end{minipage}
\begin{minipage}{0.4375\textwidth}
\begin{center}
   \includegraphics[width=1\textwidth]{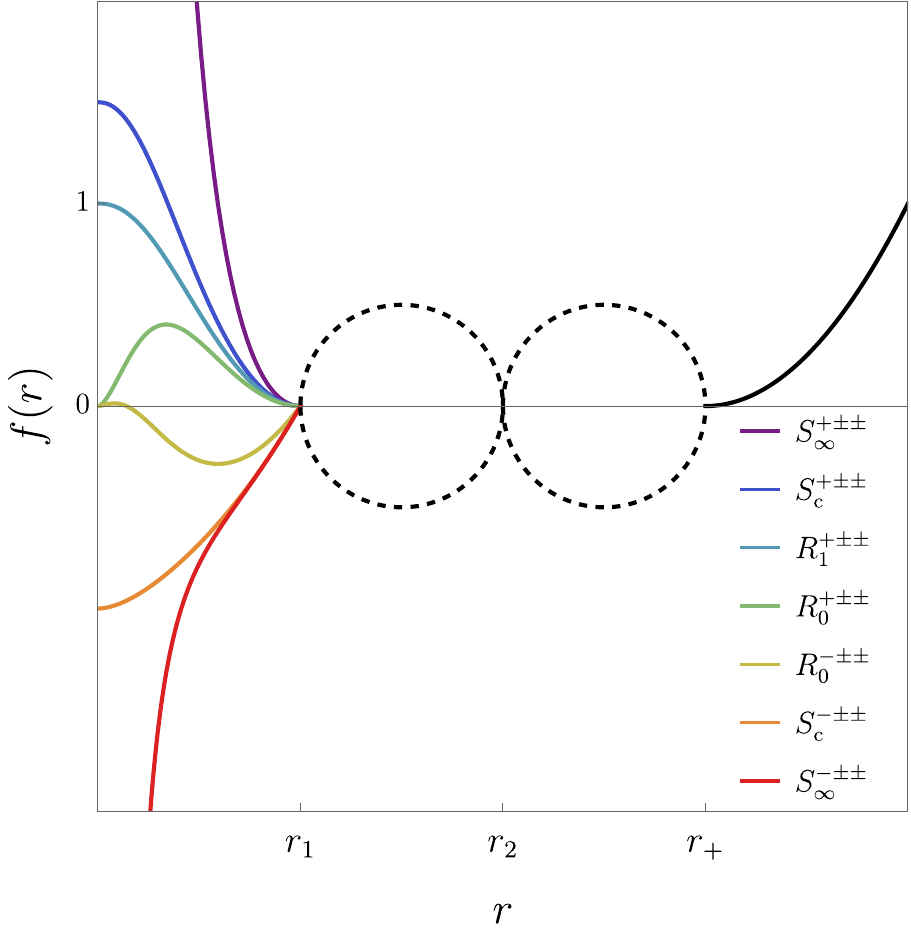}       
\end{center} 
\end{minipage}
\caption{(Left). We plot black holes with a single horizon possessing distinct causal structures associating to each of them a different color. Since the behavior in region II $ (r > r_+)$ is always asymptotically AdS, we assign the seven labels $S_\infty^\pm$, $S_\text{c}^\pm$, $R_1^+$, and $R_0^\pm$, which characterize region I ($r < r_+$) to the entire spacetime. (Center) While region I ($r < r_-$) and region III ($r > r_+$) remain unchanged with respect to the left plot, the new region II ($r_- < r < r_+$) accounts for the possible signatures between the roots, shown using the two dashed lines between $r_-$ and $r_+$. We denote the fourteen possible branches, as $S_\infty^{\pm\pm}$, $S_\text{c}^{\pm\pm}$, $R_1^{+\pm}$, and $R_0^{\pm\pm}$. (Right). In this plot, the asymptotic region is labeled as IV ($r>r_+$) while the near horizon is again labeled as I ($r<r_-$) while the two intermediate regions II and III are located in $r_1<r<r_2$ and $r_2<r<r_+$. As each of the intermediate region can possess either positive or negative signature we identify them with labels $S_\infty^{\pm\pm\pm}$, $S_\text{c}^{\pm\pm\pm}$, $R_1^{+\pm\pm}$, and $R_0^{\pm\pm\pm}$ and represent them in the plot with dashed lines. \label{fig:gsplots}}
\end{figure}

As  $r \to \infty $, the spacetime always asymptotically approaches AdS, which means that there is a unique block diagram representing $f(r)$ in the region $r_+<r<\infty$---that we shall name region II.\footnote{Notice that the usual convention is to assign ``I" to the asymptotic region. Here we choose to assign it to the innermost region, as the variability for the latter is greater and makes our discussion easier to follow.} Such block diagram includes two null boundaries and a timelike (vertical) boundary for the asymptotic region, this is
\begin{equation}
f(r_+<r<\infty)=
\vcenter{\hbox{\begin{tikzpicture}[scale=0.85,
point/.style={circle,fill,inner sep=1.3pt}
]
\draw[line width=0.5mm,egyptianblue,fill=egyptianblue,fill opacity=0.1] (0,0) -- (1,-1) -- (1,1) -- cycle;  
\node[inner sep=2] at (0.6,0) {II};
\node[egyptianblue,above,rotate=45] at (0.5,0.5) {$r=r_+$};
\node[egyptianblue,below,rotate=-45] at (0.5,-0.5) { $r=r_+$};
\node[egyptianblue,above,rotate=-90] at (1,0) { $r\rightarrow\infty$};
\end{tikzpicture}}}\,.
\end{equation}
However, the behavior of  $f(r)$ in region I, \ie when  $0 < r < r_+$, is significantly richer. It is determined not only by its limit as $r \to 0$, but also by its sign in this region. Attending to the different global structures, we may distinguish seven possible situations and their associated block diagrams: i) $S^\pm_\infty$: the metric function diverges positively or negatively as we approach the origin, $f(r\rightarrow0)\rightarrow\pm\infty$, leading to divergent curvature invariants. Diagrammatically, we represent this divergence with a zigzag line. ii) $S^\pm_\text{c}$: the metric function asymptotes to a constant value, $f(r \to 0) \in (-\infty, 0)\cup(0, 1) \cup (1, \infty)$. In this case the spacetime develops a conical singularity for positive constant values, $f(r \to 0) \in (0, 1) \cup (1, \infty)$ or a BTZ-like singularity for negative values, $f(r \to 0) \in (-\infty, 0)$, which are represented in the diagram with a dashed line. iii) $R^+_1$: the metric function tends to one as we approach the origin, $f(r\rightarrow0)\rightarrow1$. This case corresponds to a regular solution, with finite curvature invariants, and thus, we represent it with a solid line in the diagram. iv) $R^\pm_0$: the metric function approaches zero positively or negatively $f(r\rightarrow0)\rightarrow0^\pm$, defining a new horizon-like and regular structure that is represented with a null boundary.\footnote{We sometimes refer to the resulting rotated square block as diamond block.} Based on this, we assign to the seven cases the following block diagrams:
\begin{align}
S^+_\infty:\  f(r<r_+)=&\vcenter{\hbox{\begin{tikzpicture}[scale=0.85,
point/.style={circle,fill,inner sep=1.3pt}
]
\draw[line width=0.5mm,carmine,fill=carmine,fill opacity=0.1] (0,0) -- (-1,-1) decorate [singularity]{-- (-1,1)}-- cycle;  
\node[inner sep=2] at (-0.6,0) {I};
\node[carmine,above,rotate=-45] at (-0.5,0.5) {$r=r_+$};
\node[carmine,below,rotate=45] at (-0.5,-0.5) { $r=r_+$};
\node[carmine,above,rotate=90] at (-1,0) {$r=0$};
\end{tikzpicture}}} &\text{ if } f(r\rightarrow0)=+\infty\, ,\label{eq:S+infb}
\\
S^+_\text{c}: \ f(r<r_+)=&\vcenter{\hbox{\begin{tikzpicture}[scale=0.85,
point/.style={circle,fill,inner sep=1.3pt}
]
\draw[line width=0.5mm,carmine] (-1,1)--(0,0) -- (-1,-1);
\draw[line width=0.5mm,carmine,dashed] (-1,1) -- (-1,-1);  
\fill[carmine,opacity=0.1] (0,0) -- (-1,-1) -- (-1,1) -- cycle;  
\node[inner sep=2] at (-0.6,0) {I};
\node[carmine,above,rotate=-45] at (-0.5,0.5) {$r=r_+$};
\node[carmine,below,rotate=45] at (-0.5,-0.5) { $r=r_+$};
\node[carmine,above,rotate=90] at (-1,0) {$r=0$};
\end{tikzpicture}}}&\text{ if }f(r \to 0) \in (0, 1) \cup (1, \infty)\, ,\\
R^+_1:\ f(r<r_+)=&\vcenter{\hbox{\begin{tikzpicture}[scale=0.85,
point/.style={circle,fill,inner sep=1.3pt}
]
\draw[line width=0.5mm,carmine,fill=carmine,fill opacity=0.1] (0,0) -- (-1,-1) -- (-1,1)-- cycle;  
\node[inner sep=2] at (-0.6,0) {I};
\node[carmine,above,rotate=-45] at (-0.5,0.5) {$r=r_+$};
\node[carmine,below,rotate=45] at (-0.5,-0.5) { $r=r_+$};
\node[carmine,above,rotate=90] at (-1,0) {$r=0$};
\end{tikzpicture}}} &\text{ if } f(r\rightarrow0)=1\,\\
R^+_0:\ f(r<r_+)=&\vcenter{\hbox{\begin{tikzpicture}[scale=0.85,
point/.style={circle,fill,inner sep=1.3pt}
]
\draw[line width=0.5mm,carmine,fill=carmine,fill opacity=0.1] (0,0) -- (1,-1) -- (2,0)-- (1,1)  -- cycle;  
\node[inner sep=2] at (1,0) {I};
\node[carmine,above,rotate=45] at (0.5,0.5) {$r=0$};
\node[carmine,below,rotate=-45] at (0.5,-0.5) { $r=0$};
\node[carmine,above,rotate=-45] at (1.5,0.5) { $r=r_+$};
\node[carmine,below,rotate=45] at (1.5,-0.5) { $r=r_+$};
\end{tikzpicture}}} &\text{ if }f(r\rightarrow0)=0^+\,,\\
R^-_0:\ f(r<r_+)=&\vcenter{\hbox{\begin{tikzpicture}[scale=0.85,
point/.style={circle,fill,inner sep=1.3pt}
]
\draw[line width=0.5mm,carmine,fill=carmine,fill opacity=0.1] (0,0) -- (1,-1) -- (2,0)-- (1,1)  -- cycle;  
\node[inner sep=2] at (1,0) {I};
\node[carmine,above,rotate=45] at (0.5,0.5) {$r=r_+$};
\node[carmine,below,rotate=-45] at (0.5,-0.5) { $r=0$};
\node[carmine,above,rotate=-45] at (1.5,0.5) { $r=r_+$};
\node[carmine,below,rotate=45] at (1.5,-0.5) { $r=0$};
\end{tikzpicture}}} &\text{ if } f(r\rightarrow0)=0^-\, ,\\
S^-_\text{c}:\ f(r<r_+)=&\vcenter{\hbox{\begin{tikzpicture}[scale=0.85,
point/.style={circle,fill,inner sep=1.3pt}
]
\fill[carmine,opacity=0.1] (0,-1) -- (-1,0) -- (1,0)-- cycle;  
\draw[line width=0.5mm,carmine] (-1,0)-- (0,-1)  -- (1,0);  
\draw[line width=0.5mm,carmine,dashed] (-1,0) -- (1,0);  
\node[inner sep=2] at (0,-0.4) {I};
\node[carmine,below,rotate=-45] at (-0.5,-0.5) {$r=r_+$};
\node[carmine,below,rotate=45] at (0.5,-0.5) { $r=r_+$};
\node[carmine,above] at (0,0) {$r=0$};
\end{tikzpicture}}} &\text{ if } f(r\rightarrow0)\in(0,-\infty)\,,\\
S^-_\infty:\ f(r<r_+)=&\vcenter{\hbox{\begin{tikzpicture}[scale=0.85,
point/.style={circle,fill,inner sep=1.3pt}
]
\draw[line width=0.5mm,carmine,fill=carmine,fill opacity=0.1] (0,-1) -- (-1,0) decorate [singularity]{-- (1,0)}-- cycle;  
\node[inner sep=2] at (0,-0.4) {I};
\node[carmine,below,rotate=-45] at (-0.5,-0.5) {$r=r_+$};
\node[carmine,below,rotate=45] at (0.5,-0.5) { $r=r_+$};
\node[carmine,above] at (0,0) {$r=0$};
\end{tikzpicture}}} &\text{ if }f(r\rightarrow0)=-\infty\,. \label{eq:S-infb}
\end{align}
 A schematic plot of the global behavior of $f(r)$ for each of the seven possibilities in region I, together with the unique behavior region II, is shown in Figure \ref{fig:gsplots} (Left).
As is customary, the vertical direction represents the timelike coordinate. In region I, whenever $ f(r) $ is positive---namely in $ S_\infty^+ $, $ S_\text{c}^+ $, $ R_1^+ $, and $ R_0^+ $---the $ r $-coordinate is spacelike, while the $ t $-coordinate is timelike. Conversely, when $ f(r) $ is negative---as in $ S_0^- $, $ S_\text{c}^- $, and $ S_\infty^- $---the roles of the coordinates are reversed. To account for this, we rotate the block diagram, treating $ r $ as the timelike coordinate and $ t $ as the spacelike one.
With all possible block diagrams for a black hole with a single root---\ie one horizon---at hand, we can now construct the corresponding maximally extended Penrose diagrams. To do so, we have to assemble the block diagrams gluing the compatible radii $r=r_+$ and account for the symmetry in the timelike coordinate, which flips the block diagrams horizontally or vertically, depending on which coordinate is spacelike and which is timelike---\ie $t\rightarrow -t$ if $t$ is the timelike coordinate or $r\rightarrow -r$ otherwise.

Below, we present the seven possible Penrose diagrams, specifying the values of $\alpha_i$ and $\beta_j$ with the parameters of the solutions fixed as $p=1$, $L=1$, $\alpha_1=1/2$ and $\lambda=1+\sum_{i=2}\frac{\alpha_i}{2(i-1)}$. Whenever the value of any $\alpha_i$, $\beta_j$ is not explicitly specified it is set to zero. Since the region II is always the same, we can use the seven labels introduced above to identify the maximally extended Penrose diagrams constructed from the block diagrams presented in equations \eqref{eq:S+infb} to \eqref{eq:S-infb}, namely:
\begin{align}
S^+_\infty=&\ \vcenter{\hbox{\begin{tikzpicture}[scale=0.85,
point/.style={circle,fill,inner sep=1.3pt}
]
\draw[line width=0.5mm,carmine,fill=carmine,fill opacity=0.1] (0,0) -- (-1,-1) decorate [singularity]{-- (-1,1)}-- cycle;  
\draw[line width=0.5mm,egyptianblue,fill=egyptianblue,fill opacity=0.1] (0,0) -- (0,2) -- (-1,1)-- cycle; 
\draw[line width=0.5mm,frenchlilac] (-1,1) -- (0,0); 
\node[inner sep=2] at (-0.6,0) {I};
\node[inner sep=2] at (-0.4,1) {II};
\filldraw (-0.5,2) circle (0.5pt);
\filldraw (-0.5,1.8) circle (0.5pt);
\filldraw (-0.5,2.2) circle (0.5pt);
\filldraw (-0.5,-1) circle (0.5pt);
\filldraw (-0.5,-0.8) circle (0.5pt);
\filldraw (-0.5,-1.2) circle (0.5pt);
\end{tikzpicture}}}\, , & \text{with }\alpha_2=\frac{3}{2}\,, \label{eq:S+inf}\\
S^+_\text{c}=&\ \vcenter{\hbox{\begin{tikzpicture}[scale=0.85,
point/.style={circle,fill,inner sep=1.3pt}
]
\fill[carmine,opacity=0.1] (0,0) -- (-1,-1) -- (-1,1)-- cycle;  
\draw[line width=0.5mm,carmine]  (-1,-1) --(0,0) -- (-1,1);  
\draw[line width=0.5mm,carmine,dashed]  (-1,-1) -- (-1,1);  
\draw[line width=0.5mm,egyptianblue,fill=egyptianblue,fill opacity=0.1] (0,0) -- (0,2) -- (-1,1)-- cycle; 
\draw[line width=0.5mm,frenchlilac,] (-1,1) -- (0,0); 
\node[inner sep=2] at (-0.6,0) {I};
\node[inner sep=2] at (-0.4,1) {II};
\filldraw (-0.5,2) circle (0.5pt);
\filldraw (-0.5,1.8) circle (0.5pt);
\filldraw (-0.5,2.2) circle (0.5pt);
\filldraw (-0.5,-1) circle (0.5pt);
\filldraw (-0.5,-0.8) circle (0.5pt);
\filldraw (-0.5,-1.2) circle (0.5pt);
\end{tikzpicture}}}\, ,& \text{with }\alpha_2=\frac{3}{2}\,,\ \beta_0=0.1\, ,\label{eq:S+c}\\
R^+_1=&\ \vcenter{\hbox{\begin{tikzpicture}[scale=0.85,
point/.style={circle,fill,inner sep=1.3pt}
]
\draw[line width=0.5mm,carmine,fill=carmine,fill opacity=0.1] (0,0) -- (-1,-1) -- (-1,1)-- cycle;  
\draw[line width=0.5mm,egyptianblue,fill=egyptianblue,fill opacity=0.1] (0,0) -- (0,2) -- (-1,1)-- cycle; 
\draw[line width=0.5mm,frenchlilac,] (-1,1) -- (0,0);
\node[inner sep=2] at (-0.6,0) {I};
\node[inner sep=2] at (-0.4,1) {II};
\filldraw (-0.5,2) circle (0.5pt);
\filldraw (-0.5,1.8) circle (0.5pt);
\filldraw (-0.5,2.2) circle (0.5pt);
\filldraw (-0.5,-1) circle (0.5pt);
\filldraw (-0.5,-0.8) circle (0.5pt);
\filldraw (-0.5,-1.2) circle (0.5pt);
\end{tikzpicture}}}\,, & \text{with } \alpha_2=\frac{3}{2}\,,\ \beta_0=\frac{3}{4}\, ,\label{eq:R+1}\\
R^+_0=&\ \vcenter{\hbox{\begin{tikzpicture}[scale=0.85,
point/.style={circle,fill,inner sep=1.3pt}
]
\draw[line width=0.5mm,carmine,fill=carmine,fill opacity=0.1] (0,0) -- (-1,-1)-- (-2,0) -- (-1,1) -- cycle;  
\draw[line width=0.5mm,egyptianblue,fill=egyptianblue,fill opacity=0.1] (0,0) -- (0,2) -- (-1,1)-- cycle; 
\draw[line width=0.5mm,frenchlilac,] (-1,1) -- (0,0); 
\node[inner sep=2] at (-1,0) {I};
\node[inner sep=2] at (-0.4,1) {II};
\filldraw (-1,2) circle (0.5pt);
\filldraw (-1,1.8) circle (0.5pt);
\filldraw (-1,1.6) circle (0.5pt);
\filldraw (-0.5,-1) circle (0.5pt);
\filldraw (-0.5,-0.8) circle (0.5pt);
\filldraw (-0.5,-1.2) circle (0.5pt);
\end{tikzpicture}}}\,, &\text{with } \alpha_2=\frac{3}{2}\, ,\ \beta_1=0.1\, , \label{eq:R+0}\\
R^-_0=&\ \vcenter{\hbox{\begin{tikzpicture}[scale=0.85,
point/.style={circle,fill,inner sep=1.3pt}
]
\draw[line width=0.5mm,carmine,fill=carmine,fill opacity=0.1] (0,0) -- (-1,-1)-- (-2,0) -- (-1,1) -- cycle;  
\draw[line width=0.5mm,carmine,fill=carmine,fill opacity=0.1] (0,2) -- (-1,1)-- (-2,2) -- (-1,3) -- cycle;  
\draw[line width=0.5mm,egyptianblue,fill=egyptianblue,fill opacity=0.1] (0,0) -- (0,2) -- (-1,1)-- cycle;
\draw[line width=0.5mm,egyptianblue,fill=egyptianblue,fill opacity=0.1] (-2,0) -- (-2,2) -- (-1,1)-- cycle; 
\draw[line width=0.5mm,frenchlilac,] (0,0) -- (-2,2); 
\draw[line width=0.5mm,frenchlilac,] (-2,0) -- (0,2); 
\node[inner sep=2] at (-1,0) {I};
\node[inner sep=2] at (-0.4,1) {II};
\node[inner sep=2] at (-1,2) {I};
\node[inner sep=2] at (-1.6667,1) {II};
\end{tikzpicture}}}\, ,& \text{wtih } \alpha_2=-1,\, \beta_1=0.1\, ,\label{eq:R-0}\\
S^-_\text{c}=&\ \vcenter{\hbox{\begin{tikzpicture}[scale=0.85,
point/.style={circle,fill,inner sep=1.3pt}
]
\fill[carmine,opacity=0.1] (0,0) -- (-1,1) -- (1,1)-- cycle;
\fill[carmine,opacity=0.1] (0,0) -- (-1,-1) -- (1,-1)-- cycle;
\draw[line width=0.5mm,carmine,dashed] (-1,1) -- (1,1);
\draw[line width=0.5mm,carmine,dashed] (-1,-1) -- (1,-1);
\draw[line width=0.5mm,egyptianblue,fill=egyptianblue,fill opacity=0.1] (0,0) -- (1,1) -- (1,-1)-- cycle; 
\draw[line width=0.5mm,egyptianblue,fill=egyptianblue,fill opacity=0.1] (0,0) -- (-1,1) -- (-1,-1)-- cycle; 
\draw[line width=0.5mm,frenchlilac,] (1,1) -- (-1,-1); 
\draw[line width=0.5mm,frenchlilac,] (-1,1) -- (1,-1); 
\node[inner sep=2] at (0,0.6) {I};
\node[inner sep=2] at (0.6,0) {II};
\node[inner sep=2] at (0,-0.6) {I};
\node[inner sep=2] at (-0.6,0) {II};
\end{tikzpicture}}}\, ,& \text{with } \alpha_2=-1\,,\ \beta_0=0.1\,,\label{eq:S-c}\\
S^-_\infty=&\ \vcenter{\hbox{\begin{tikzpicture}[scale=0.85,
point/.style={circle,fill,inner sep=1.3pt}
]
\draw[line width=0.5mm,carmine,fill=carmine,fill opacity=0.1] (-1,1) decorate [singularity]{-- (1,1)} -- (0,0) -- cycle;
\draw[line width=0.5mm,carmine,fill=carmine,fill opacity=0.1] (-1,-1) decorate [singularity]{-- (1,-1)} -- (0,0) -- cycle;
\draw[line width=0.5mm,egyptianblue,fill=egyptianblue,fill opacity=0.1] (0,0) -- (1,1) -- (1,-1)-- cycle; 
\draw[line width=0.5mm,egyptianblue,fill=egyptianblue,fill opacity=0.1] (0,0) -- (-1,1) -- (-1,-1)-- cycle; 
\draw[line width=0.5mm,frenchlilac,] (1,1) -- (-1,-1); 
\draw[line width=0.5mm,frenchlilac,] (-1,1) -- (1,-1); 
\node[inner sep=2] at (0,0.6) {I};
\node[inner sep=2] at (0.6,0) {II};
\node[inner sep=2] at (0,-0.6) {I};
\node[inner sep=2] at (-0.6,0) {II};
\end{tikzpicture}}}\, ,& \text{with } \alpha_2=-1\,.\label{eq:S-inf}
\end{align}
The ellipsis mean that the maximally extended Penrose diagram is composed of an infinite number of block diagrams, with periodicity given by the regions represented. At the same time, we stress that here the equality between labels and diagrams means that they belong to the same equivalence class of global spacetimes. 

When specifying the values of the couplings $\alpha_i$ and $\beta_j$, we use both fractional and decimal notations. Fractional values are exact, and even a slight variation can significantly alter the global structure. In this sense, the black hole's causal structure can be considered fine-tuned. Conversely, when using decimal notation with only one significant figure, the value retains some flexibility, allowing for variations while still preserving the same causal structure. Observe that the Penrose diagrams $ S^+_\infty $, $ S^+_\text{c} $, $ R_1^+ $, and $ R_0^+ $ can be extended infinitely in the timelike direction, alternating between regions I and II. All of these correspond to extremal black holes, as they satisfy $ f'(r_+) = 0 $. Notably, $ S^+_\infty $ features a timelike singularity, necessitating a cosmic censor to prevent an infalling observer from having it in its causal past.   As mentioned earlier, some diagrams arise only after fine-tuning the values of the coupling constants, such as $ S_\infty^+ $ and $ R^+_1 $. In contrast, the remaining cases allow some freedom in choosing the values of $ \alpha_i $ or $ \beta_j $.   On the other hand, whenever $ f(r\rightarrow0) $ is negative---specifically for $ R_0^- $, $ S_\text{c}^- $, and $ S_{\infty}^- $---the diagrams do not alternate between two different regions. In particular, the regular diagram acquires a null block that behaves like a horizon as we approach $ r=0 $, while the conical singularity instead features a spacelike surface at the same point. 

We opted to present the Penrose diagrams for the nonsingular black holes in a compact way. However, the fact that a radially infalling observer crosses $r=0$ in finite proper time, as we will discuss in Section~\ref{sec:geodesics}, anticipates that we might consider diagrams including this continuation. Following the discussion presented for the Simpson-Visser regular black hole \cite{Simpson:2018tsi},\footnote{We thank Pedro G. S. Fernandes for pointing out the resemblance with the Penrose diagrams presented in that article.} we may identify the future null bounce at $r=0$ in $R^{\pm}_0$  to construct ``looped'' Penrose diagrams, this is
\begin{equation}\label{Penrose_geodesics1}
\mathring R^+_0=\vcenter{\hbox{\begin{tikzpicture}[scale=0.85,
point/.style={circle,fill,inner sep=1.3pt}
]
\draw[line width=0.5mm,egyptianblue,fill=egyptianblue,fill opacity=0.1] (0,0) -- (0,2) -- (-1,1)-- cycle; 
\node[inner sep=2] at (-1,0) {I};
\node[inner sep=2] at (-0.4,1) {II};
\filldraw (-1,2) circle (0.5pt);
\filldraw (-1,1.8) circle (0.5pt);
\filldraw (-1,1.6) circle (0.5pt);
\filldraw (-0.5,-1) circle (0.5pt);
\filldraw (-0.5,-0.8) circle (0.5pt);
\filldraw (-0.5,-1.2) circle (0.5pt);
\draw[stealth-stealth, thick] (-1.5,0.5) arc[start angle=45, end angle=315, radius=0.707107cm];
\draw[line width=0.5mm,carmine,fill=carmine,fill opacity=0.1] (0,0) -- (-1,-1)-- (-2,0) -- (-1,1) -- cycle; 
\draw[line width=0.5mm,frenchlilac,] (-1,1) -- (0,0); 
\end{tikzpicture}}}\,,\quad 
\vcenter{\hbox{\includegraphics[scale=1]{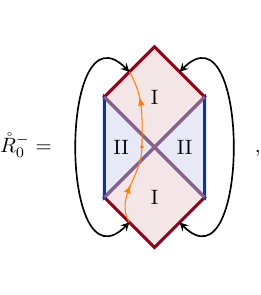}}}
\end{equation}
where $\mathring D$ denotes the looped version of diagram $D$. In this looped version an observer is subject to periodic boundary conditions when reaching $r=0$ and, thus, this spacetime contains closed timelike curves. On the other hand, we may interpret that at $r=0$ an infalling observer experiences a black hole to white hole transition, bouncing between copies of asymptotic regions II, this is
\begin{equation}\label{Penrose_geodesics2}
\bar R^+_0=\vcenter{\hbox{\begin{tikzpicture}[scale=0.85,
point/.style={circle,fill,inner sep=1.3pt}
]
\draw[line width=0.5mm,carmine,fill=carmine,fill opacity=0.1] (-2,0) -- (-3,1) -- (-2,2) -- (-1,1)-- cycle; 
\draw[line width=0.5mm,carmine,fill=carmine,fill opacity=0.1] (-2,-2) -- (-3,-1) -- (-2,0) -- (-1,-1)-- cycle;
\draw[line width=0.5mm,carmine,fill=carmine,fill opacity=0.1] (0,0) -- (-1,-1)-- (-2,0) -- (-1,1) -- cycle;
\draw[line width=0.5mm,egyptianblue,fill=egyptianblue,fill opacity=0.1] (0,0) -- (0,2) -- (-1,1)-- cycle; 
\draw[line width=0.5mm,egyptianblue,fill=egyptianblue,fill opacity=0.1] (0,0) -- (0,-2) -- (-1,-1)-- cycle; 
\draw[line width=0.5mm,egyptianblue,fill=egyptianblue,fill opacity=0.1] (-2,0) -- (-3,-1) -- (-3,1)-- cycle; 
\draw[line width=0.5mm,frenchlilac] (-1,1) -- (0,0) -- (-1,-1); 
\draw[line width=0.5mm,frenchlilac] (-3,1) -- (-2,0) -- (-3,-1); 
\node[inner sep=2] at (-1,0) {I};
\node[inner sep=2] at (-2,1) {I};
\node[inner sep=2] at (-2,-1) {I};
\node[inner sep=2] at (-0.4,1) {II};
\node[inner sep=2] at (-0.4,-1) {II};
\node[inner sep=2] at (-2.6,0) {II};
\filldraw (-1.5,2.2) circle (0.5pt);
\filldraw (-1.5,2.4) circle (0.5pt);
\filldraw (-1.5,2.6) circle (0.5pt);
\filldraw (-1.5,-2.2) circle (0.5pt);
\filldraw (-1.5,-2.4) circle (0.5pt);
\filldraw (-1.5,-2.6) circle (0.5pt);
\end{tikzpicture}}}\,,\quad 
\vcenter{\hbox{\includegraphics[scale=1]{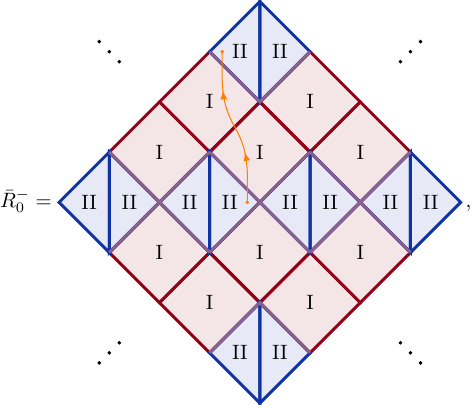}}}
\end{equation}
Notably, after gluing the copies of region I along their null horizons, the diagram for $\bar R^-_0$ tessellates the two-dimensional plane. In this case there are no closed timelike curves. We will discuss further on Penrose diagram tessellation when analyzing black holes with three horizons in Section~\ref{ref:Pd3hor} in which we will not have to extend the trajectories beyond $r=0$. In addition, we included orange trajectories in $\mathring R_0^-$ and $\bar R_0^-$ that will be addressed in Section \ref{sec:geodesics}.

The same discussion about extending $R^\pm_{0}$ applies to $R^+_1$, this is
\begin{equation}
\mathring R^+_1=\ \vcenter{\hbox{\begin{tikzpicture}[scale=0.85,
point/.style={circle,fill,inner sep=1.3pt}
]
\draw[stealth-stealth, thick] (-1,2) arc[start angle=90, end angle=270, radius=1cm];
\fill[carmine,opacity=0.1] (0,2) -- (-1,1) -- (-1,3)-- cycle;  
\fill[carmine,opacity=0.1] (0,0) -- (-1,-1) -- (-1,1)-- cycle; 
\draw[line width=0.5mm,carmine]  (-1,1) --(0,2) -- (-1,3) -- cycle;  
\draw[line width=0.5mm,carmine]  (-1,-1) --(0,0) -- (-1,1) --cycle;  
\draw[line width=0.5mm,egyptianblue,fill=egyptianblue,fill opacity=0.1] (0,0) -- (0,2) -- (-1,1)-- cycle; 
\draw[line width=0.5mm,frenchlilac,] (0,2)--(-1,1) -- (0,0); 
\node[inner sep=2] at (-0.6,0) {I};
\node[inner sep=2] at (-0.6,2) {I};
\node[inner sep=2] at (-0.4,1) {II};
\filldraw (-0.5,3) circle (0.5pt);
\filldraw (-0.5,2.8) circle (0.5pt);
\filldraw (-0.5,3.2) circle (0.5pt);
\filldraw (-0.5,-1) circle (0.5pt);
\filldraw (-0.5,-0.8) circle (0.5pt);
\filldraw (-0.5,-1.2) circle (0.5pt);
\end{tikzpicture}}}\, ,\quad 
\bar R^+_1=\ \vcenter{\hbox{\begin{tikzpicture}[scale=0.85,
point/.style={circle,fill,inner sep=1.3pt}
] 
\draw[line width=0.5mm,carmine,fill=carmine,fill opacity=0.1]  (-1,-1) --(0,0) -- (-1,1) --cycle;
\draw[line width=0.5mm,carmine,fill=carmine,fill opacity=0.1]  (-1,-1) --(-2,0) -- (-1,1) --cycle;  
\draw[line width=0.5mm,egyptianblue,fill=egyptianblue,fill opacity=0.1] (0,0) -- (0,2) -- (-1,1)-- cycle; 
\draw[line width=0.5mm,egyptianblue,fill=egyptianblue,fill opacity=0.1] (-2,0) -- (-2,2) -- (-1,1)-- cycle; 
\draw[line width=0.5mm,egyptianblue,fill=egyptianblue,fill opacity=0.1] (0,-2) -- (0,0) -- (-1,-1)-- cycle; 
\draw[line width=0.5mm,egyptianblue,fill=egyptianblue,fill opacity=0.1] (-2,-2) -- (-2,0) -- (-1,-1)-- cycle; 
\draw[line width=0.5mm,frenchlilac,] (-2,0) -- (-1,1) -- (0,0) -- (-1,-1) -- cycle; 
\node[inner sep=2] at (-0.6,0) {I};
\node[inner sep=2] at (-1.4,0) {I};
\node[inner sep=2] at (-0.4,1) {II};
\node[inner sep=2] at (-0.4,-1) {II};
\node[inner sep=2] at (-1.6,-1) {II};
\node[inner sep=2] at (-1.6,1) {II};
\filldraw (-1,2) circle (0.5pt);
\filldraw (-1,1.8) circle (0.5pt);
\filldraw (-1,2.2) circle (0.5pt);
\filldraw (-1,-2) circle (0.5pt);
\filldraw (-1,-1.8) circle (0.5pt);
\filldraw (-1,-2.2) circle (0.5pt);
\end{tikzpicture}}}
\end{equation}
where both geometries display a naked black hole-to-white hole bounce. In the remainder of this section, we will focus on the compact versions of the diagrams enclosed by $r=0$ and $r\rightarrow\infty$.

Before concluding our discussion on one-horizon black holes, we introduce an operation that facilitates the study of cases with multiple horizons in a more systematic manner. By examining the diagrams in \eqref{eq:S+inf} to \eqref{eq:S-inf}, we observe that the blocks can be classified into two distinct categories based on the sign of their contributions in the region $r < r_+$. Specifically, these categories correspond to blocks with either a positive or negative sign.  For a fixed sign we define a \textit{diagram regularization} operation that transforms a singular diagram block into a regular one. This transformation simplifies the analysis, reducing the general study from seven to two Penrose diagrams, allowing to obtain the rest from such transformation. Mathematically, this transformation corresponds to turning on or off appropriate values of $ \beta_j $ that modify the behavior near the origin of coordinates, while keeping $ \alpha_i $ unchanged, this is
\begin{align}
&S^+_\infty\left[\vcenter{\hbox{\begin{tikzpicture}[scale=0.85,
point/.style={circle,fill,inner sep=1.3pt}]
\draw[line width=0.5mm,carmine,fill=carmine,fill opacity=0.1] (0,0) -- (-1,-1) decorate [singularity]{-- (-1,1)}-- cycle;   
\node[inner sep=2] at (-0.6,0) {I};
\end{tikzpicture}}}\rightarrow \vcenter{\hbox{\begin{tikzpicture}[scale=0.85,
point/.style={circle,fill,inner sep=1.3pt}
]
\fill[carmine,opacity=0.1] (0,0) -- (-1,-1) -- (-1,1)-- cycle;  
\draw[line width=0.5mm,carmine]  (-1,-1) -- (0,0) -- (-1,1);  
\draw[line width=0.5mm,carmine,dashed]  (-1,-1) -- (-1,1);  
\node[inner sep=2] at (-0.6,0) {I};
\end{tikzpicture}}}\right]=S^+_\text{c}\quad \iff\quad \beta_{j=i_{\text{max}-2}}>0\,,\label{eq:pentr1} \\
&S^+_\infty\left[\vcenter{\hbox{\begin{tikzpicture}[scale=0.85,
point/.style={circle,fill,inner sep=1.3pt}]
\draw[line width=0.5mm,carmine,fill=carmine,fill opacity=0.1] (0,0) -- (-1,-1) decorate [singularity]{-- (-1,1)}-- cycle;   
\node[inner sep=2] at (-0.6,0) {I};
\end{tikzpicture}}}\rightarrow \vcenter{\hbox{\begin{tikzpicture}[scale=0.85,
point/.style={circle,fill,inner sep=1.3pt}
]
\draw[line width=0.5mm,carmine,fill=carmine,fill opacity=0.1] (0,0) -- (-1,-1) -- (-1,1)-- cycle;  
\node[inner sep=2] at (-0.6,0) {I};
\end{tikzpicture}}}\right]=R^+_1\quad \iff\quad \beta_{j=i_{\text{max}-2}}=\frac{\alpha_{i_\text{max}}}{2(i_\text{max}-1)(2i_\text{max}-3)}\,, \label{eq:pentr2}\\ &S^+_\infty\left[\vcenter{\hbox{\begin{tikzpicture}[scale=0.85,
point/.style={circle,fill,inner sep=1.3pt}
]
\draw[line width=0.5mm,carmine,fill=carmine,fill opacity=0.1] (0,0) -- (-1,-1) decorate [singularity]{-- (-1,1)}-- cycle;  
\node[inner sep=2] at (-0.6,0) {I};
\end{tikzpicture}}}\rightarrow \vcenter{\hbox{\begin{tikzpicture}[scale=0.85,
point/.style={circle,fill,inner sep=1.3pt}
]
\draw[line width=0.5mm,carmine,fill=carmine,fill opacity=0.1] (0,0) -- (-1,-1) -- (-2,0)-- (-1,1)-- cycle;  
\node[inner sep=2] at (-1,0) {I};
\end{tikzpicture}}}\right]=R_0^+\quad \iff\quad \beta_{j=i_{\text{max}-1}}>0\,,\label{eq:pentr3}\\
&S^-_\infty\left[\vcenter{\hbox{\begin{tikzpicture}[scale=0.85,
point/.style={circle,fill,inner sep=1.3pt}
]
\draw[line width=0.5mm,carmine,fill=carmine,fill opacity=0.1] (0,0) -- (-1,1) decorate [singularity]{-- (1,1)}-- cycle;    
\node[inner sep=2] at (0,0.6) {I};
\end{tikzpicture}}}\rightarrow \vcenter{\hbox{\begin{tikzpicture}[scale=0.85,
point/.style={circle,fill,inner sep=1.3pt}
]
\draw[line width=0.5mm,carmine,fill=carmine,fill opacity=0.1] (0,0) -- (-1,-1) -- (-2,0)-- (-1,1)-- cycle;  
\node[inner sep=2] at (-1,0) {I};
\end{tikzpicture}}}\right]=R_0^-\quad \iff\quad \beta_{j=i_{\text{max}-1}}>0\,,\label{eq:pentr4}\\
&S^-_\infty\left[\vcenter{\hbox{\begin{tikzpicture}[scale=0.85,
point/.style={circle,fill,inner sep=1.3pt}
]
\draw[line width=0.5mm,carmine,fill=carmine,fill opacity=0.1] (0,0) -- (-1,1) decorate [singularity]{-- (1,1)}-- cycle;    
\node[inner sep=2] at (0,0.6) {I};
\end{tikzpicture}}}\rightarrow \vcenter{\hbox{\begin{tikzpicture}[scale=0.85,
point/.style={circle,fill,inner sep=1.3pt}
]
\fill[carmine,fill opacity=0.1] (0,0) -- (-1,1) -- (1,1)-- cycle;  
\draw[line width=0.5mm,carmine] (-1,1) -- (0,0) -- (1,1);   
\draw[line width=0.5mm,carmine,dashed] (-1,1)  -- (1,1);   
\node[inner sep=2] at (0,0.6) {I};
\end{tikzpicture}}}\right]=S_\text{c}^-\quad \iff\quad \beta_{j=i_{\text{max}-2}}>0\,.\label{eq:pentr5}
\end{align}
This can be observed, for example, in $ S_\infty^+ $ with $ \alpha_2 = \frac{3}{2} $, as shown in \eqref{eq:S+inf}. The operation described in \eqref{eq:pentr1} involves setting $ \beta_0 $ to a positive value, such as $ \beta_0 = 0.1 $. This realizes the black hole $ S^+_c $, as described in \eqref{eq:S+c}.

\subsection{Black holes with two horizons}

Suppose now that the metric function $f(r)$ has two real roots, $r_-$ and $r_+$, this is, $f(r_-)=f(r_+)=0$ and assume, without loss of generality, that $r_- < r_+$. We can extend the discussion presented in the previous subsection to enumerate the distinct global spacetimes, taking into account the following considerations: i) The region II, defined in the interval $r_- < r < r_+$, can now have either a positive or negative sign for $f(r)$. ii) There is a new region III at $r > r_+$, which corresponds to a relabeling of the asymptotically AdS region II described in the previous subsection. Aside from these modifications, all aspects of the discussion regarding region I remain unchanged. We illustrate these new cases schematically in Figure~\ref{fig:gsplots} (Center). From the block diagrams perspective, this means that, besides the previous ones for region I, we now have
\begin{equation}
f(r_-<r<r_+)=\begin{cases}
\vcenter{\hbox{\begin{tikzpicture}[scale=0.85,
point/.style={circle,fill,inner sep=1.3pt}
]
\draw[line width=0.5mm,egyptianblue,fill=egyptianblue,fill opacity=0.1] (0,0) -- (1,-1) -- (2,0)-- (1,1)  -- cycle;  
\node[inner sep=2] at (1,0) {II};
\node[egyptianblue,above,rotate=45] at (0.5,0.5) {$r=r_-$};
\node[egyptianblue,below,rotate=-45] at (0.5,-0.5) { $r=r_-$};
\node[egyptianblue,above,rotate=-45] at (1.5,0.5) { $r=r_+$};
\node[egyptianblue,below,rotate=45] at (1.5,-0.5) { $r=r_+$};
\end{tikzpicture}}}\quad \text{if positive,} \\
\vcenter{\hbox{\begin{tikzpicture}[scale=0.85,
point/.style={circle,fill,inner sep=1.3pt}
]

\draw[line width=0.5mm,egyptianblue,fill=egyptianblue,fill opacity=0.1] (0,0) -- (1,-1) -- (2,0)-- (1,1)  -- cycle;  

\node[inner sep=2] at (1,0) {II};
\node[egyptianblue,above,rotate=45] at (0.5,0.5) {$r=r_-$};
\node[egyptianblue,below,rotate=-45] at (0.5,-0.5) { $r=r_+$};

\node[egyptianblue,above,rotate=-45] at (1.5,0.5) {$r=r_-$};
\node[egyptianblue,below,rotate=45] at (1.5,-0.5) {$r=r_+$};

\end{tikzpicture}}}\quad \text{if negative,}
\end{cases}\quad f(r_+<r<\infty)=
\vcenter{\hbox{\begin{tikzpicture}[scale=0.85,
point/.style={circle,fill,inner sep=1.3pt}
]

\draw[line width=0.5mm,darkspringgreen,fill=darkspringgreen,fill opacity=0.1] (0,0) -- (1,-1) -- (1,1) -- cycle;  

\node[inner sep=2] at (0.6,0) {III};
\node[darkspringgreen,above,rotate=45] at (0.5,0.5) {$r=r_+$};
\node[darkspringgreen,below,rotate=-45] at (0.5,-0.5) { $r=r_+$};

\node[darkspringgreen,above,rotate=-90] at (1,0) { $r\rightarrow\infty$};
\end{tikzpicture}}}\,.
\end{equation}

If we restrict ourselves to Penrose diagrams that include a singular block with $ f(r \to 0) = \pm \infty $ in region I, we observe that the blocks can be combined in four possible ways, which we denote explicitly as $ S_\infty^{\pm\pm} $, namely
\begin{align}
&S^{++}_\infty=\vcenter{\hbox{\begin{tikzpicture}[scale=0.85,
point/.style={circle,fill,inner sep=1.3pt}
]
\draw[line width=0.5mm,carmine,fill=carmine,fill opacity=0.1] (0,0) -- (-1,-1) decorate [singularity]{-- (-1,1)}-- cycle; 
\draw[line width=0.5mm,carmine,fill=carmine,fill opacity=0.1] (0,2) -- (-1,1) decorate [singularity]{-- (-1,3)}-- cycle;   
\draw[line width=0.5mm,darkspringgreen,fill=darkspringgreen,fill opacity=0.1] (0,0) -- (1,-1) -- (1,1)-- cycle; 
\draw[line width=0.5mm,darkspringgreen,fill=darkspringgreen,fill opacity=0.1] (0,2) -- (1,1) -- (1,3)-- cycle; 
 \draw[line width=0.5mm,frenchlilac,fill=egyptianblue,fill opacity=0.1] (0,0)--  (-1,1) -- (0,2) -- (1,1) -- cycle;
\node[inner sep=2] at (-0.6,0) {I};
\node[inner sep=2] at (0.6,0) {III};
\node[inner sep=2] at (0,1) {II};
\node[inner sep=2] at (-0.6,2) {I};
\node[inner sep=2] at (0.6,2) {III};
\filldraw (0,3) circle (0.5pt);
\filldraw (0,2.8) circle (0.5pt);
\filldraw (0,2.6) circle (0.5pt);
\filldraw (0,-1) circle (0.5pt);
\filldraw (0,-0.8) circle (0.5pt);
\filldraw (0,-0.6) circle (0.5pt);
\end{tikzpicture}}}\, ,& \text{ if } \begin{cases}
    \alpha_2=3.2828\,,\\ 
    \alpha_3=-2.1035\,,\\
    \alpha_4=0.32070,
\end{cases}\\
&S^{+-}_\infty=\vcenter{\hbox{\begin{tikzpicture}[scale=0.85,
point/.style={circle,fill,inner sep=1.3pt}
]
\draw[line width=0.5mm,carmine,fill=carmine,fill opacity=0.1] (0,0) -- (-1,-1) decorate [singularity]{-- (-1,1)}-- cycle; 
\draw[line width=0.5mm,carmine,fill=carmine,fill opacity=0.1] (0,0) -- (1,-1) decorate [singularity]{-- (1,1)}-- cycle;   
\draw[line width=0.5mm,darkspringgreen,fill=darkspringgreen,fill opacity=0.1] (0,2) -- (1,1) -- (1,3)-- cycle; 
\draw[line width=0.5mm,darkspringgreen,fill=darkspringgreen,fill opacity=0.1] (0,2) -- (-1,1) -- (-1,3)-- cycle; 
\draw[line width=0.5mm,frenchlilac,fill=egyptianblue,fill opacity=0.1] (0,0)--  (-1,1) -- (0,2) -- (1,1) -- cycle; 
\node[inner sep=2] at (-0.6,0) {I};
\node[inner sep=2] at (0.6,0) {I};
\node[inner sep=2] at (0,1) {II};
\node[inner sep=2] at (-0.6,2) {III};
\node[inner sep=2] at (0.6,2) {III};
\filldraw (0,3) circle (0.5pt);
\filldraw (0,2.8) circle (0.5pt);
\filldraw (0,2.6) circle (0.5pt);
\filldraw (0,-1) circle (0.5pt);
\filldraw (0,-0.8) circle (0.5pt);
\filldraw (0,-0.6) circle (0.5pt);
\end{tikzpicture}}}\,,& \text{ if }\alpha_2=0.5\,,\\
&S^{-+}_\infty=\vcenter{\hbox{\begin{tikzpicture}[scale=0.85,
point/.style={circle,fill,inner sep=1.3pt}
]
\draw[line width=0.5mm,carmine,fill=carmine,fill opacity=0.1] (0,0) -- (-1,-1) decorate [singularity]{-- (1,-1)}-- cycle; 
\draw[line width=0.5mm,carmine,fill=carmine,fill opacity=0.1] (0,0) -- (-1,1) decorate [singularity]{-- (1,1)}-- cycle;
\draw[line width=0.5mm,darkspringgreen,fill=darkspringgreen,fill opacity=0.1] (1,1) -- (2,2) -- (2,0)-- cycle; 
\draw[line width=0.5mm,darkspringgreen,fill=darkspringgreen,fill opacity=0.1] (-1,1) -- (-2,2) -- (-2,0)-- cycle; 
\draw[line width=0.5mm,darkspringgreen,fill=darkspringgreen,fill opacity=0.1] (1,-1) -- (2,-2) -- (2,0)-- cycle; 
\draw[line width=0.5mm,darkspringgreen,fill=darkspringgreen,fill opacity=0.1] (-1,-1) -- (-2,-2) -- (-2,0)-- cycle; 
\draw[line width=0.5mm,frenchlilac,fill=egyptianblue,fill opacity=0.1] (0,0)--  (1,1) -- (2,0) -- (1,-1) -- cycle;
\draw[line width=0.5mm,frenchlilac,fill=egyptianblue,fill opacity=0.1] (0,0)--  (-1,1) -- (-2,0) -- (-1,-1) -- cycle;
\filldraw (0,2) circle (0.5pt);
\filldraw (0,1.8) circle (0.5pt);
\filldraw (0,1.6) circle (0.5pt);
\filldraw (0,-2) circle (0.5pt);
\filldraw (0,-1.8) circle (0.5pt);
\filldraw (0,-1.6) circle (0.5pt);
\node[inner sep=2] at (0,-0.6) {I};
\node[inner sep=2] at (0,0.6) {I};
\node[inner sep=2] at (1,0) {II};
\node[inner sep=2] at (-1,0) {II};
\node[inner sep=2] at (1.6,1) {III};
\node[inner sep=2] at (-1.6,1) {III};
\node[inner sep=2] at (1.6,-1) {III};
\node[inner sep=2] at (-1.6,-1) {III};
\end{tikzpicture}}}\,,&\text{ if } \alpha_2=\frac{5}{2}\,, \alpha_3=-1\,,\label{eq:S--inf}\\
&
S^{--}_\infty=\vcenter{\hbox{\begin{tikzpicture}[scale=0.85,
point/.style={circle,fill,inner sep=1.3pt}
]
\draw[line width=0.5mm,carmine,fill=carmine,fill opacity=0.1] (-1,-1) --(0,-2) decorate [singularity]{-- (-2,-2)}-- cycle; 
\draw[line width=0.5mm,carmine,fill=carmine,fill opacity=0.1] (1,-1) --(0,-2) decorate [singularity]{-- (2,-2)}-- cycle; 
\draw[line width=0.5mm,carmine,fill=carmine,fill opacity=0.1] (-1,1) --(-2,2) decorate [singularity]{-- (0,2)}-- cycle; 
\draw[line width=0.5mm,carmine,fill=carmine,fill opacity=0.1] (1,1) --(2,2) decorate [singularity]{-- (0,2)}-- cycle; 
\draw[line width=0.5mm,carmine,fill=carmine,fill opacity=0.1] (-3,-1) --(-2,-2) decorate [singularity]{-- (-4,-2)}-- cycle; 
\draw[line width=0.5mm,carmine,fill=carmine,fill opacity=0.1] (-3,1) --(-4,2) decorate [singularity]{-- (-2,2)}-- cycle; 
\draw[line width=0.5mm,carmine,fill=carmine,fill opacity=0.1] (3,-1) --(4,-2) decorate [singularity]{-- (2,-2)}-- cycle; 
\draw[line width=0.5mm,carmine,fill=carmine,fill opacity=0.1] (3,1) --(2,2) decorate [singularity]{-- (4,2)}-- cycle; 
\draw[line width=0.5mm,darkspringgreen,fill=darkspringgreen,fill opacity=0.1] (0,0) -- (-1,-1) -- (-1,1)-- cycle; 
\draw[line width=0.5mm,darkspringgreen,fill=darkspringgreen,fill opacity=0.1] (0,0) -- (1,-1) -- (1,1)-- cycle; 
\draw[line width=0.5mm,darkspringgreen,fill=darkspringgreen,fill opacity=0.1] (2,0) -- (1,-1) -- (1,1)-- cycle; 
\draw[line width=0.5mm,darkspringgreen,fill=darkspringgreen,fill opacity=0.1] (-2,0) -- (-1,-1) -- (-1,1)-- cycle; 
\draw[line width=0.5mm,darkspringgreen,fill=darkspringgreen,fill opacity=0.1] (-2,0) -- (-3,-1) -- (-3,1)-- cycle; 
\draw[line width=0.5mm,darkspringgreen,fill=darkspringgreen,fill opacity=0.1] (2,0) -- (3,-1) -- (3,1)-- cycle; 
\draw[line width=0.5mm,frenchlilac,fill=egyptianblue,fill opacity=0.1] (0,0)--  (1,1) -- (0,2) -- (-1,1) -- cycle;
\draw[line width=0.5mm,frenchlilac,fill=egyptianblue,fill opacity=0.1] (0,0)--  (1,-1) -- (0,-2) -- (-1,-1) -- cycle;
\draw[line width=0.5mm,frenchlilac,fill=egyptianblue,fill opacity=0.1] (2,0)--  (3,1) -- (2,2) -- (1,1) -- cycle;
\draw[line width=0.5mm,frenchlilac,fill=egyptianblue,fill opacity=0.1] (2,0)--  (3,-1) -- (2,-2) -- (1,-1) -- cycle;
\draw[line width=0.5mm,frenchlilac,fill=egyptianblue,fill opacity=0.1] (-2,0)--  (-1,1) -- (-2,2) -- (-3,1) -- cycle;
\draw[line width=0.5mm,frenchlilac,fill=egyptianblue,fill opacity=0.1] (-2,0)--  (-1,-1) -- (-2,-2) -- (-3,-1) -- cycle;
\node[inner sep=2] at (-1,-1.6) {I};
\node[inner sep=2] at (1,-1.6) {I};
\node[inner sep=2] at (-1,1.6) {I};
\node[inner sep=2] at (1,1.6) {I};
\node[inner sep=2] at (-3,-1.6) {I};
\node[inner sep=2] at (3,-1.6) {I};
\node[inner sep=2] at (-3,1.6) {I};
\node[inner sep=2] at (3,1.6) {I};
\node[inner sep=2] at (0,1) {II};
\node[inner sep=2] at (0,-1) {II};
\node[inner sep=2] at (2,1) {II};
\node[inner sep=2] at (2,-1) {II};
\node[inner sep=2] at (-2,1) {II};
\node[inner sep=2] at (-2,-1) {II};
\node[inner sep=2] at (0.6,0) {III};
\node[inner sep=2] at (-0.6,0) {III};
\node[inner sep=2] at (1.4,0) {III};
\node[inner sep=2] at (-1.4,0) {III};
\node[inner sep=2] at (2.6,0) {III};
\node[inner sep=2] at (-2.6,0) {III};
\filldraw (4,0) circle (0.5pt);
\filldraw (3.8,0) circle (0.5pt);
\filldraw (3.6,0) circle (0.5pt);
\filldraw (-4,0) circle (0.5pt);
\filldraw (-3.8,0) circle (0.5pt);
\filldraw (-3.6,0) circle (0.5pt);
\end{tikzpicture}}}\,,& \text{ if } \alpha_2=-0.003490\,.
\end{align}
Similarly to the previous subsection, when the values of the couplings are expressed as fractions, they represent exact values. Decimals with five significant figures also imply an exact number. In contrast, quantities given with only one significant figure indicate some degree of flexibility---in the sense that we can locally modify its value and still obtain the same global structure. If we compare the new diagrams to those with only one horizon in the previous subsection---\eqref{eq:S+inf} to \eqref{eq:S-inf}, we now observe that all of them extend infinitely either along the timelike direction or along the spacelike one. The diagrams $S^{\pm+}_\infty$ represent extremal black holes and so are their transformed versions $S^{\pm+}_\text{c}$, $R^{++}_1$ and $R^{\pm+}_{0}$.

By performing the operations described in equations \eqref{eq:pentr1} to \eqref{eq:pentr5}---this is, turning on the $\beta_j$ described in the equations, we are able to find the other ten Penrose diagrams, namely, $S^{\pm\pm}_\text{c}$, $R^{\pm\pm}_\text{0}$ and $R^{+\pm}_\text{1}$. These only differ in region I, which is replaced with its corresponding regularized version. 

\begin{figure}
    \centering
 \includegraphics[scale=1]{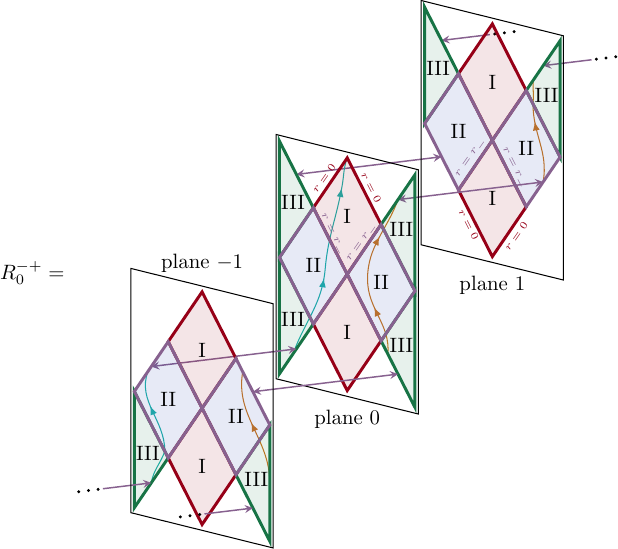}   
    \caption{Nonplanar Penrose diagram of the  $R^{-+}_0$  spacetime. We depict three planes---specifically, planes $-1$, 0, and 1---out of the infinitely many, along with the trajectories of two massive observers, which are shown in brown and aquamarine. For clarity, in planes 0 and 1, we show the radii corresponding to the null boundaries of region I, highlighting that they cannot be glued in the same plane.}
    \label{fig:nonplanar}
\end{figure}

Let us examine the solution $R^{-+}_0$  in detail. Its Penrose diagram resembles the one presented in \eqref{eq:S--inf}, but with its singular block diagrams replaced by diamond-shaped blocks. Interestingly, when attempting to represent it in two dimensions, an overlap arises between two distinct regions  I  whose boundaries correspond to different radii. To resolve this, we may choose to use multiple planes, as illustrated in Figure~\ref{fig:nonplanar}. Consequently, we refer to this representation as a \textit{nonplanar Penrose diagram}.

Observe that in this diagram we replaced the infinite extension in the timelike direction with an infinite sequence of planes, each containing compact diagrams. Focusing on the internal bifurcation point in plane 0, the past asymptotic regions III are connected to regions II of plane $-1$, while the future ones are connected to plane 1. 

To illustrate the global structure in this example, we have included two possible geodesics that massive observers may follow. The left aquamarine observer begins their journey in region III of plane $-1$, crosses into plane 0, and continues towards $r=0$ in region I. In contrast, the right brown observer starts in a parallel region III of plane $-1$. Upon crossing into plane 0, they avoid the aquamarine observer and instead continue their trajectory into plane 1. From there, they proceed to a subsequent plane where their path extends beyond the scope of this diagram. Spacetime $R^{-+}_0$ turns out to be the only one displaying this nonplanar structure if we restrict ourselves to black holes with two horizons. However, as the number of horizons increases, the number of nonplanar diagrams also grows. We are not aware of previous examples of nonplanar Penrose diagrams.

\subsection{Black holes with three horizons and beyond}\label{ref:Pd3hor}

At this point, it is important to recall that the metric function $f(r)$ can describe black holes with an arbitrary number of horizons. This is enabled by turning on as many $\alpha_i$ coefficients as we desire and later on regularize them by turning on $\beta_j$ described in the diagram regularization scheme presented in \eqref{eq:pentr1} to \eqref{eq:pentr1}. Therefore, we shall discuss black holes with three horizons, which are less explored in the literature, and conclude the section with some remarks on black holes with even more horizons.

In this case, we denote the roots of $f(r)$ as $r_1$, $r_2$ and $r_+$, taking $r_1<r_2<r_+$ so that $r_+$ represents the outermost horizon, following the previous convention. As before, we label the region near the origin as I, while regions II and III correspond to the intervals between $r_1$ and $r_2$, and $r_2$ and $r_+$, respectively. The asymptotic region is labeled IV. Depending on the sign of $f(r)$, the corresponding block diagrams for each region must be oriented so that the timelike coordinate runs vertically, this is
\begin{equation}
f(r_1<r<r_2)=\begin{cases}
\vcenter{\hbox{\begin{tikzpicture}[scale=0.85,
point/.style={circle,fill,inner sep=1.3pt}
]

\draw[line width=0.5mm,egyptianblue,fill=egyptianblue,fill opacity=0.1] (0,0) -- (1,-1) -- (2,0)-- (1,1)  -- cycle;  

\node[inner sep=2] at (1,0) {II};
\node[egyptianblue,above,rotate=45] at (0.5,0.5) {$r=r_1$};
\node[egyptianblue,below,rotate=-45] at (0.5,-0.5) { $r=r_1$};

\node[egyptianblue,above,rotate=-45] at (1.5,0.5) { $r=r_2$};
\node[egyptianblue,below,rotate=45] at (1.5,-0.5) { $r=r_2$};

\end{tikzpicture}}}\quad \text{if positive,} \\
\vcenter{\hbox{\begin{tikzpicture}[scale=0.85,
point/.style={circle,fill,inner sep=1.3pt}
]

\draw[line width=0.5mm,egyptianblue,fill=egyptianblue,fill opacity=0.1] (0,0) -- (1,-1) -- (2,0)-- (1,1)  -- cycle;  

\node[inner sep=2] at (1,0) {II};
\node[egyptianblue,above,rotate=45] at (0.5,0.5) {$r=r_1$};
\node[egyptianblue,below,rotate=-45] at (0.5,-0.5) { $r=r_2$};

\node[egyptianblue,above,rotate=-45] at (1.5,0.5) {$r=r_1$};
\node[egyptianblue,below,rotate=45] at (1.5,-0.5) {$r=r_2$};

\end{tikzpicture}}}\quad \text{if negative,}
\end{cases}
\end{equation}
\begin{equation}
f(r_2<r<r_+)=\begin{cases}
\vcenter{\hbox{\begin{tikzpicture}[scale=0.85,
point/.style={circle,fill,inner sep=1.3pt}
]

\draw[line width=0.5mm,darkspringgreen,fill=darkspringgreen,fill opacity=0.1] (0,0) -- (1,-1) -- (2,0)-- (1,1)  -- cycle;  

\node[inner sep=2] at (1,0) {III};
\node[darkspringgreen,above,rotate=45] at (0.5,0.5) {$r=r_2$};
\node[darkspringgreen,below,rotate=-45] at (0.5,-0.5) { $r=r_2$};

\node[darkspringgreen,above,rotate=-45] at (1.5,0.5) { $r=r_+$};
\node[darkspringgreen,below,rotate=45] at (1.5,-0.5) { $r=r_+$};

\end{tikzpicture}}}\quad \text{if positive,} \\
\vcenter{\hbox{\begin{tikzpicture}[scale=0.85,
point/.style={circle,fill,inner sep=1.3pt}
]

\draw[line width=0.5mm,darkspringgreen,fill=darkspringgreen,fill opacity=0.1] (0,0) -- (1,-1) -- (2,0)-- (1,1)  -- cycle;  

\node[inner sep=2] at (1,0) {III};
\node[darkspringgreen,above,rotate=45] at (0.5,0.5) {$r=r_2$};
\node[darkspringgreen,below,rotate=-45] at (0.5,-0.5) { $r=r_+$};

\node[darkspringgreen,above,rotate=-45] at (1.5,0.5) {$r=r_2$};
\node[darkspringgreen,below,rotate=45] at (1.5,-0.5) {$r=r_+$};

\end{tikzpicture}}}\quad \text{if negative,}
\end{cases}\quad f(r_+<r<\infty)=
\vcenter{\hbox{\begin{tikzpicture}[scale=0.85,
point/.style={circle,fill,inner sep=1.3pt}
]

\draw[line width=0.5mm,ochre,fill=ochre,fill opacity=0.1] (0,0) -- (1,-1) -- (1,1) -- cycle;  

\node[inner sep=2] at (0.6,0) {IV};
\node[ochre,above,rotate=45] at (0.5,0.5) {$r=r_+$};
\node[ochre,below,rotate=-45] at (0.5,-0.5) { $r=r_+$};

\node[ochre,above,rotate=-90] at (1,0) { $r\rightarrow\infty$};
\end{tikzpicture}}}\,,
\end{equation}
while we can reuse the block diagrams \eqref{eq:S+infb} to \eqref{eq:S-infb} relabeling $r_-$ as $r_1$.

\begin{figure}
\centering
\includegraphics[width=0.5\textwidth]{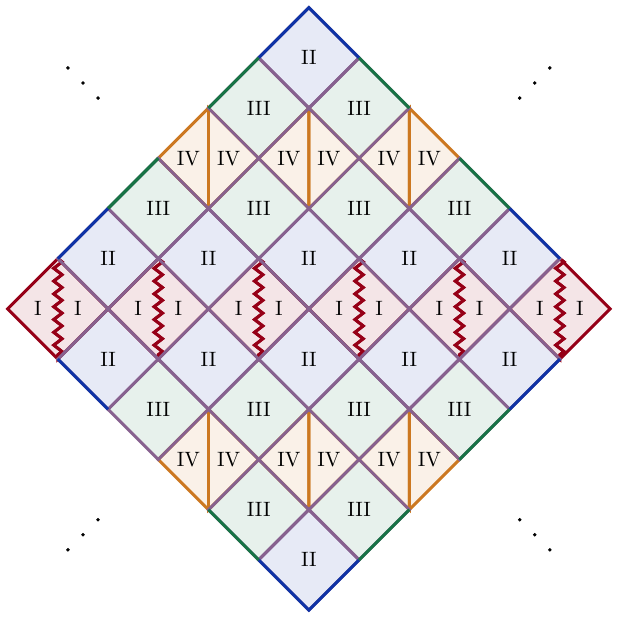}       
\caption{We present an alternative but equivalent diagram to \eqref{eq:S+--inf}, in which it becomes manifest that $S^{+--}_\infty$ tessellates the two-dimensional plane. This tessellation occurs as the square can be translated diagonally, with each side seamlessly joining the opposite of an adjacent one.}
\label{fig:tes}
\end{figure}

Again, all possible combinations of block diagrams account for the different global structures existing with three horizons. In this case we have twice as many possibilities with respect of having two horizons. This can be easily seen from Figure~\ref{fig:gsplots}, as an additional root to $f(r)$ implies having two new possibilities in the new region, either $f(r)$ is positive, or negative. As a  consequence, for three horizons we identify twenty-eight distinct Penrose diagrams, or global structures. Below we list the eight corresponding ones to singular black holes, together with the choices of the coupling constants that realize them,
\begin{align}
&S^{+++}_\infty=\vcenter{\hbox{\begin{tikzpicture}[scale=0.85,
point/.style={circle,fill,inner sep=1.3pt}
]
\draw[line width=0.5mm,carmine,fill=carmine,fill opacity=0.1] (0,0) -- (-1,1) decorate [singularity] {-- (-1,-1)} -- cycle;
\draw[line width=0.5mm,carmine,fill=carmine,fill opacity=0.1] (0,2) -- (-1,3) decorate [singularity] {-- (-1,1)} -- cycle;
\draw[line width=0.5mm,carmine,fill=carmine,fill opacity=0.1] (0,-2) -- (-1,-3) decorate [singularity] {-- (-1,-1)} -- cycle;
\draw[line width=0.5mm,ochre,fill=ochre,fill opacity=0.1] (1,-1) -- (2,-2) -- (2,0) -- cycle;
\draw[line width=0.5mm,ochre,fill=ochre,fill opacity=0.1] (1,1) -- (2,0) -- (2,2) -- cycle;
\draw[line width=0.5mm,ochre,fill=ochre,fill opacity=0.1] (1,3) -- (2,2) -- (2,4) -- cycle;
\draw[line width=0.5mm,frenchlilac,fill=egyptianblue,fill opacity=0.1] (0,0) -- (1,-1) -- (0,-2) -- (-1,-1) -- cycle;
\draw[line width=0.5mm,frenchlilac,fill=egyptianblue,fill opacity=0.1] (0,2) -- (1,1) -- (0,0) -- (-1,1) -- cycle;
\draw[line width=0.5mm,frenchlilac,fill=egyptianblue,fill opacity=0.1] (0,4) -- (1,3) -- (0,2) -- (-1,3) -- cycle;
\draw[line width=0.5mm,frenchlilac,fill=darkspringgreen,fill opacity=0.1] (0,0) -- (1,1) -- (2,0) -- (1,-1) -- cycle;
\draw[line width=0.5mm,frenchlilac,fill=darkspringgreen,fill opacity=0.1] (0,2) -- (1,3) -- (2,2) -- (1,1) -- cycle;
\draw[line width=0.5mm,frenchlilac,fill=darkspringgreen,fill opacity=0.1] (0,-2) -- (1,-1) -- (2,-2) -- (1,-3) -- cycle;
\draw[line width=0.5mm,egyptianblue] (-1,3) -- (0,4) -- (1,3);
\draw[line width=0.5mm,darkspringgreen] (0,-2) -- (1,-3) -- (2,-2);
\node[inner sep=2] at (-0.6,0) {I};
\node[inner sep=2] at (-0.6,2) {I};
\node[inner sep=2] at (-0.6,-2) {I};
\node[inner sep=2] at (0,-1) { II};
\node[inner sep=2] at (0,3) {II};
\node[inner sep=2] at (0,1) {II};
\node[inner sep=2] at (1,0) {III};
\node[inner sep=2] at (1,2) {III};
\node[inner sep=2] at (1,-2) {III};
\node[inner sep=2] at (1.6,-1) {IV};
\node[inner sep=2] at (1.6,1) {IV};
\node[inner sep=2] at (1.6,3) {IV};
\filldraw (0.5,4) circle (0.5pt);
\filldraw (0.5,4.2) circle (0.5pt);
\filldraw (0.5,4.4) circle (0.5pt);
\filldraw (0.5,-3) circle (0.5pt);
\filldraw (0.5,-3.2) circle (0.5pt);
\filldraw (0.5,-3.4) circle (0.5pt);
\end{tikzpicture}}}\, ,&\text{ if }
\begin{cases}
\alpha_2=6.0786\,,\\
\alpha_3=-6.9424\,,\\
\alpha_4=2.7689\,,\\
\alpha_5=-0.42687\,,\\
\alpha_6=0.021838\,,
\end{cases}\\
&
S^{++-}_\infty=\vcenter{\hbox{\begin{tikzpicture}[scale=0.85,
point/.style={circle,fill,inner sep=1.3pt}
]
\draw[line width=0.5mm,carmine,fill=carmine,fill opacity=0.1] (-1,-1) -- (-2,-2) decorate [singularity] {-- (-2,0)} -- cycle;
\draw[line width=0.5mm,carmine,fill=carmine,fill opacity=0.1] (-1,1) -- (-2,2) decorate [singularity] {-- (-2,0)} -- cycle;
\draw[line width=0.5mm,carmine,fill=carmine,fill opacity=0.1] (1,-1) -- (2,-2) decorate [singularity] {-- (2,0)} -- cycle;
\draw[line width=0.5mm,carmine,fill=carmine,fill opacity=0.1] (1,1) -- (2,2) decorate [singularity] {-- (2,0)} -- cycle;
\draw[line width=0.5mm,ochre,fill=ochre,fill opacity=0.1] (-1,1) -- (-1,3) -- (0,2) -- cycle;
\draw[line width=0.5mm,ochre,fill=ochre,fill opacity=0.1] (1,1) -- (1,3) -- (0,2) -- cycle;
\draw[line width=0.5mm,ochre,fill=ochre,fill opacity=0.1] (-1,-1) -- (-1,-3) -- (0,-2) -- cycle;
\draw[line width=0.5mm,ochre,fill=ochre,fill opacity=0.1] (1,-1) -- (1,-3) -- (0,-2) -- cycle;
\draw[line width=0.5mm,frenchlilac,fill=egyptianblue,fill opacity=0.1] (0,0) -- (-1,1) -- (-2,0)-- (-1,-1) -- cycle;
\draw[line width=0.5mm,frenchlilac,fill=egyptianblue,fill opacity=0.1] (0,0) -- (1,1) -- (2,0)-- (1,-1) -- cycle;
\draw[line width=0.5mm,frenchlilac,fill=darkspringgreen,fill opacity=0.1] (0,0) -- (-1,1) -- (0,2)-- (1,1) -- cycle;
\draw[line width=0.5mm,frenchlilac,fill=darkspringgreen,fill opacity=0.1] (0,0) -- (-1,-1) -- (0,-2)-- (1,-1) -- cycle;
\node[inner sep=2] at (-1.6,1) {I};
\node[inner sep=2] at (-1.6,-1) {I};
\node[inner sep=2] at (1.6,1) {I};
\node[inner sep=2] at (1.6,-1) {I};
\node[inner sep=2] at (0,1) {III};
\node[inner sep=2] at (0,-1) {III};
\node[inner sep=2] at (1,0) {II};
\node[inner sep=2] at (-1,0) {II};
\node[inner sep=2] at (-0.6,2) {IV};
\node[inner sep=2] at (-0.6,-2) {IV};
\node[inner sep=2] at (0.6,2) {IV};
\node[inner sep=2] at (0.6,-2) {IV};
\filldraw (0,3) circle (0.5pt);
\filldraw (0,2.8) circle (0.5pt);
\filldraw (0,2.6) circle (0.5pt); 
\filldraw (0,-3) circle (0.5pt);
\filldraw (0,-2.8) circle (0.5pt);
\filldraw (0,-2.6) circle (0.5pt);
\filldraw (1.4,1.6) circle (0.5pt);
\filldraw (1.3,1.7) circle (0.5pt);
\filldraw (1.2,1.8) circle (0.5pt);
\filldraw (-1.4,1.6) circle (0.5pt);
\filldraw (-1.3,1.7) circle (0.5pt);
\filldraw (-1.2,1.8) circle (0.5pt);
\filldraw (1.4,-1.6) circle (0.5pt);
\filldraw (1.3,-1.7) circle (0.5pt);
\filldraw (1.2,-1.8) circle (0.5pt);
\filldraw (-1.4,-1.6) circle (0.5pt);
\filldraw (-1.3,-1.7) circle (0.5pt);
\filldraw (-1.2,-1.8) circle (0.5pt);
\end{tikzpicture}}}\, ,&\text{ if }
\begin{cases}
\alpha_2=1.0595\,,\\
\alpha_3=-0.34749\,,\\
\alpha_4=0.025149\,,\\
\end{cases}\\
&
S^{+-+}_\infty=\vcenter{\hbox{\begin{tikzpicture}[scale=0.85,
point/.style={circle,fill,inner sep=1.3pt}
]
\draw[line width=0.5mm,carmine,fill=carmine,fill opacity=0.1] (0,0) -- (-1,1) decorate [singularity] {-- (-1,-1)} -- cycle;
\draw[line width=0.5mm,carmine,fill=carmine,fill opacity=0.1] (0,0) -- (1,1) decorate [singularity] {-- (1,-1)} -- cycle;
\draw[line width=0.5mm,ochre,fill=ochre,fill opacity=0.1] (1,1) -- (2,2) -- (2,0) -- cycle;
\draw[line width=0.5mm,ochre,fill=ochre,fill opacity=0.1] (-1,1) -- (-2,2) -- (-2,0) -- cycle;
\draw[line width=0.5mm,ochre,fill=ochre,fill opacity=0.1] (1,-1) -- (2,-2) -- (2,0) -- cycle;
\draw[line width=0.5mm,ochre,fill=ochre,fill opacity=0.1] (-1,-1) -- (-2,-2) -- (-2,0) -- cycle;
\draw[line width=0.5mm,ochre,fill=ochre,fill opacity=0.1] (-1,-3) -- (-2,-4) -- (-2,-2) -- cycle;
\draw[line width=0.5mm,ochre,fill=ochre,fill opacity=0.1] (-1,3) -- (-2,4) -- (-2,2) -- cycle;
\draw[line width=0.5mm,ochre,fill=ochre,fill opacity=0.1] (1,-3) -- (2,-4) -- (2,-2) -- cycle;
\draw[line width=0.5mm,ochre,fill=ochre,fill opacity=0.1] (1,3) -- (2,4) -- (2,2) -- cycle;
\draw[line width=0.5mm,frenchlilac,fill=darkspringgreen,fill opacity=0.1] (0,2) -- (1,1) -- (2,2) -- (1,3) -- cycle;
\draw[line width=0.5mm,frenchlilac,fill=darkspringgreen,fill opacity=0.1] (0,2) -- (-1,1) -- (-2,2) -- (-1,3) -- cycle;
\draw[line width=0.5mm,frenchlilac,fill=darkspringgreen,fill opacity=0.1] (0,-2) -- (1,-1) -- (2,-2) -- (1,-3) -- cycle;
\draw[line width=0.5mm,frenchlilac,fill=darkspringgreen,fill opacity=0.1] (0,-2) -- (-1,-1) -- (-2,-2) -- (-1,-3) -- cycle;
\draw[line width=0.5mm,darkspringgreen] (-1,-3) -- (0,-2) -- (1,-3);
\draw[line width=0.5mm,darkspringgreen] (-1,3) -- (0,2) -- (1,3);
\draw[line width=0.5mm,frenchlilac,fill=egyptianblue,fill opacity=0.1] (0,0) -- (1,1) -- (0,2) -- (-1,1) -- cycle;
\draw[line width=0.5mm,frenchlilac,fill=egyptianblue,fill opacity=0.1] (0,0) -- (1,-1) -- (0,-2) -- (-1,-1) -- cycle;
\node[inner sep=2] at (0.6,0) {I};
\node[inner sep=2] at (-0.6,0) {I};
\node[inner sep=2] at (0,1) {II};
\node[inner sep=2] at (0,-1) {II};
\node[inner sep=2] at (1,2) {III};
\node[inner sep=2] at (1,-2) {III};
\node[inner sep=2] at (-1,2) {III};
\node[inner sep=2] at (-1,-2) {III};
\node[inner sep=2] at (1.6,1) {IV};
\node[inner sep=2] at (1.6,-1) {IV};
\node[inner sep=2] at (-1.6,1) {IV};
\node[inner sep=2] at (-1.6,-1) {IV};
\node[inner sep=2] at (1.6,3) {IV};
\node[inner sep=2] at (1.6,-3) {IV};
\node[inner sep=2] at (-1.6,3) {IV};
\node[inner sep=2] at (-1.6,-3) {IV};
\filldraw (0,3) circle (0.5pt);
\filldraw (0,3.2) circle (0.5pt);
\filldraw (0,2.8) circle (0.5pt);
\filldraw (0,-3) circle (0.5pt);
\filldraw (0,-3.2) circle (0.5pt);
\filldraw (0,-2.8) circle (0.5pt);
\filldraw (1+0.303553,0+0.303553) circle (0.5pt);
\filldraw (1+0.203553,0+0.203553) circle (0.5pt);
\filldraw (1+0.403553,0+0.403553) circle (0.5pt);
\filldraw (-1-0.303553,0+0.303553) circle (0.5pt);
\filldraw (-1-0.203553,0+0.203553) circle (0.5pt);
\filldraw (-1-0.403553,0+0.403553) circle (0.5pt);
\filldraw (1+0.303553,0-0.303553) circle (0.5pt);
\filldraw (1+0.203553,0-0.203553) circle (0.5pt);
\filldraw (1+0.403553,0-0.403553) circle (0.5pt);
\filldraw (-1-0.303553,0-0.303553) circle (0.5pt);
\filldraw (-1-0.203553,0-0.203553) circle (0.5pt);
\filldraw (-1-0.403553,0-0.403553) circle (0.5pt);
\end{tikzpicture}}}\, ,&\text{ if }
\begin{cases}
\alpha_2=3.1801\,,\\
\alpha_3=-1.7801\,,\\
\alpha_4=0.1\,,\\
\end{cases}\label{eq:S+-+inf}\\
&
\label{eq:S+--inf} S^{+--}_\infty=\vcenter{\hbox{\begin{tikzpicture}[scale=0.85,
point/.style={circle,fill,inner sep=1.3pt}
]
\draw[line width=0.5mm,carmine,fill=carmine,fill opacity=0.1] (1,1) decorate [singularity] {-- (1,3)} -- (0,2) -- cycle;
\draw[line width=0.5mm,carmine,fill=carmine,fill opacity=0.1] (-1,1) decorate [singularity] {-- (-1,3)} -- (0,2) -- cycle;
\draw[line width=0.5mm,carmine,fill=carmine,fill opacity=0.1] (1,-3) decorate [singularity] {-- (1,-5)} -- (0,-4) -- cycle;
\draw[line width=0.5mm,carmine,fill=carmine,fill opacity=0.1] (-1,-3) decorate [singularity] {-- (-1,-5)} -- (0,-4) -- cycle;
\draw[line width=0.5mm,ochre,fill=ochre,fill opacity=0.1] (-1,-1) -- (-2,-2) -- (-2,0) -- cycle;
\draw[line width=0.5mm,ochre,fill=ochre,fill opacity=0.1] (1,-1) -- (2,-2) -- (2,0) -- cycle;
\draw[line width=0.5mm,ochre,fill=ochre,fill opacity=0.1] (1,-1) -- (0,-2) -- (0,0) -- cycle;
\draw[line width=0.5mm,ochre,fill=ochre,fill opacity=0.1] (-1,-1) -- (0,-2) -- (0,0) -- cycle;
\draw[line width=0.5mm,frenchlilac,fill=darkspringgreen,fill opacity=0.1] (0,0) -- (-1,1) -- (-2,0)-- (-1,-1) -- cycle;
\draw[line width=0.5mm,frenchlilac,fill=darkspringgreen,fill opacity=0.1] (0,0) -- (1,1) -- (2,0)-- (1,-1) -- cycle;
\draw[line width=0.5mm,frenchlilac,fill=darkspringgreen,fill opacity=0.1] (0,-2) -- (-1,-1) -- (-2,-2)-- (-1,-3) -- cycle;
\draw[line width=0.5mm,frenchlilac,fill=darkspringgreen,fill opacity=0.1] (0,-2) -- (1,-1) -- (2,-2)-- (1,-3) -- cycle;
\draw[line width=0.5mm,frenchlilac,fill=egyptianblue,fill opacity=0.1] (0,0) -- (-1,1) -- (0,2)-- (1,1) -- cycle;
\draw[line width=0.5mm,frenchlilac,fill=egyptianblue,fill opacity=0.1] (0,-2) -- (-1,-3) -- (0,-4)-- (1,-3) -- cycle;
\node[inner sep=2] at (-1.6,-1) {IV};
\node[inner sep=2] at (1.6,-1) {IV};
\node[inner sep=2] at (-0.4,-1) {IV};
\node[inner sep=2] at (0.4,-1) {IV};
\node[inner sep=2] at (0,1) {II};
\node[inner sep=2] at (0,-3) {II};
\node[inner sep=2] at (1,0) {III};
\node[inner sep=2] at (-1,0) {III};
\node[inner sep=2] at (1,-2) {III};
\node[inner sep=2] at (-1,-2) {III};
\node[inner sep=2] at (-0.6,2) {I};
\node[inner sep=2] at (0.6,2) {I};
\node[inner sep=2] at (-0.6,-4) {I};
\node[inner sep=2] at (0.6,-4) {I};
\filldraw (0,3) circle (0.5pt);
\filldraw (0,2.8) circle (0.5pt);
\filldraw (0,2.6) circle (0.5pt); 
\filldraw (0,-5) circle (0.5pt);
\filldraw (0,-4.8) circle (0.5pt);
\filldraw (0,-4.6) circle (0.5pt); 
\filldraw (-2.2,-1) circle (0.5pt);
\filldraw (-2.4,-1) circle (0.5pt);
\filldraw (-2.6,-1) circle (0.5pt); 
\filldraw (2.2,-1) circle (0.5pt);
\filldraw (2.4,-1) circle (0.5pt);
\filldraw (2.6,-1) circle (0.5pt); 
\end{tikzpicture}}}\, ,&\text{ if }
\begin{cases}
\alpha_2=1.9016\,,\\
\alpha_3=-0.99335\,,\\
\alpha_4=0.1\,,\\
\end{cases}\\
&
S^{-++}_\infty=\vcenter{\hbox{\begin{tikzpicture}[scale=0.85,
point/.style={circle,fill,inner sep=1.3pt}
]
\draw[line width=0.5mm,carmine,fill=carmine,fill opacity=0.1] (0,0) -- (-1,1) decorate [singularity] {-- (1,1)} -- cycle;
\draw[line width=0.5mm,carmine,fill=carmine,fill opacity=0.1] (0,0) -- (-1,-1) decorate [singularity] {-- (1,-1)} -- cycle;
\draw[line width=0.5mm,ochre,fill=ochre,fill opacity=0.1] (-3,1) -- (-2,2) -- (-3,3) -- cycle;
\draw[line width=0.5mm,ochre,fill=ochre,fill opacity=0.1] (-3,-1) -- (-2,0) -- (-3,1) -- cycle;
\draw[line width=0.5mm,ochre,fill=ochre,fill opacity=0.1] (-3,-1) -- (-2,-2) -- (-3,-3) -- cycle;
\draw[line width=0.5mm,ochre,fill=ochre,fill opacity=0.1] (3,1) -- (2,2) -- (3,3) -- cycle;
\draw[line width=0.5mm,ochre,fill=ochre,fill opacity=0.1] (3,-1) -- (2,-2) -- (3,-3) -- cycle;
\draw[line width=0.5mm,ochre,fill=ochre,fill opacity=0.1] (3,1) -- (2,0) -- (3,-1) -- cycle;
\draw[line width=0.5mm,frenchlilac,fill=darkspringgreen,fill opacity=0.1] (-2,0) -- (-1,1) -- (-2,2) -- (-3,1)-- cycle;
\draw[line width=0.5mm,frenchlilac,fill=darkspringgreen,fill opacity=0.1] (-2,0) -- (-1,-1) -- (-2,-2) -- (-3,-1)-- cycle;
\draw[line width=0.5mm,frenchlilac,fill=darkspringgreen,fill opacity=0.1] (2,0) --(1,1) -- (2,2) -- (3,1)-- cycle;
\draw[line width=0.5mm,frenchlilac,fill=darkspringgreen,fill opacity=0.1] (2,0) -- (1,-1) -- (2,-2) -- (3,-1) -- cycle;
\draw[line width=0.5mm,frenchlilac,fill=egyptianblue,fill opacity=0.1] (0,0) -- (-1,1) -- (-2,0) -- (-1,-1) -- cycle;
\draw[line width=0.5mm,frenchlilac,fill=egyptianblue,fill opacity=0.1] (0,0) -- (1,1) -- (2,0) -- (1,-1) -- cycle;
\draw[line width=0.5mm,darkspringgreen] (-2,2) -- (-1,1);
\draw[line width=0.5mm,darkspringgreen] (2,2) -- (1,1);
\draw[line width=0.5mm,darkspringgreen] (-2,-2) -- (-1,-1);
\draw[line width=0.5mm,darkspringgreen] (2,-2) -- (1,-1);
\node[inner sep=2] at (0,0.6) {I};
\node[inner sep=2] at (0,-0.6) {I};
\node[inner sep=2] at (1,0) {II};
\node[inner sep=2] at (-1,0) {II};
\node[inner sep=2] at (2,1) {III};
\node[inner sep=2] at (2,-1) {III};
\node[inner sep=2] at (-2,1) {III};
\node[inner sep=2] at (-2,-1) {III};
\node[inner sep=2] at (2.6,2) {IV};
\node[inner sep=2] at (2.6,0) {IV};
\node[inner sep=2] at (-2.6,0) {IV};
\node[inner sep=2] at (-2.6,2) {IV};
\node[inner sep=2] at (2.6,-2) {IV};
\node[inner sep=2] at (-2.6,-2) {IV};
\filldraw (0,-1.8) circle (0.5pt);
\filldraw (0,-2.2) circle (0.5pt);
\filldraw (0,-2) circle (0.5pt); 
\filldraw (0,1.8) circle (0.5pt);
\filldraw (0,2.2) circle (0.5pt);
\filldraw (0,2) circle (0.5pt); 
\end{tikzpicture}}}\, ,&\text{ if }
\begin{cases}
\alpha_2=5.2766\,,\\
\alpha_3=-5.2082\,,\\
\alpha_4=1.5316\,,\\
\alpha_5=-0.1\,,
\end{cases}\\
&
S^{-+-}_\infty=\vcenter{\hbox{\begin{tikzpicture}[scale=0.85,
point/.style={circle,fill,inner sep=1.3pt}
]
\draw[line width=0.5mm,carmine,fill=carmine,fill opacity=0.1] (0,0) -- (-1,1) decorate [singularity] {-- (1,1)} -- cycle;
\draw[line width=0.5mm,carmine,fill=carmine,fill opacity=0.1] (0,0) -- (-1,-1) decorate [singularity] {-- (1,-1)} -- cycle;
\draw[line width=0.5mm,ochre,fill=ochre,fill opacity=0.1] (-1,1) -- (-2,2) -- (-1,3) -- cycle;
\draw[line width=0.5mm,ochre,fill=ochre,fill opacity=0.1] (-3,1) -- (-2,2) -- (-3,3) -- cycle;
\draw[line width=0.5mm,ochre,fill=ochre,fill opacity=0.1] (-1,-1) -- (-2,-2) -- (-1,-3) -- cycle;
\draw[line width=0.5mm,ochre,fill=ochre,fill opacity=0.1] (-3,-1) -- (-2,-2) -- (-3,-3) -- cycle;
\draw[line width=0.5mm,ochre,fill=ochre,fill opacity=0.1] (1,1) -- (2,2) -- (1,3) -- cycle;
\draw[line width=0.5mm,ochre,fill=ochre,fill opacity=0.1] (3,1) -- (2,2) -- (3,3) -- cycle;
\draw[line width=0.5mm,ochre,fill=ochre,fill opacity=0.1] (1,-1) -- (2,-2) -- (1,-3) -- cycle;
\draw[line width=0.5mm,ochre,fill=ochre,fill opacity=0.1] (3,-1) -- (2,-2) -- (3,-3) -- cycle;
\draw[line width=0.5mm,darkspringgreen] (-3,1) -- (-2,0);
\draw[line width=0.5mm,frenchlilac,fill=darkspringgreen,fill opacity=0.1] (-2,0) -- (-1,1) -- (-2,2) -- (-3,1);
\draw[line width=0.5mm,darkspringgreen] (-3,-1) -- (-2,0);
\draw[line width=0.5mm,frenchlilac,fill=darkspringgreen,fill opacity=0.1] (-2,0) -- (-1,-1) -- (-2,-2) -- (-3,-1);
\draw[line width=0.5mm,darkspringgreen] (3,1) -- (2,0);
\draw[line width=0.5mm,frenchlilac,fill=darkspringgreen,fill opacity=0.1] (2,0) --(1,1) -- (2,2) -- (3,1);
\draw[line width=0.5mm,darkspringgreen] (3,-1) -- (2,0);
\draw[line width=0.5mm,frenchlilac,fill=darkspringgreen,fill opacity=0.1] (2,0) -- (1,-1) -- (2,-2) -- (3,-1);
\draw[line width=0.5mm,frenchlilac,fill=egyptianblue,fill opacity=0.1] (0,0) -- (-1,1) -- (-2,0) -- (-1,-1) -- cycle;
\draw[line width=0.5mm,frenchlilac,fill=egyptianblue,fill opacity=0.1] (0,0) -- (1,1) -- (2,0) -- (1,-1) -- cycle;
\node[inner sep=2] at (0,0.6) {I};
\node[inner sep=2] at (0,-0.6) {I};
\node[inner sep=2] at (1,0) {II};
\node[inner sep=2] at (-1,0) {II};
\node[inner sep=2] at (2,1) {III};
\node[inner sep=2] at (2,-1) {III};
\node[inner sep=2] at (-2,1) {III};
\node[inner sep=2] at (-2,-1) {III};
\node[inner sep=2] at (1.3,2) {IV};
\node[inner sep=2] at (2.6,2) {IV};
\node[inner sep=2] at (-1.3,2) {IV};
\node[inner sep=2] at (-2.6,2) {IV};
\node[inner sep=2] at (1.3,-2) {IV};
\node[inner sep=2] at (2.6,-2) {IV};
\node[inner sep=2] at (-1.3,-2) {IV};
\node[inner sep=2] at (-2.6,-2) {IV};
\filldraw (-2.6,0) circle (0.5pt);
\filldraw (-2.8,0) circle (0.5pt);
\filldraw (-3,0) circle (0.5pt); 
\filldraw (2.6,0) circle (0.5pt);
\filldraw (2.8,0) circle (0.5pt);
\filldraw (3,0) circle (0.5pt); 
\filldraw (-2,3) circle (0.5pt);
\filldraw (-2,2.8) circle (0.5pt);
\filldraw (-2,2.6) circle (0.5pt); 
\filldraw (-2,-3) circle (0.5pt);
\filldraw (-2,-2.8) circle (0.5pt);
\filldraw (-2,-2.6) circle (0.5pt); 
\filldraw (2,3) circle (0.5pt);
\filldraw (2,2.8) circle (0.5pt);
\filldraw (2,2.6) circle (0.5pt); 
\filldraw (2,-3) circle (0.5pt);
\filldraw (2,-2.8) circle (0.5pt);
\filldraw (2,-2.6) circle (0.5pt); 
\end{tikzpicture}}}\, ,&\text{ if }
\begin{cases}
\alpha_2=0.72702\,,\\
\alpha_3=-0.1\,,
\end{cases}\\
&
S^{--+}_\infty=\vcenter{\hbox{\begin{tikzpicture}[scale=0.85,
point/.style={circle,fill,inner sep=1.3pt}
]
\draw[line width=0.5mm,carmine,fill=carmine,fill opacity=0.1] (0,2) decorate [singularity] {-- (2,2)} -- (1,1) -- cycle;
\draw[line width=0.5mm,carmine,fill=carmine,fill opacity=0.1] (0,2) decorate [singularity] {-- (-2,2)} -- (-1,1) -- cycle;
\draw[line width=0.5mm,carmine,fill=carmine,fill opacity=0.1] (0,-4) decorate [singularity] {-- (2,-4)} -- (1,-3) -- cycle;
\draw[line width=0.5mm,carmine,fill=carmine,fill opacity=0.1] (0,-4) decorate [singularity] {-- (-2,-4)} -- (-1,-3) -- cycle;
\draw[line width=0.5mm,ochre,fill=ochre,fill opacity=0.1] (-1,-1) -- (-2,-2) -- (-2,0) -- cycle;
\draw[line width=0.5mm,ochre,fill=ochre,fill opacity=0.1] (1,-1) -- (2,-2) -- (2,0) -- cycle;
\draw[line width=0.5mm,ochre,fill=ochre,fill opacity=0.1] (1,-1) -- (0,-2) -- (0,0) -- cycle;
\draw[line width=0.5mm,ochre,fill=ochre,fill opacity=0.1] (-1,-1) -- (0,-2) -- (0,0) -- cycle;
\draw[line width=0.5mm,frenchlilac,fill=darkspringgreen,fill opacity=0.1] (0,0) -- (-1,1) -- (-2,0)-- (-1,-1) -- cycle;
\draw[line width=0.5mm,frenchlilac,fill=darkspringgreen,fill opacity=0.1] (0,0) -- (1,1) -- (2,0)-- (1,-1) -- cycle;
\draw[line width=0.5mm,frenchlilac,fill=darkspringgreen,fill opacity=0.1] (0,-2) -- (-1,-1) -- (-2,-2)-- (-1,-3) -- cycle;
\draw[line width=0.5mm,frenchlilac,fill=darkspringgreen,fill opacity=0.1] (0,-2) -- (1,-1) -- (2,-2)-- (1,-3) -- cycle;
\draw[line width=0.5mm,frenchlilac,fill=egyptianblue,fill opacity=0.1] (0,0) -- (-1,1) -- (0,2)-- (1,1) -- cycle;
\draw[line width=0.5mm,frenchlilac,fill=egyptianblue,fill opacity=0.1] (0,-2) -- (-1,-3) -- (0,-4)-- (1,-3) -- cycle;
\node[inner sep=2] at (-1.6,-1) {IV};
\node[inner sep=2] at (1.6,-1) {IV};
\node[inner sep=2] at (-0.4,-1) {IV};
\node[inner sep=2] at (0.4,-1) {IV};
\node[inner sep=2] at (0,1) {II};
\node[inner sep=2] at (0,-3) {II};
\node[inner sep=2] at (1,0) {III};
\node[inner sep=2] at (-1,0) {III};
\node[inner sep=2] at (1,-2) {III};
\node[inner sep=2] at (-1,-2) {III};
\node[inner sep=2] at (1,1.6) {I};
\node[inner sep=2] at (-1,1.6) {I};
\node[inner sep=2] at (1,-3.6) {I};
\node[inner sep=2] at (-1,-3.6) {I};
\filldraw (-2.2,-1) circle (0.5pt);
\filldraw (-2.4,-1) circle (0.5pt);
\filldraw (-2.6,-1) circle (0.5pt); 
\filldraw (2.2,-1) circle (0.5pt);
\filldraw (2.4,-1) circle (0.5pt);
\filldraw (2.6,-1) circle (0.5pt); 
\end{tikzpicture}}}\, ,&\text{ if }
\begin{cases}
\alpha_2=3.9559\,,\\
\alpha_3=-3.0508\,,\\
\alpha_4=0.63303\,,\\
\alpha_5=-0.038142\,,\\
\end{cases}\\
&
S^{---}_\infty=\vcenter{\hbox{\begin{tikzpicture}[scale=0.85,
point/.style={circle,fill,inner sep=1.3pt}
]
\draw[line width=0.5mm,carmine,fill=carmine,fill opacity=0.1] (0,2) -- (-1,3) decorate [singularity] {-- (1,3)} -- cycle;
\draw[line width=0.5mm,carmine,fill=carmine,fill opacity=0.1] (2,2) -- (1,3) decorate [singularity] {-- (3,3)} -- cycle;
\draw[line width=0.5mm,carmine,fill=carmine,fill opacity=0.1] (-2,2) -- (-1,3) decorate [singularity] {-- (-3,3)} -- cycle;
\draw[line width=0.5mm,carmine,fill=carmine,fill opacity=0.1] (0,-2) -- (-1,-3) decorate [singularity] {-- (1,-3)} -- cycle;
\draw[line width=0.5mm,carmine,fill=carmine,fill opacity=0.1] (2,-2) -- (1,-3) decorate [singularity] {-- (3,-3)} -- cycle;
\draw[line width=0.5mm,carmine,fill=carmine,fill opacity=0.1] (-2,-2) -- (-1,-3) decorate [singularity] {-- (-3,-3)} -- cycle;
\draw[line width=0.5mm,ochre,fill=ochre,fill opacity=0.1] (0,0) -- (-1,1) -- (-1,-1) -- cycle;
\draw[line width=0.5mm,ochre,fill=ochre,fill opacity=0.1] (0,0) -- (1,1) -- (1,-1) -- cycle;
\draw[line width=0.5mm,ochre,fill=ochre,fill opacity=0.1] (-1,-1) -- (-2,0) -- (-1,1) -- cycle;
\draw[line width=0.5mm,ochre,fill=ochre,fill opacity=0.1] (1,-1) -- (2,0) -- (1,1) -- cycle;
\draw[line width=0.5mm,ochre,fill=ochre,fill opacity=0.1] (-2,0) -- (-3,-1) -- (-3,1) -- cycle;
\draw[line width=0.5mm,ochre,fill=ochre,fill opacity=0.1] (2,0) -- (3,-1) -- (3,1) -- cycle;
\draw[line width=0.5mm,frenchlilac,fill=darkspringgreen,fill opacity=0.1] (0,0) -- (-1,1) -- (0,2) -- (1,1) -- cycle;
\draw[line width=0.5mm,frenchlilac,fill=darkspringgreen,fill opacity=0.1] (0,0) -- (-1,-1) -- (0,-2) -- (1,-1) -- cycle;
\draw[line width=0.5mm,frenchlilac,fill=darkspringgreen,fill opacity=0.1] (-2,0) -- (-3,1) -- (-2,2) -- (-1,1) -- cycle;
\draw[line width=0.5mm,frenchlilac,fill=darkspringgreen,fill opacity=0.1] (2,0) -- (3,1) -- (2,2) -- (1,1) -- cycle;
\draw[line width=0.5mm,frenchlilac,fill=darkspringgreen,fill opacity=0.1] (-2,0) -- (-3,-1) -- (-2,-2) -- (-1,-1) -- cycle;
\draw[line width=0.5mm,frenchlilac,fill=darkspringgreen,fill opacity=0.1] (2,0) -- (3,-1) -- (2,-2) -- (1,-1) -- cycle;
\draw[line width=0.5mm,frenchlilac,fill=egyptianblue,fill opacity=0.1] (1,1) -- (2,2) -- (1,3) -- (0,2) -- cycle;
\draw[line width=0.5mm,frenchlilac,fill=egyptianblue,fill opacity=0.1] (-1,1) -- (-2,2) -- (-1,3) -- (0,2) -- cycle;
\draw[line width=0.5mm,frenchlilac,fill=egyptianblue,fill opacity=0.1] (1,-1) -- (2,-2) -- (1,-3) -- (0,-2) -- cycle;
\draw[line width=0.5mm,frenchlilac,fill=egyptianblue,fill opacity=0.1] (-1,-1) -- (-2,-2) -- (-1,-3) -- (0,-2) -- cycle;
\draw[line width=0.5mm,darkspringgreen] (-3,1) -- (-2,2);
\draw[line width=0.5mm,darkspringgreen] (-3,-1) -- (-2,-2);
\draw[line width=0.5mm,darkspringgreen] (3,1) -- (2,2);
\draw[line width=0.5mm,darkspringgreen] (3,-1) -- (2,-2);
\node[inner sep=2] at (0,2.6) {I};
\node[inner sep=2] at (-2,2.6) {I};
\node[inner sep=2] at (2,2.6) {I};
\node[inner sep=2] at (0,-2.6) {I};
\node[inner sep=2] at (-2,-2.6) {I};
\node[inner sep=2] at (2,-2.6) {I};
\node[inner sep=2] at (1,2) {	II};
\node[inner sep=2] at (1,-2) {II};
\node[inner sep=2] at (-1,2) {II};
\node[inner sep=2] at (-1,-2) {II};
\node[inner sep=2] at (0,1) {III};
\node[inner sep=2] at (0,-1) {III};
\node[inner sep=2] at (0,1) {III};
\node[inner sep=2] at (0,-1) {III};
\node[inner sep=2] at (2,1) {III};
\node[inner sep=2] at (2,-1) {III};
\node[inner sep=2] at (-2,1) {III};
\node[inner sep=2] at (-2,-1) {III};
\node[inner sep=2] at (0.6,0) {IV};
\node[inner sep=2] at (-0.6,0) {IV};
\node[inner sep=2] at (1.3,0) {IV};
\node[inner sep=2] at (-1.3,0) {IV};
\node[inner sep=2] at (2.6,0) {IV};
\node[inner sep=2] at (-2.6,0) { IV};
\filldraw (-3.3,0) circle (0.5pt);
\filldraw (-3.5,0) circle (0.5pt);
\filldraw (-3.7,0) circle (0.5pt);
\filldraw (3.3,0) circle (0.5pt);
\filldraw (3.5,0) circle (0.5pt);
\filldraw (3.7,0) circle (0.5pt);
\end{tikzpicture}}}\, ,&\text{ if }
\begin{cases}
\alpha_2=2.2975\,,\\
\alpha_3=-1.5030\,,\\
\alpha_4=0.28498\,,\\
\alpha_5=-0.016303\,.
\end{cases}
\end{align}
We remind that whenever the values of the coupling constant include five significant figures the solution requires fine tuning. We observe that all black holes with three horizons extend infinitely either in the spacelike, timelike directions or in both.\footnote{Some of the exotic global structures we observe resemble those found in topologically Lifshitz black holes within four-dimensional Einstein-Yang-Mills-dilaton theory \cite{Canfora:2021nca}.} Among them, four correspond to extremal black holes, $S^{\pm\pm+}_\infty$. We can also see that the complicated structure of these black holes allows for non-planar diagrams even before replacing singular regions I with its regularized version, this is the case of $S^{+-+}_\infty$ in \eqref{eq:S+-+inf}. Another particularly interesting diagram is given by $S^{+--}_\infty$ in \eqref{eq:S+--inf} which perfectly tessellates the two-dimensional plane. This may not be immediately evident in \eqref{eq:S+--inf}, so we present an equivalent diagram in Figure \ref{fig:tes}.\footnote{For a previous example of a black hole Penrose diagram tessellating the plane, see Figure~6 in \cite{Lemos:1993py}. Also, although it would have been somewhat funny, these ``Penrose-diagram tilings'' of the plane are not ``Penrose tilings'' \cite{PT}.}


The remaining twenty diagrams, $S^{\pm\pm\pm}_\text{c}$, $R^{+\pm\pm}_1$ and $R^{\pm\pm\pm}_0$, with distinct behavior in region I are generated using the diagram regularization procedure outlined above. To avoid redundancy, we omit the explicit values of these diagrams. The transformation process is straightforward, and we have already demonstrated how it is applied explicitly for black holes with one and two horizons.

To conclude the section, let us discuss the number of possible global structures $\#_{\text{gs}}$ existing with a higher number of horizons. As we can turn an infinite number of coupling constants $\alpha_i$, there is in principle no obstruction to include as many horizons as we desire, and then, we can regularize them by turning on suitable $\beta_j$. Including a new root will always increase the number of global spacetimes with a factor of two. As a consequence, the number of distinct global structures is given by
\begin{equation}
\#_{\text{gs}}=7\times 2^{\text{card}(Z)-1}\,,
\end{equation} 
where $\text{card}(Z)$ denotes the cardinal of the set $Z=\left\{r|f(r)=0\right\}$. 

\subsection{``Black holes'' with ``zero'' horizons}\label{zeroh}
Thus far, we have classified the different types of solutions and their diagrams using the number of positive roots of $f(r)$ as a criterion. The above analysis reveals that in the case of regular black holes with $f(r)\overset{(r\rightarrow 0)}{\sim} \mathcal{O}(r^{2s})$, $r=0$ is a null surface which separates distinct causal regions and therefore plays the role of a horizon---see Section \ref{conc}. Hence, it is natural to consider the case of regular solutions of this type for which no positive roots of $f(r)$ exist. We succinctly address them here. 

Let us label this ``zero-horizon black holes'' as $R_0$ in our classification.
Their metric function satisfies $f(r) > 0$ $\forall r>0$, and satisfies $f(0) = 0$ and $f'(0) = 0$. 
Such a solution can be realized for $f(r)$ using the parameter values defined earlier in this section—\textit{i.e.}, $p=1$, $L=1$, $\alpha_1=1/2$, and $\lambda=\sum_{i=2}\alpha_i/[2(i-1)]$—together with $\alpha_2=1$ and $\beta_1=1$, with all other $\alpha_{i\geq2}=0$ and $\beta_{j\neq1}=0$, that is,
\begin{equation}
\vcenter{\hbox{\includegraphics[width=0.25\textwidth]{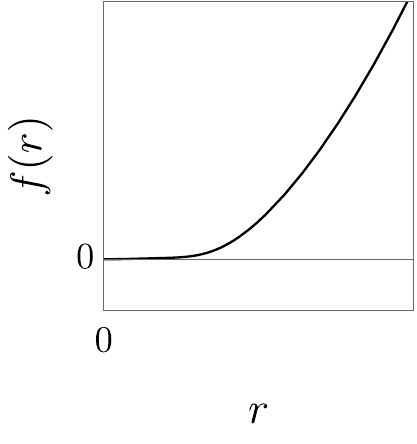}}}\quad R_0=\vcenter{\hbox{\begin{tikzpicture}[scale=0.85,
point/.style={circle,fill,inner sep=1.3pt}
]
\draw[line width=0.5mm,carmine,fill=carmine,fill opacity=0.1] (0,0) -- (1,-1) -- (1,1) -- cycle;  
\node[inner sep=2] at (0.6,0) {I};
\node[carmine,above,rotate=45] at (0.5,0.5) {$r=0$};
\node[carmine,below,rotate=-45] at (0.5,-0.5) { $r=0$};
\node[carmine,above,rotate=-90] at (1,0) { $r\rightarrow\infty$};
\end{tikzpicture}}}\,,\quad \text{with } \alpha_2=1\,,\ \beta_1=1\,,
\end{equation}
Of course the resulting diagram, which consists of a single region, is extremely simple. It is reminiscent of the one associated with Minkowski spacetime in spherical coordinates in which we have exchanged $(r=0)\leftrightarrow (r\rightarrow\infty)$. The key difference is that the null boundary at $r=0$ allows continuation beyond $r=0$. Replicating the argument presented above regarding the extension beyond $r=0$ we may find the extended diagrams 
\begin{equation}
\mathring R_0=\vcenter{\hbox{\begin{tikzpicture}[scale=0.85,
point/.style={circle,fill,inner sep=1.3pt}
]
\node[inner sep=2] at (-1.4,0) {I};
\draw[stealth-stealth, thick] (-1.5,0.5) arc[start angle=45, end angle=315, radius=0.707107cm];
\draw[line width=0.5mm,carmine,fill=carmine,fill opacity=0.1] (-1,-1)-- (-2,0) -- (-1,1) -- cycle;  
\end{tikzpicture}}}\, ,\quad 
\bar R_0= \vcenter{\hbox{\begin{tikzpicture}[scale=0.85,
point/.style={circle,fill,inner sep=1.3pt}
]
\draw[line width=0.5mm,carmine,fill=carmine,fill opacity=0.1] (0,0) -- (-1,-1) -- (-1,1)-- cycle;  
\draw[line width=0.5mm,carmine,fill=carmine,fill opacity=0.1] (0,0) -- (0,2) -- (-1,1)-- cycle; 
\node[inner sep=2] at (-0.6,0) {I};
\node[inner sep=2] at (-0.4,1) {I};
\filldraw (-0.5,2) circle (0.5pt);
\filldraw (-0.5,1.8) circle (0.5pt);
\filldraw (-0.5,2.2) circle (0.5pt);
\filldraw (-0.5,-1) circle (0.5pt);
\filldraw (-0.5,-0.8) circle (0.5pt);
\filldraw (-0.5,-1.2) circle (0.5pt);
\end{tikzpicture}}}\,,
\end{equation}
where, again $\mathring R_0$ presents closed timelike curves.

While Penrose diagrams offer valuable insights into the causal structure of black holes, they are not sufficient for a complete understanding. 
Additional things can be learnt from the study of geodesics, which we perform in the following section.

\section{Geodesics and tidal forces}\label{sec:geodesics}
In this section we study the motion of neutral free-falling particles in our black hole backgrounds (\ref{eq:SSSM}), paying special attention to the way in which the singularity resolution of the regular black holes reflects on both their trajectories and the tidal acceleration that they experience as they approach the center $r=0$.

\subsection{Geodesics}
\begin{figure}[t!]
\begin{subfigure}{1\linewidth}
\includegraphics[width=0.35\linewidth]{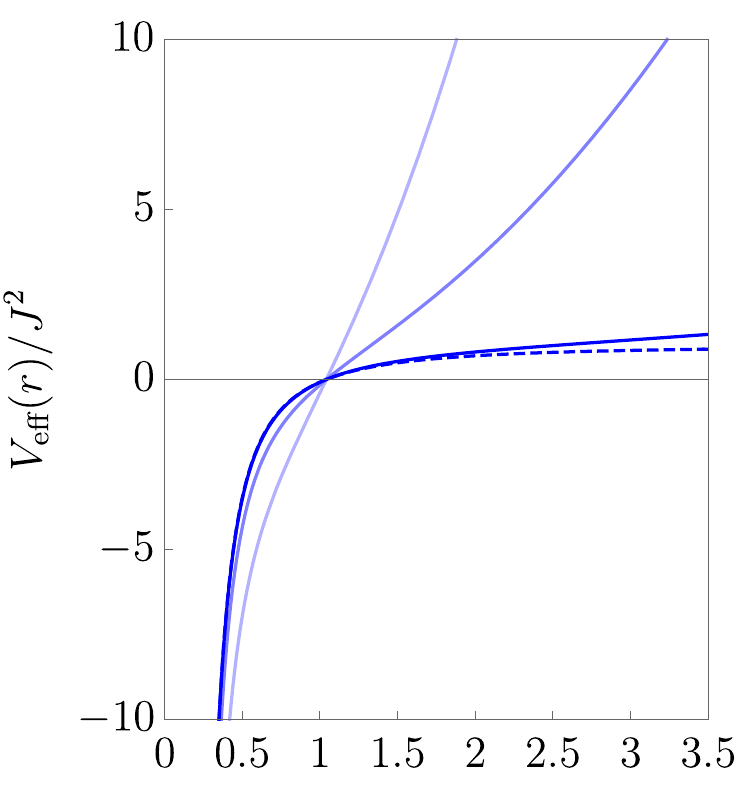}  
\includegraphics[width=0.29\linewidth]{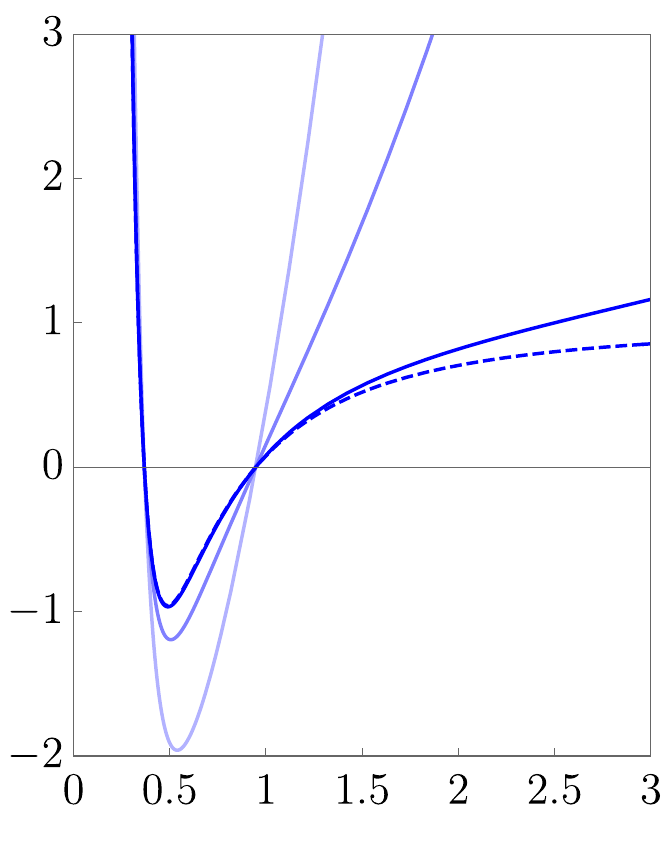}  
\includegraphics[width=0.325\linewidth]{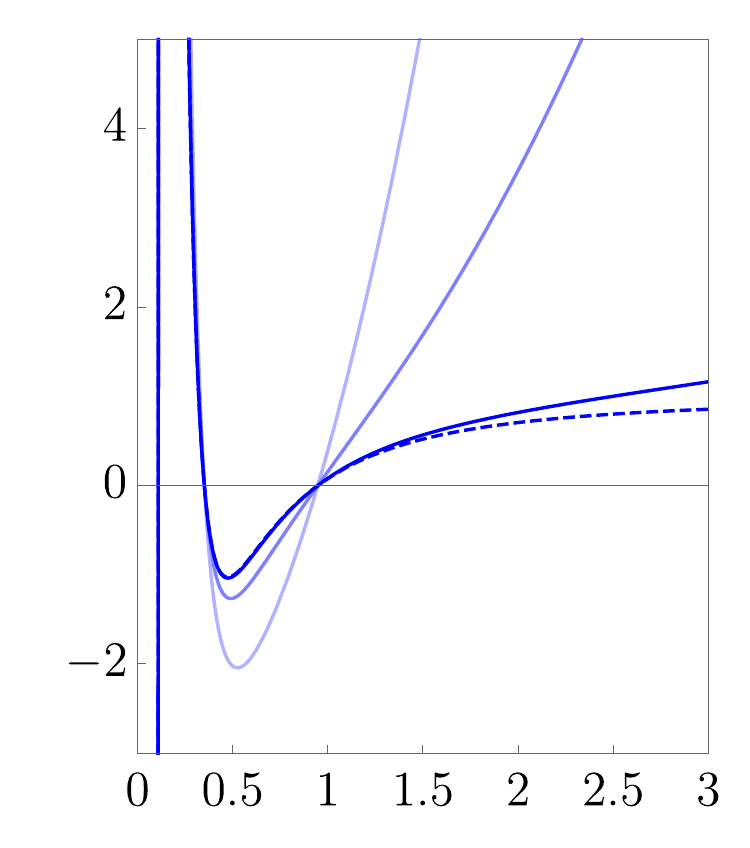}  
\caption{Black holes with a curvature singularity at $r=0$ and one, two and three horizons respectively.}
\label{fig:curvsingVeff}
\end{subfigure}\\  
\begin{subfigure}{1\linewidth}
\includegraphics[width=0.335\linewidth]{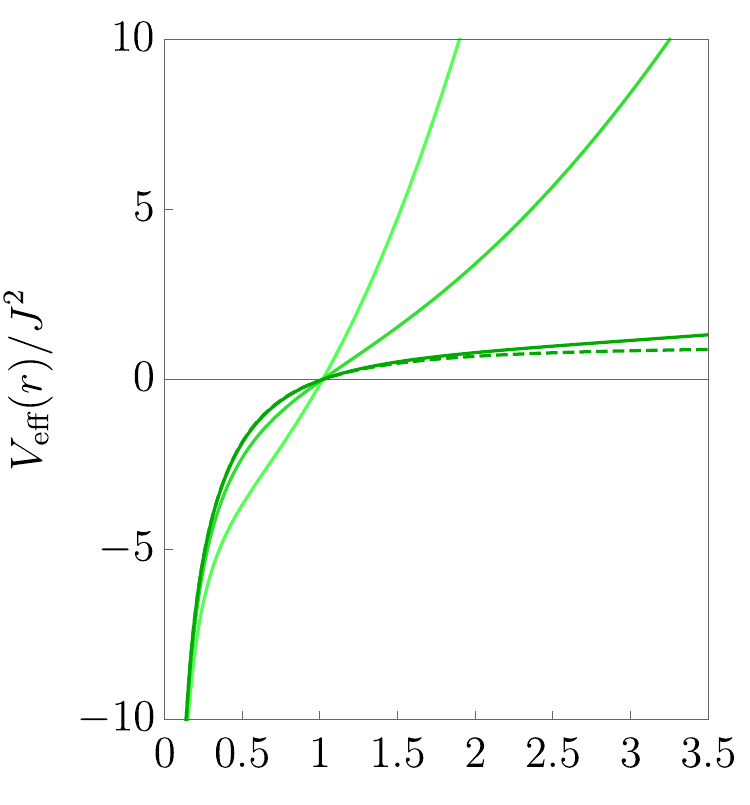}  
\includegraphics[width=0.325\linewidth]{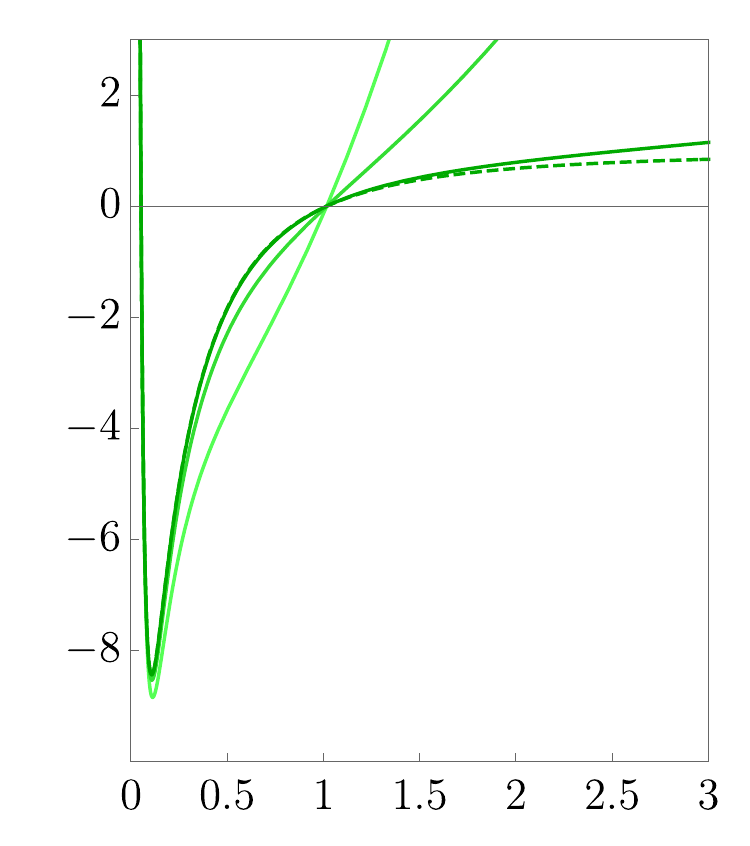}  
\includegraphics[width=0.325\linewidth]{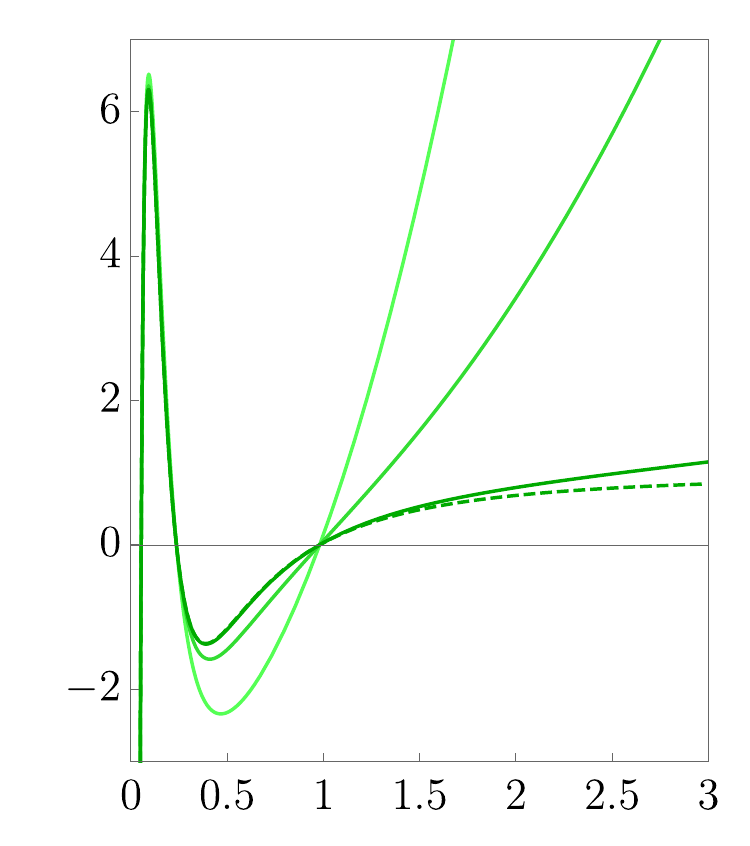}  
\caption{Black holes with a conical or pseudo-conical singularity at $r=0$ and one, two and three horizons.}
\label{fig:BTzlikeVeff}
\end{subfigure}\\
\begin{subfigure}{1\linewidth}
\includegraphics[width=0.33\linewidth]{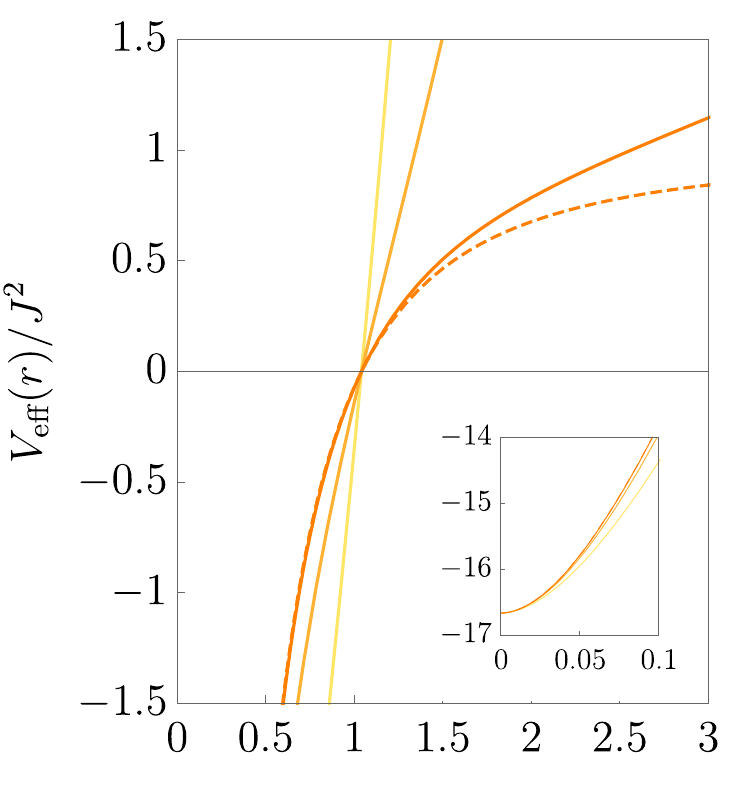}  
\includegraphics[width=0.33\linewidth]{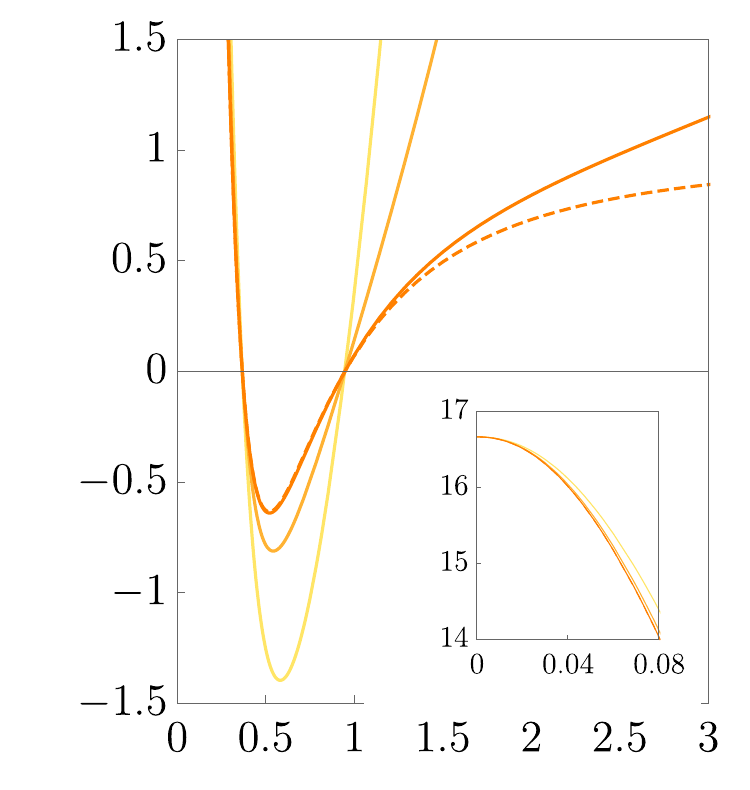}  
\includegraphics[width=0.32\linewidth]{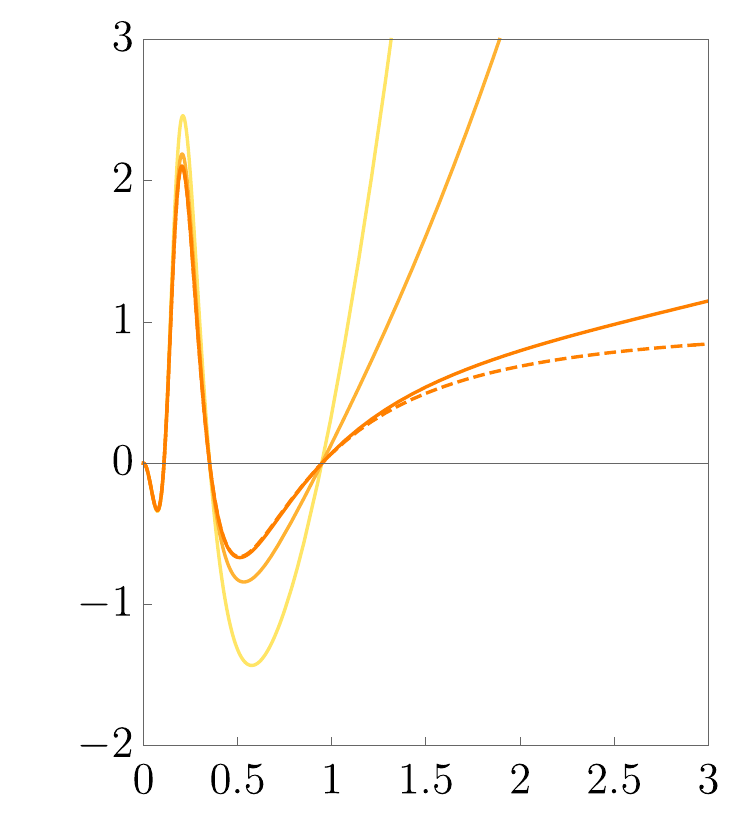}%
\caption{Regular black holes with one, two and three horizons. Note that $\vert V_\text{eff}\vert < \infty$ at $r=0$.}
\label{fig:regVeff}
\end{subfigure}
\caption{We show the effective potential over the angular momentum squared $V_{\text{eff}}(r)/J^2$. The color gradient code means that the angular momentum is larger the darker the curves are. The values chosen are $J=\{ 0.5, 1, 5, \infty\}$, the last one corresponding to null geodesics and represented as dashed lines.}
\label{Vefftypes}
\end{figure}

\begin{figure}[!t]
\begin{center}
\includegraphics[scale=0.6]{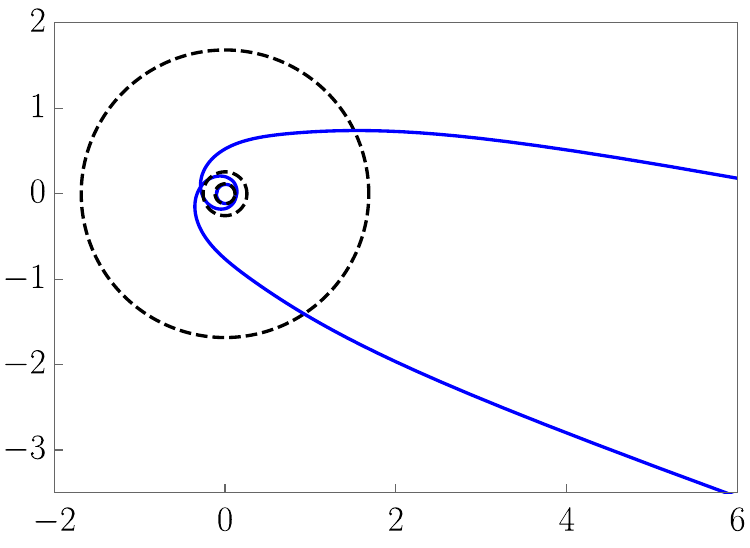}
\end{center}  
\caption{Unbound null geodesic in a singular black hole background with non-vanishing coupling constants: $\alpha_1=1\, , \alpha_2=4\, , \alpha_3=-0.6\, , \alpha_4=0.01\, , \beta_1=1$. Other parameters are $J=p=1\, , \lambda=3$, and $E^2=1.5$. }
    \label{fig:flyby}
\end{figure}

Since (\ref{eq:SSSM}) are static and spherically symmetric backgrounds, $\partial_t$ and $\partial_\varphi$ are obvious Killing vectors. Hence, the particles will have their energy and angular momentum conserved along their trajectories,
\begin{align}
\label{E}
-E\equiv p_t&=-f(r)\frac{\diff t}{\diff \sigma}\, ,\\\label{J}
J\equiv p_{\varphi}&=r^2\frac{\diff \varphi}{\diff \sigma}\, ,
\end{align}
where $\sigma$ is an affine parameter that, as usual, we choose to coincide with the proper time when the geodesics are timelike. 
On the other hand, the mass-shell condition reads
\begin{align}\label{m}
-\epsilon =p^{\mu}p_{\mu}&=-f(r)\left(\frac{\diff t}{\diff \sigma}\right)^2+\frac{1}{f(r)}\left(\frac{\diff r}{\diff \sigma}\right)^2+r^2\left(\frac{\diff \varphi}{\diff \sigma}\right)^2\, ,
\end{align}
with $\epsilon$ either equal to 1 for massive particles or zero for light rays. Substituting (\ref{J}) and (\ref{E}) into (\ref{m}), we can derive an equation for the $r$ coordinate
\begin{eqnarray}\label{geodesic_eq}
\left(\frac{\diff r}{\diff \sigma}\right)^2=E^2 -V_{\text{eff}}(r)\, ,\quad \text{where} \quad
V_{\text{eff}}(r)\equiv f(r)\left(\epsilon+\frac{J^2}{r^2}\right).
\end{eqnarray}
In Figure~\ref{Vefftypes} we plot $V_{\text{eff}}(r)/J^2$ for the metric functions shown in Figures~\ref{fig:curvsing}, \ref{fig:BTZlike} and \ref{fig:reg}, respectively. Note that the asymptotia of the effective potential determines the radial turning points, whose values define qualitatively different orbits. For example, from the behavior of the potential at infinity
\begin{equation}
 V_{\text{eff}} \xrightarrow[r \to \infty]{}\begin{cases}
    \frac{r^2}{L^2}\, , & \text{if $\epsilon=1$}\, , \\
    \frac{J^2}{L^2}\, , & \text{if $\epsilon=0$}\, ,
  \end{cases}
\end{equation}
it follows that only massless particles can escape the AdS barrier, having unbound orbits for high enough energies, \textit{i.e.} $E^2>\frac{J^2}{L^2}$ ---see Figure \ref{fig:flyby} for an instance of this kind of ``flyby" orbit. Moreover, for non-radial trajectories
\begin{equation}
 V_{\text{eff}} \xrightarrow[r \to 0]{}\frac{J^2 f(r)}{r^2}\, .
\end{equation}
Therefore, when $f(r)\rightarrow ar^{2s}$ with $a>0$ and $s<1$ at the origin, a particle falling into the black hole will unavoidably reach a periapsis, that is, a minimal radius $r_{\text{min}}>0$. This also happens when $f(r)\rightarrow -a\log{r}$, such as in the charged BTZ background. In all other cases, if the energy is sufficiently high the orbit could end at the singularity or the particle might cross the regular center and follow its journey. Note that (\ref{geodesic_eq}) implies that the proper time it takes to reach the origin from $r_{\max}<\infty$ is finite, save the very fine-tuned case in which $f(r)\rightarrow a r^2$ and $J^2a=E^2$.

\begin{figure}[!t]
  \begin{subfigure}{1\linewidth}
 \centering
 \includegraphics[width=0.48\linewidth]{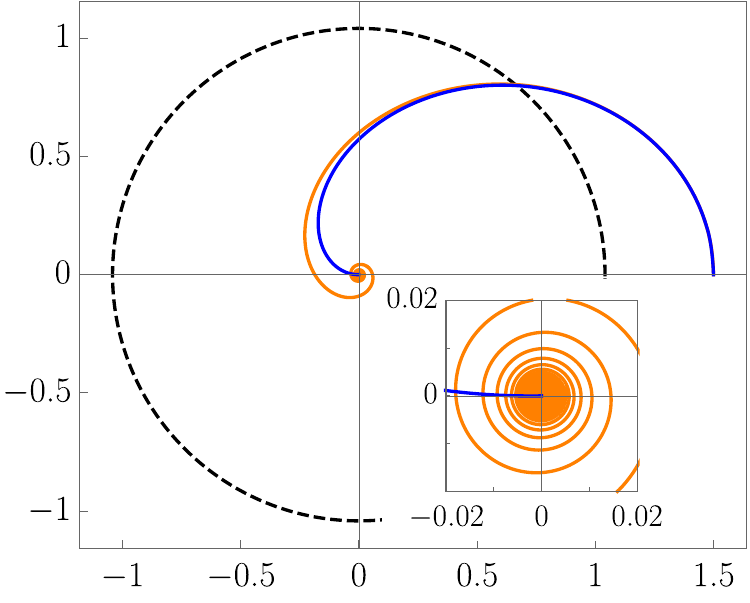}  
\includegraphics[width=0.48\linewidth]{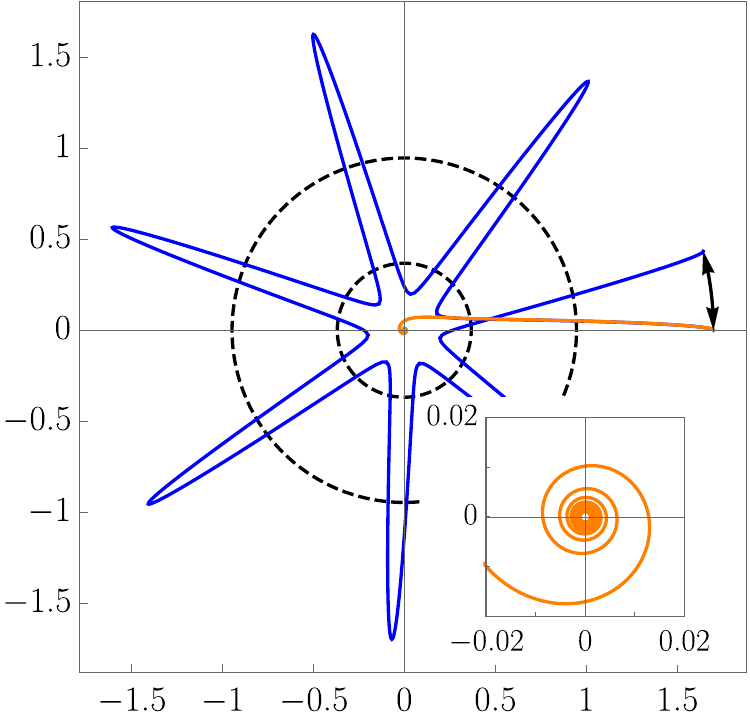}  
    \caption{Left: black hole with one horizon. The initial conditions are $\dot{r}=0$ and $r=1.5$, with $J=10$. Right: black hole with two horizons. The initial conditions are $\dot{r}=0$ and $r=1.7$, with $J=0.1$. Note that there is an apsidal precession, represented by the black arrow, and only when $2\pi$ is a rational multiple of the precession angle the orbit is closed.}
    \label{SUBFIGURE LABEL X1}
\end{subfigure} \\
 \begin{subfigure}{1\linewidth}
 \centering
 \includegraphics[width=0.48\linewidth]{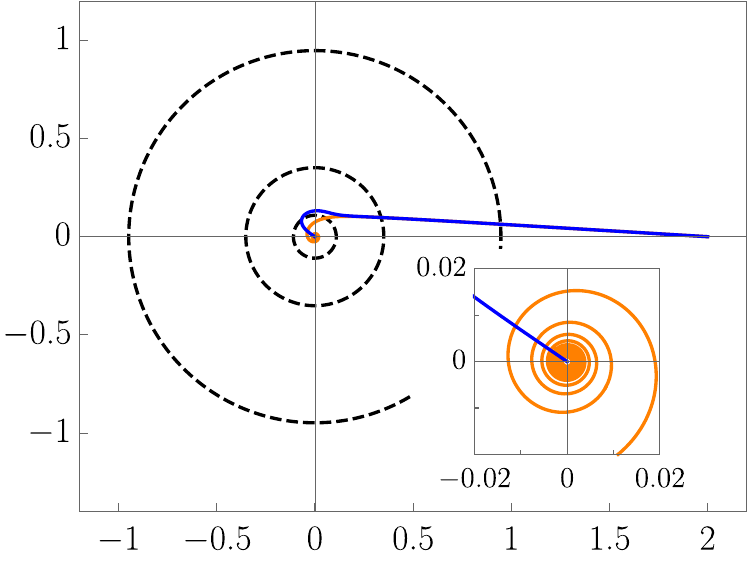}  
   \caption{Black holes with three horizons. The initial conditions are $\dot{r}=0$ and $r=83$, with $J=10$.}
    \label{SUBFIGURE LABEL X2}
\end{subfigure}
\caption{Timelike geodesics on the background solution (\ref{eq:SSSM}) for the metric functions $f(r)$ displayed in Figures~\ref{fig:curvsing}, \ref{fig:BTZlike} and \ref{fig:reg}. The blue trajectories correspond to the singular black holes, while the orange ones correspond to the regular black holes with $f(r)\overset{(r \rightarrow 0)}{\sim} \mathcal{O}(r^{2s}),\quad s\geq 1$. The black dashed curves represent the horizons, and the inset figures zoom in around the origin.}
\label{geodesics_plot}
\end{figure}

In Figure~\ref{geodesics_plot} we compare the timelike geodesics on the singular and regular black holes corresponding to the effective potentials of Figure~\ref{Vefftypes}. Curiously, the trajectories on the regular black holes look like spiral paths. This is because
\begin{equation}
\left(\frac{\diff r}{\diff \varphi}\right)^2=\frac{r^4}{J^2}\left(E^2-V_{\text{eff}}(r)\right)\, ,
\end{equation}
so if $f(r)\sim r^{2 s}$ with $s\geq 0$ at the center, $\varphi\rightarrow \infty$ as $r\rightarrow 0$. Hence, the same spirals would also be present on black holes with BTZ-like singularities.\footnote{Exact analytic solutions to the geodesic equations were found for the BTZ in \cite{Cruz:1994ar}. Although the angular coordinate $\varphi$ explicitly diverges as $r\rightarrow 0$, this spiral orbit was not emphasized by the authors in the corresponding plot.}

\FloatBarrier

The geodesics corresponding to regular black holes on the left panel of Figure \ref{SUBFIGURE LABEL X1} and Figure \ref{SUBFIGURE LABEL X2} should be extended to represent a periodical bounce from a maximum radius $r=r_{\text{max}}$ to the center $r=0$, where the tidal forces remain finite, as thoroughly explained in the next subsection\footnote{For probe particles with non-vanishing angular momentum, such a periodic extension of the trajectories becomes more subtle, because the angular coordinate diverges at $r=0$, in agreement with Equation (\ref{J}).}. The extension of the trajectory past $r=0$ is also depicted in the Penrose diagram (\ref{Penrose_geodesics1}), where a loop identification in the null surface corresponding to the center of a single-horizon black hole allows for a continuation of the geodesics. For an alternative construction, achieved by gluing infinitely many copies of an elementary block, see (\ref{Penrose_geodesics2}). In turn, the orange trajectory shown in the right panel of Figure \ref{SUBFIGURE LABEL X1} terminates at $r=0$ due to a logarithmic divergence of the geodesic deviation---see (\ref{K_special}).

Another noteworthy feature is that bound geodesics may cross the horizons periodically, each time entering a causally disconnected universe. 

Finally, from the effective potentials represented by dashed lines in Figure~\ref{Vefftypes} it is apparent that for the chosen metric functions there are no photon spheres. Stated differently, the effective potentials for null geodesics in Figure~\ref{Vefftypes} have no local maximum outside the horizon $r_{\text{ph}}>r_{\text{h}}$. This is due to the AdS asymptotics, which makes the effective potential monotonically approach $V_{\text{eff}}\rightarrow \frac{J^2}{L^2}$.  The only way for there to be a photon sphere is for the metric itself to have a local maximum. According to the analysis in Section~\ref{sec:number_roots}, this might happen when $n_{\text{max}}\geq 2$, and for particular ranges of the parameters--- see Figure \ref{fig:f1roots}.


\subsection{Tidal forces}
An issue that may certainly worry the free-falling observer is whether they would be eventually teared apart by an infinite gravitational force when reaching the center of the black hole. One would expect that if the black hole was regular they would be able to endure the journey, passing through the origin and later traversing the horizon to enter a different, causally disconnected, universe. If that was the case, they would oscillate between two antipodal points in a ``many-world'' orbit, as they cannot escape the AdS barrier, no matter how large their energy is. In order to check if this expectation is correct, in this section we compute the tidal forces and compare the results near the center for the different black holes, according to their singularity structures.

The tidal force is the relative acceleration between two infinitesimally close geodesics, measured in the co-moving reference frame. Let $\eta^{\mu}$ be the spacelike vector representing the geodesic deviation and $\frac{D (\cdot)}{D\sigma}\equiv \frac{\diff x^\mu}{\diff\sigma}\nabla_\mu (\cdot) $ represent the covariant derivative along the geodesics, then \cite{Hartle:2003yu, Lima:2020wcb}
\begin{equation}\label{tidal_tensor}
\frac{D^2 \eta^{\hat{\alpha}}}{D\sigma^2}=K^{\hat{\alpha}}_{\hat{\beta}}\eta^{\hat{\beta}}\, ,\quad \text{with} \quad
K^{\hat{\alpha}}_{\hat{\beta}}=R^{\mu}_{\nu\rho\sigma}e^{\hat{\alpha}}_{\mu}e^{\nu}_{\hat{0}}e^{\rho}_{\hat{0}}e^{\sigma}_{\hat{\beta}}\, .
\end{equation} $K^{\hat{\alpha}}_{\hat{\beta}}$ is called the tidal tensor, and the vectors $e^{\mu}_{\hat{\alpha}}$ form the tetrad basis, satisfying $e^{\mu}_{\hat{\alpha}}e^{\nu}_{\hat{\beta}}g_{\mu\nu}=\eta_{\hat{\alpha}\hat{\beta}}$. For arbitrary timelike motion, these are 
\begin{equation}
\begin{split}
e^{\mu}_{\hat{0}}&=\frac{\diff x^\mu}{\diff\sigma}=\left(\frac{E}{f(r)}, -V_{\text{eff}}^{1/2},\frac{J}{r^2}\right)\, ,\\
e^{\mu}_{\hat{r}}&=\frac{r}{\sqrt{r^2+J^2}}\left(\frac{-V_{\text{eff}}^{1/2}}{f(r)},E,0\right)\, ,\\
e^{\mu}_{\hat{\varphi}}&=\frac{1}{\sqrt{r^2+J^2}}\left(\frac{J E}{f(r)}, -J~ V_{\text{eff}}^{1/2} , 1+\frac{J^2}{r^2}\right)\, ,
\end{split}
\end{equation}
where $V_{\text{eff}}$ is the efective potential defined in (\ref{geodesic_eq}). Note that, as customary, we denote the vectors by $e_{\hat{r}}$ and $e_{\hat{\varphi}}$ because for an observer at rest these lie along the radial and angular directions, respectively. Therefore, the components of the tidal tensor read
\begin{equation}
K^{\hat{r}}_{\hat{r}}=-\frac{1}{2}f''(r)+ \frac{J^2}{2 r^2}\left( -f''(r)+\frac{f'(r)}{r}\right)\, , \quad 
K^{\hat{\varphi}}_{\hat{\varphi}}=-\frac{1}{2 r}f'(r)\, .
\end{equation}
\subsubsection{Radial motion}
We first consider the case with no angular momentum, that is, we first analyze the force measured by an observer falling along the radial direction. Expanding the general solution (\ref{fEQT}) near the origin, we get that for both black holes with curvature singularities and regular black holes with $f(r)\overset{r\sim 0}{=} \mathcal{O}(r^{2s})$\footnote{We exclude the special cases with ($n_{\text{max}}=1$, $m_{\text{max}}=0$) and ($n_{\text{max}}=2$, $m_{\text{max}}=0$), for which the tidal forces have a logarithmic dependence.}
\begin{eqnarray}\label{K_radial_motion}
K^{\hat{r}}_{\hat{r}}&\approx&-\frac{ (n_{\rm max}-m_{\rm max}-2)(n_{\rm max}-m_{\rm max}-3/2)\alpha_{n_{\rm max}}}{(n_{\rm max}-1)(2 m_{\rm max}+1)L^2 \beta_{m_{\rm max}}}\left(\frac{L p}{r}\right)^{2(n_{\rm max}-m_{\rm max}-1)}\, ,\\
K^{\hat{\varphi}}_{\hat{\varphi}}&\approx&\frac{ (n_{\rm max}-m_{\rm max}-2)\alpha_{n_{\rm max}}}{2(n_{\rm max}-1)(2 m_{\rm max}+1)L^2 \beta_{m_{\rm max}}}\left(\frac{L p}{r}\right)^{2(n_{\rm max}-m_{\rm max}-1)}\, .
\end{eqnarray}
This implies that an observer approaching the center of a singular black hole with $n_{\rm max}>m_{\rm max}+2$ would experience infinite stretching in the radial direction and infinite compressing in the perpendicular direction, which is usually known as spaghettification. However, when $n_{\rm max}<m_{\rm max}+2$ the tidal forces go to zero at $r=0$, so the observer would be able to pass through the center with no fuss.

As made apparent from the pole in (\ref{K_radial_motion}), the tidal forces in the background of regular black holes with $n_{\text{max}}=1$ must be dealt with separately. This might have also been expected from the behavior of the Ricci scalar at $r=0$, which in that case does not go to zero as a positive power of $r$, but rather like $R\sim \log(r) r^{2 m_{\text{max}}}$. The components of the tidal tensor are
\begin{eqnarray}\label{K_radial_motion_log}
K^{\hat{r}}_{\hat{r}}&\approx&\frac{\alpha_1 p^2 (m_{\text{max}}+1)}{(L p)^{2(m_{\text{max}}+1)}\beta_{m_{\text{max}}}}\log{(r)} ~r^{2 m_{\text{max}}}\, ,\\
K^{\hat{\varphi}}_{\hat{\varphi}}&\approx&\frac{\alpha_1 p^2}{(2 m_{\text{max}}+1)(L p)^{2(m_{\text{max}}+1)}\beta_{m_{\text{max}}}}\log{(r)} ~r^{2 m_{\text{max}}}\, ,
\end{eqnarray}
so the force vanishes as long as $m_{\text{max}}\geq 1$.
\subsubsection{Non-radial motion}
Now we turn to the case of non-vanishing angular momentum. While the angular part of the tidal tensor remains unchanged, the radial component is multiplied by a pole of second order at the origin, changing its behavior  close to the center. More precisely, we get
\begin{align}\nonumber \label{K_non_radial}
K^{\hat{r}}_{\hat{r}}\approx &-\frac{J^2}{L^4 p^2} \left[ \frac{(n_{\text{max}}-m_{\text{max}}-2)(n_{\text{max}}-m_{\text{max}}-1)}{(n_{\text{max}}-1)(2 m_{\text{max}}+1)}\frac{\alpha_{n_{\text{max}}}}{ \beta_{m_{\text{max}}}}\right]\left(\frac{L p}{r}\right)^{2(n_{\text{max}}-m_{\text{max}})}\\
&+\mathcal{O}\left(r^{-2(n_{\text{max}}-m_{\text{max}}-1)}\right)\, .
\end{align}
Therefore, in singular black holes the geodesic deviation diverges faster than for radial motion (\ref{K_radial_motion}). On the other hand, in regular black hole backgrounds the tidal tensor still goes to zero at $r=0$, although at a slower rate. However, when $n_{\text{max}}=m_{\text{max}}+1$, which corresponds to a metric function $f(r)\sim r^2$, the leading term in (\ref{K_non_radial}) vanishes, so we need to take into account the next-to leading order. Schematically, 
\begin{equation}\label{K_special}
K^{\hat{r}}_{\hat{r}}\approx \begin{cases}
\text{const.} + J^2 \log{(r)} & \text{if} \quad n_{\text{max}}=2\, ,\\
\text{const.} &\text{if} \quad n_{\text{max}}>2\, .
\end{cases}   
\end{equation}
This means that an observer falling to a regular black hole with $n_{\text{max}}=2, m_{\text{max}}=1$, and with non-vanishing angular velocity will experience infinite tidal forces. This is precisely the scenario corresponding to the right panel of Figure \ref{SUBFIGURE LABEL X1}. For all other regular black holes the tidal forces do not diverge. In particular, observers following the geodesics represented by orange curves in the left panel of Figure \ref{SUBFIGURE LABEL X1} and Figure \ref{SUBFIGURE LABEL X2} pass through the center and continue to positive values of the radial coordinate.

Again, the situation with $n_{\text{max}}=1$ must be considered separately. For these kind of solutions we get
\begin{equation}
K^{\hat{r}}_{\hat{r}}\approx J^2\frac{\alpha_1 p^2}{(L p)^{2(m_{\text{max}}+1)}\beta_{m_{\text{max}}}}\frac{2 m_{\text{max}}(m_{\text{max}}+1)}{2 m_{\text{max}}+1}\log{(r)} ~r^{2( m_{\text{max}}-1)}+ \mathcal{O}\left( r^{2(m_{\text{max}}-1)}\right)\, ,
\end{equation}
which for regular backgrounds goes to zero at $r=0$ as well.



\section{Kasner Eons}\label{eons}
In this section we have a look at the deep interior region of our black holes yet from another perspective. 

Let us start with some relevant introductory comments. Einstein gravity in $D$ spacetime dimensions admits an anisotropic family of solutions whose metric can be written as
\be \label{Ka}
\mathrm{d}s^2 = - \mathrm{d}\tau^2 + \sum_{i=1}^{D-1}\tau^{2p_i}  \mathrm{d} x_i^2 \, .
\ee
These ``Kasner metrics'' solve Einstein's equations whenever the following two conditions hold:
\begin{equation}\label{condE}
\sum_{i=1}^{D-1} p_i= 1\, , \quad \sum_{i=1}^{D-1} p_i^2=1\, .
\end{equation}
These constraints fix two of the $(D-1)$ Kasner exponents $p_i$, corresponding to a $(D-3)$-parametric family of solutions \cite{Belinski:2017fas}. This class of metrics plays a prominent role in the study of space-like singularities. Indeed, as shown by Belinski, Khalatnikov and Lifshitz (BKL), in the context of Einstein gravity in $D\geq 4$, the metric in the vicinity of a space-like singularity precisely takes the Kasner form \req{Ka} at each spatial point \cite{Belinsky:1970ew}. The Kasner exponents are constrained by the conditions \req{condE} but otherwise have an undetermined dependence on the spatial coordinates, $p_i=p_i(x)$. In other words, the dynamics of every spatial point is decoupled from the one of the rest---hence, the evolution becomes \emph{ultralocal}---and, for each point, it is given by some Kasner metric. As the singularity is approached, a ``Kasner epoch'' is defined as the period during which some generalized Kasner metric is a good approximation. Ultimately, this approximation breaks down due to spatial curvature effects, giving rise to a transition into a new epoch characterized by a new set of exponents. In each transition between epochs, one of the directions along which spacetime is contracting switches its role with the one along which it is expanding. A ``Kasner era'' is then defined as the period during which the transitions between Kasner epochs are such that these swaps occur repeatedly between the same two directions. 

As pointed out recently in \cite{Bueno:2024fzg}, this chaotic sequence of epochs and eras is terminated when the dynamical description provided by Einstein gravity is no longer valid due to quantum/higher-curvature effects. The period during which such description holds defines a ``Kasner Eon'', which may include many eras. As putative higher-curvature effects kick in, additional Kasner eons may appear, for which some kind of modified BKL description may hold, and the relevant Kasner metrics would satisfy different conditions from \req{condE} as well as modified transition rules \cite{Bueno:2024qhh}. During each additional eon, a particular higher-curvature density---or set of densities---controls the dynamics, and a cascade of eons may occur as the singularity is approached provided certain hierarchy of scales among the various densities exists. Thus far, these expectations have been verified in the simplified context of spherically symmetric black holes  in $D\geq 4$ \cite{Bueno:2024fzg,Bueno:2024qhh,Caceres:2024edr}. For those, neither chaos nor ultralocality feature, and the Einsteinian eon involves both a single epoch and a single era. However, cascades of Kanser eons do arise.

In this section we study the appearance of Kasner eons in the deep interior of the black hole solutions of EQT. Before doing so, let us a make a few general comments.  Note first that in the case of Einstein gravity (without matter) in $D=3$, all freedom is gone and the two Kasner exponents which exist in that case are fixed to specific values, namely,\footnote{Or $p_1 \leftrightarrow p_2$, which is equivalent.}
\begin{equation}
    p_1=0\, , \quad p_2=1\, .
\end{equation}
Hence, in this case there is no room for any sort of ultralocality in the form of spatial dependence of the exponents: they are simply fixed to the above values. This is a manifestation of the fact that the BKL analysis does not hold for three-dimensional Einstein gravity. Indeed, in that case all solutions are locally equivalent to maximally symmetric solutions---except at specific points corresponding to BTZ-like or conical singularities\footnote{For a recent interesting discussion about the nature of the singularities at the center of BTZ spacetimes, see \cite{Briceno:2024ddc}.}---and no local degrees of freedom propagate. The situation changes in the presence of matter. In that case, non-trivial near-singularity dynamics is possible, and hints of partial BKL-like behavior in $D=3$ have been reported---see \eg \cite{Gao:2023rqc,Fleig:2018djg}.

Consider now a three-dimensional static black hole with a single horizon, a spacelike curvature singularity at $r=0$ and with an interior metric described in Schwarzschild coordinates as
\begin{align} \label{bhfN}
\mathrm{d} s^2 =&\, \frac{\mathrm{d}r^2}{f(r)} - f(r) \mathrm{d} z^2 + r^2 \mathrm{d} \varphi^2 \, .
\end{align}
Obviously, this is the case of most of the solutions described in Section~\ref{sec:black holes in EQTG} with metric function \req{fEQT2} provided the single-horizon condition holds, but the following description is more general.
Whenever $f(r)$ behaves as
\begin{equation}
f(r) \overset{r\rightarrow 0}{\sim} \, - r^{-s} \, ,
\end{equation}
near the singularity, we can make the coordinate transformation 
\be 
\mathrm{d} \tau \equiv \frac{\mathrm{d} r}{\sqrt{-f}} \quad \Rightarrow \quad \tau \sim r^{(s+2)/2} \, .
\ee
Then, the metric becomes 
\be \label{taup}
\mathrm{d}s^2 = - \mathrm{d}\tau^2 + \tau^{2p_1} \mathrm{d} z^2 + \tau^{2p_2} \mathrm{d} \varphi^2 \, ,
\ee
where the two independent exponents read
\be \label{p1p2}
p_1 = - \frac{s}{(2+s)} \, , \quad p_2 = \frac{2}{(2+s)} \, .
\ee
Hence, in that case, the metric takes the form of a Kasner spacetime  \req{Ka}. 
Following \cite{Bueno:2024fzg}, we can introduce an ``effective'' Kasner exponent $p_{\rm eff}$ for the $\mathrm{d}z^2$ component of the metric,
\begin{align}\label{effp}
\peff(r) \equiv \frac{r f'(r)}{[2 f(r) - r f'(r)]} \, ,
\end{align}
so that any time that $f(r) \sim r^{-s}$, $\peff(r)$ becomes constant and the metric is locally Kasner with exponents
\be
\label{peffs0} 
p_1 = \peff \, , \quad p_2 = \peff + 1\, .
\ee
In this context, a Kasner eon would correspond to a period during which the black hole interior in the vicinity of the singularity behaves, approximately, as a Kasner metric with exponents satisfying \req{peffs0}. In each case, the exponents may or may not match the ones of exact Kasner solutions of the corresponding densities, as we explain below.

\begin{figure}[!t]
  \begin{subfigure}{1\linewidth}
 \includegraphics[width=0.51\linewidth]{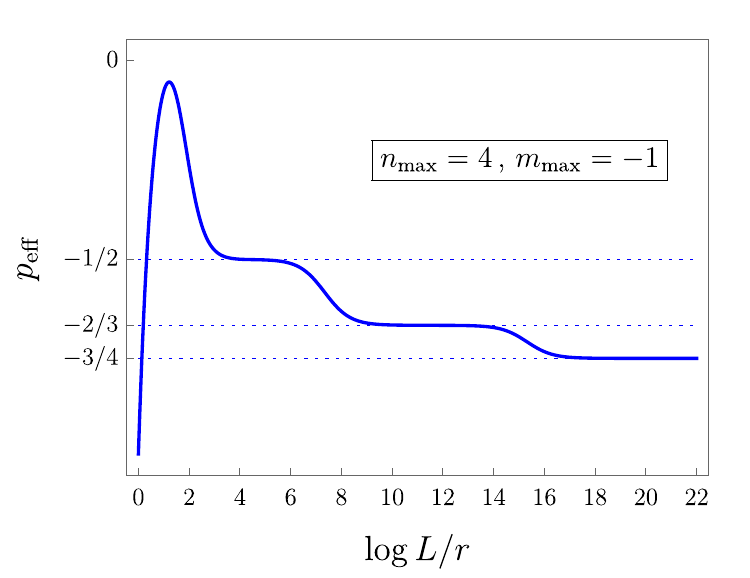}  
    \includegraphics[width=0.51\linewidth]{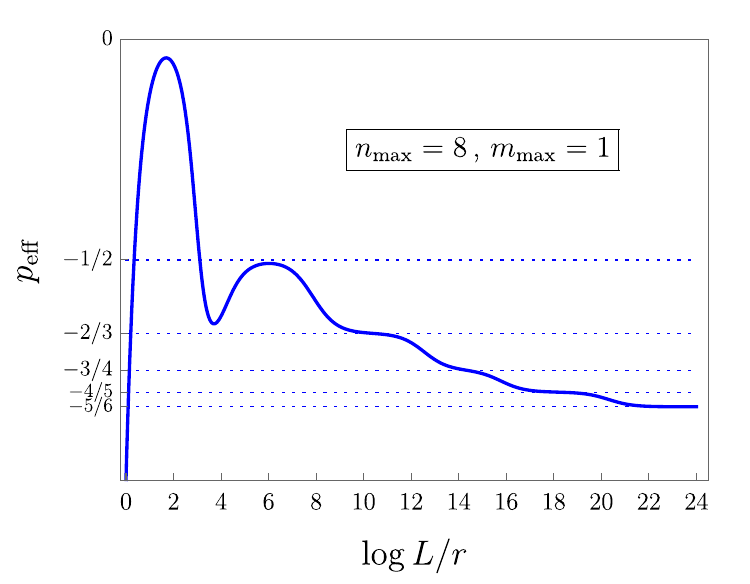}  
    \caption{(Left) A black hole with a curvature singularity. Three eons are visible. The transitions (from right to left) are: $(n=4,m=-1)\rightarrow (n=3,m=-1)\rightarrow(n=2,m=-1)$. (Right) Another black hole with a curvature singularity. Five eons are visible. The transitions (from right to left) are: $(n=8,m=1)\rightarrow (n=7,m=1)\rightarrow(n=6,m=1)\rightarrow(n=5,m=1)\rightarrow(n=4,m=1)$. Many putative eons are skipped and observe that the one with $(n=4,m=0)$, corresponding to $p_1=-2/3$, is actually about to form after the $(n=4,m=1)$ one.   }
    \label{sub}
\end{subfigure} \\  
\begin{subfigure}{1\linewidth}
 \includegraphics[width=0.51\linewidth]{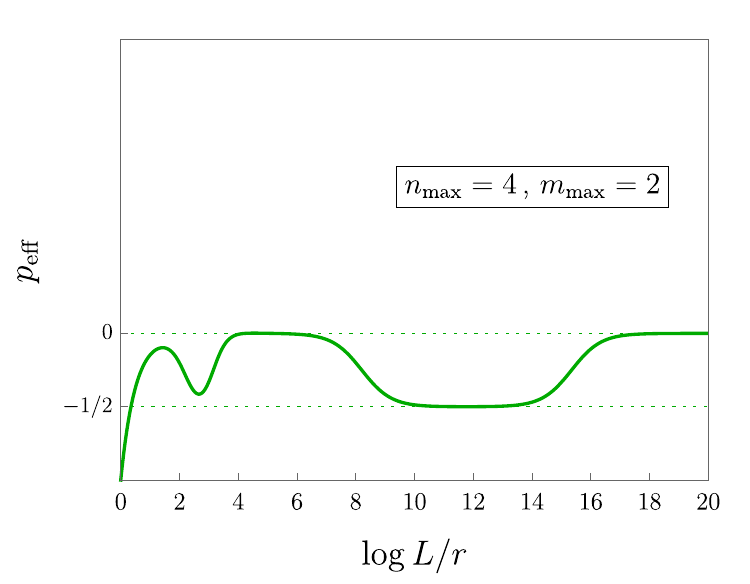}  
   \includegraphics[width=0.51\linewidth]{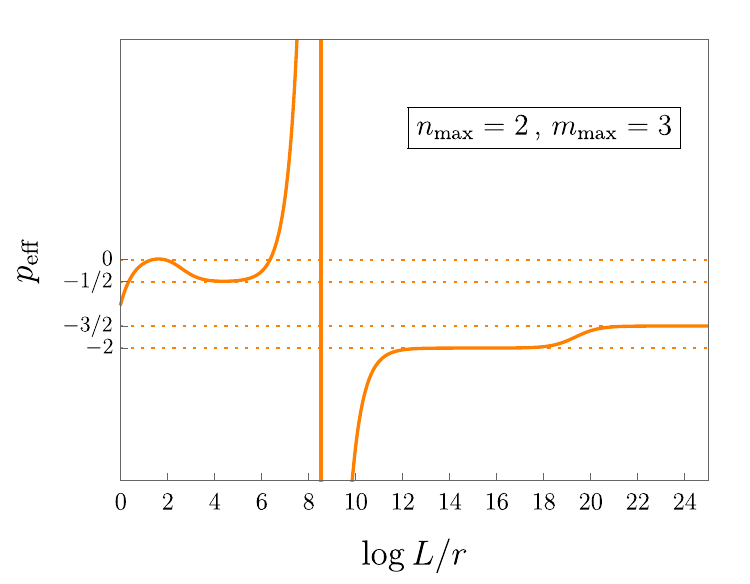}  
       \caption{(Left) A black hole with a BTZ-like singularity. Three eons are clearly visible, corresponding to $(n=4,m=2)$, $(n=4,m=1)$ and $(n=3,m=1)$, respectively. Observe that the effective Kasner exponent returns to the same value, $p_1=0$, after the second eon. The putative eon with $(n=3,m=0)$, corresponding to $p_1=-1/2$, is about to form after the $(n=3,m=1)$ one. (Right) A regular black hole---see the main text. 
}
\end{subfigure}
\caption{We plot the effective Kasner exponent as defined in \req{effp} for single-horizon black holes displaying different types of behaviors at $r=0$. }
\label{fig:eons}
\end{figure}

Consider first the case of a particular density of order $n\geq 2$ within the $Q_{\alpha}$ class in \req{qb}. From \req{fEQT2} it is obvious that this admits exact Kasner solutions. Indeed, $ f(r)\propto r^{-2(n-1)}$
with the same azimuthal ansatz for the scalar solves the equations of the theory. As explained above, this can be written in the form \req{taup}, with Kasner exponents
\begin{eqnarray}\label{noM}
    p_1=-\frac{(n-1)}{n}\, , \quad p_2=\frac{1}{n}\, .
\end{eqnarray}
On the other hand, densities of the $Q_{\beta}$ class do not admit exact Kasner solutions in an analogous way. In that case, we have
$
    f(r)\propto 1/(1+c_m r^{-2(m+1)})$, which is not of the Kasner type because of the ``$1+$'' appearing in the denominator. In this case, it would only be for $r\rightarrow 0$ that the metric becomes approximately Kasner with exponents
\begin{equation}\label{noN}
    p_1=-\frac{(m+1)}{m}\, , \quad p_2=-\frac{1}{m}\, .
\end{equation}

Let us now go back to our black hole solutions. Consider the general case in which both $Q_{\alpha}$ and $Q_{\beta}$ are present in the Lagrangian. 
In that case, the metric function behaves as
\begin{equation}\label{nmax2}
    f(r) \overset{(r \rightarrow 0)}{\sim} \frac{\alpha_{n_{\rm max}}}{\beta_{m_{\rm max}}} \, r^{-2(n_{\rm max}-m_{\rm max}-2)} \, , \quad (n_{\rm max}\geq 2)
\end{equation}
which may correspond to black holes with curvature or BTZ-like/conical singularities, or to regular black holes, as explained in Section~\ref{sec:black holes in EQTG}. Exceptions to this behavior take place when $n_{\rm max}=1,0$, in which cases the above expression would be replaced by 
\begin{equation}
    f(r) \overset{(r \rightarrow 0)}{\sim} \frac{(-\lambda-\alpha_{1}p^2 \log r)}{\beta_{m_{\rm max}}} \, r^{2(m_{\rm max}+1)}  \quad \text{and} \quad f(r) \overset{(r \rightarrow 0)}{\sim} \frac{\lambda}{\beta_{m_{\rm max}}} \, r^{2(m_{\rm max}+1)} \, ,
\end{equation}
respectively.\footnote{Observe that the $n_{\rm max}=0$ case, for which there is no logarithmic term, cannot be obtained naively by setting $n_{\rm max}=0$ in \req{nmax2}. The correct scaling is obtained, however, by setting $n_{\rm max}=1$ in that formula, and the same will apply for the rest of formulas of this section. The fact that we are interpreting ``$n_{\rm max}=1$'' as the case in which there is a logarithm introduces this notational nuisance.  }
As the $r=0$ locus is approached, the black hole interior may undergo a series of Kasner eons which terminate with a final eon with exponents satisfying 
\begin{equation}\label{peffnm}
    p_1=-\frac{(n_{\rm max}-m_{\rm max}-2)}{(n_{\rm max}-m_{\rm max}-1)}\, , \quad p_2=\frac{1}{(n_{\rm max}-m_{\rm max}-1)}\, ,
\end{equation}
as long as $n_{\rm max}\geq 2$ and 
\begin{equation}\label{peffnm}
    p_1=-\frac{(m_{\rm max}+1)}{(m_{\rm max})}\, , \quad p_2=-\frac{1}{m_{\rm max}}\, ,
\end{equation}
for $n_{\rm max}=0$, which is precisely the case \req{noN}. 
The case \req{noM}, on the other hand, is recovered for $m_{\rm max}=-1$. Whenever $n_{\rm max} > m_{\rm max}+2$, corresponding to a curvature singularity at the origin, $p_1<0$ and $p_2>0$, whereas for $n_{\rm max} =m_{\rm max}+2$, which is the case of BTZ-like/conical singularities, $p_1=0$, $p_2=1$---see Figure~\ref{fig:eons}. On the other hand, if $n_{\rm max} < m_{\rm max}+2$, which is the case of regular black holes, we have  $p_1<0$ and $p_2<0$ except in the particular case $n_{\rm max}=m_{\rm max}+1$, for which both diverge. This happens when $f(r)\overset{(r\rightarrow0)}{\sim} (\alpha_{m_{\rm max}+1}/\beta_{m_{\rm max}})\cdot r^2$, which corresponds to pure dS$_3$ or AdS$_3$, depending on the sign.  

Provided there exist intermediate ranges of values of $r/L$ such that certain pair of couplings $\alpha_i$, $\beta_j$ satisfy
\begin{equation}\label{eoncondi}
\frac{\alpha_i}{\alpha_n }   \frac{(n-1)}{(i-1)}\gg \left(\frac{r}{p L}\right)^{2(i-n)}\, \quad \text{and} \quad \frac{\beta_j}{\beta_m }   \frac{(2j+1)}{(2m+1)}\gg \left(\frac{r}{p L}\right)^{2(j-m)}\, \quad \forall\, \alpha_n,\beta_m\, ,
\end{equation}
intermediate eons will arise with effective exponents given by
\begin{equation}\label{peffnm}
    p_1=-\frac{(i-j-2)}{(i-j-1)}\, , \quad p_2=\frac{1}{(i-j-1)}\, .
\end{equation}
Each of these eons will terminate whenever the values of $r$ become such that the ``$\gg$'' above become ``$\sim$'' for certain $\alpha_k$, $\beta_l$. Naturally, if \req{eoncondi} is not fulfilled for any range of values of $r$, the corresponding $(i,j)$ eon will be skipped altogether.
Observe that the effective Kasner exponents may diverge for intermediate values of $r$ provided \req{eoncondi} holds for $i=j+1$---see Figure\,\ref{fig:eons}. Also, the values of the effective Kasner exponents can go back to the same values for different eons, since \req{peffnm} can hold for different pairs of values, namely, provided $(i_a-j_a)=(i_b-j_b)$. This implies, in particular, that the effective exponents may both increase and decrease  during different periods as $r=0$ is approached. From \req{peffnm} it is obvious that, as opposed to the cases previously considered in the literature 
\cite{Bueno:2024fzg,Bueno:2024qhh,Caceres:2024edr}, most of the Kasner eons arising in this context do not correspond to Kasner solutions of particular  higher-derivative densities but rather, they are a consequence of the joint effect of pairs of densities---one from $\mathcal{Q}_\alpha$ and one from $\mathcal{Q}_\beta$ in each case---with the only exceptions of those produced by particular densities of the $\mathcal{Q}_\alpha$ set, for which \req{noM} holds.

As an explicit example, consider the case depicted in the lower right plot of Figure\,\ref{fig:eons}, for which we set $\alpha_1=1/3$, $\alpha_2=-1/100$, $\beta_0=10^{-5}$, $\beta_1=10^{-11}$, $\beta_2=10^{-20}$, $\beta_3=10^{-37}$, $L=1$, $p=3/4$, $\lambda=1$. This corresponds to a regular black hole with $f(r)\overset{(r\rightarrow 0)}{\sim} \mathcal{O}(r^6)$. The final eon corresponds to $(n=2, m=3)$, for which $p_1=-3/2$, $p_2=-1/2$. From that one, moving towards the left in the plot, there is a transition to an eon with $(n=2, m=2)$, for which $p_1=-2$, $p_2=-1$. This is followed by another transition in which the exponents diverge, corresponding to  $(n=2, m=1)$. Then there is a transition into another eon with $(n=2, m=0)$, for which $p_1=-1/2$, $p_2=1/2$. Finally, there is a short eon with  $(n=1, m=0)$, for which $p_1=0$, $p_2=1$.
Additional examples both for singular black holes as well as for black holes with BTZ-like/conical singularities are presented in the same figure.


\section{Conclusions}\label{conc}
In this paper we have studied various aspects of EQT black holes. A summary of our main results can be found in the Introduction. Let us close the paper with a few comments regarding the implications of our findings and plans for future research. 

\subsection*{EGQT generalizations}
 The EQT theories considered here are the most general models of gravity non-minimally coupled to an azimuthal scalar for which the metric function can be obtained explicitly. On the other hand, as argued in \cite{Bueno:2022ewf}, infinite families of additional theories of the EGQT class exist for which the metric function satisfies a second-order differential equation. Many of the features explored in this paper could be extended to the EGQT black holes.

\subsection*{Thermodynamics and holography}
The basic thermodynamic properties of the EQT and EGQT black holes were briefly studied in \cite{Bueno:2022ewf}. It would be interesting to pursue this analysis further and explore the phase space of solutions, possibly within the framework of ``extended thermodynamics'' \cite{Kastor:2009wy,Dolan:2010ha,Cvetic:2010jb,Kubiznak:2012wp}. On the other hand, a holographic study of three-dimensional EQT theories is also lacking. In that context, these theories could be used to test properties of two-dimensional CFTs with global symmetries. As a particular application, it would be interesting to use these models to  shed light on the reported seemingly peculiar behavior of  charged R\'enyi entropies in two-dimensional CFTs \cite{Belin:2013uta, Arenas-Henriquez:2022ntz}---similarly to what was done for their higher-dimensional counterparts in \cite{Bueno:2022jbl}.

\subsection*{Inner-extremal regular black holes, stability and cosmic censorship}
Regular black holes in higher dimensions require the presence of an even number of horizons, and small perturbations on top of inner horizons have been argued to lead to an exponential backreaction on the geometry \cite{Carballo-Rubio:2018pmi,Carballo-Rubio:2024dca}. Classical stability is restored in the case of inner horizons with vanishing surface gravity  \cite{Carballo-Rubio:2022kad}, but achieving this seems to always require a certain degree of fine-tuning in the case of actual solutions to dynamical models---see \eg \cite{DiFilippo:2024mwm}. In addition, semiclassical effects may bring instabilities back in any case \cite{McMaken:2023uue}. The lack of theories in which regular black holes arise as the end-point of matter collapse has prevented reaching fully conclusive results regarding these issues, although this may be about to change \cite{Bueno:2024eig,Bueno:2024zsx}. 

On the other hand, due to their distinct nature---peculiar to $D=3$---some of the regular black holes considered here may automatically be dynamically stable generically. In order to see this, let us have a closer to look at the region $r=0$ in the case of interest corresponding to $f(r)\overset{(r\rightarrow 0)}{\sim}\mathcal{O}(r^{2s})$, $s\geq 1$. Firstly, note that $r=0$ corresponds to a Killing horizon associated to $\xi=\partial_t$, since $\xi^2\equiv g_{ab}\xi^a\xi^b=g_{tt}\overset{(r=0)}{=}0$ for those solutions.\footnote{On the other hand, it is a certainly peculiar horizon, as its cross-sections would correspond to circles of zero length. This feature deserves a better understanding.} It is also manifest that such horizons are ``extremal'' in the sense of having a vanishing surface gravity, namely, $f'(r)\overset{(r\rightarrow 0)}{\sim}2s\cdot \mathcal{O}(r^{2s-1})$ and then $\left. \kappa \right|_{r=0}= -\frac{1}{2}f'(0)=0$. Remarkably, this holds for all regular black holes considered here. Hence, all black holes of this class automatically include an inner-extremal horizon. Naturally, if the number of additional horizons is sufficiently large, there can be additional inner horizons---see Sections \ref{sec:number_roots} and \ref{penro}---with their putative instability issues. Consider however the case in which $f(r)$ has a single additional zero for some $r_H>0$---see the left plot in Figure\,\ref{fig:reg} for $f(r)$ in that case and $R_0^-$, $\mathring R_0^-$, $\bar R_0^-$  in \req{eq:R-0}, \req{Penrose_geodesics1} and \req{Penrose_geodesics2} for its Penrose diagrams. This corresponds to a black hole with an outer event horizon at $r=r_H$ and an inner horizon
with vanishing surface gravity at $r=0$. Such a solution is therefore rather similar to a standard higher-dimensional regular black hole, except for the fact that inner-horizon extremality is inherently incorporated---namely, not only it does not require any sort of fine-tuning, but it simply cannot be avoided within the model. To the best of our knowledge, these are the first examples of this class available in the literature. Therefore they warrant further study, particularly concerning their dynamical stability.

Note also that the presence of higher-curvature terms in the action may introduce additional instabilities, although this is far from evident without performing the actual computations---\eg it is not even clear that the equations for gravitational perturbations will involve more than two derivatives. Related to this, it would be interesting to analyze the quasi-normal mode spectrum of the solutions---a program started in \cite{deOliveira:2024pab}---and study possible violations of cosmic censorship---along the lines of \cite{Dias:2019ery,Emparan:2020rnp}---across all the possible classes of black holes.

\subsection*{Dynamical formation}
As we showed in Section \ref{birk}, EQT gravities in three dimensions satisfy a Birkhoff theorem. This means that the outside region of any spherically symmetric configuration involving matter will be described by the corresponding static black hole. In particular, Eqs.\,\req{eomM1} are ready to be used for different types of stress energy tensors. As a first application, it would be interesting to study the collapse of charged thin shells of dust, which should lead to the formation of the corresponding black holes, analogously to \cite{Bueno:2024eig,Bueno:2024zsx}.  Another direction would entail studying the modified conditions for stellar equilibrium \cite{Cruz:1994ar}.

\subsection*{Different asymptotes}
Throughout this paper---and also in the original one on the subject \cite{Bueno:2021krl}---EQT theories have been considered in the presence of a negative cosmological constant. However, as opposed to the three-dimensional Einstein gravity case, EQT black holes exist also for Minkowski and dS asymptotes, so it would be interesting to study the solutions further in those cases.

\section*{Acknowledgments}

We thank Pablo A. Cano, Roberto Emparan, Pedro G. S. Fernandes, Robie A. Hennigar, Ángel Murcia, Julio Oliva and Tomás Ortín for useful discussions. PB was supported by a Ramón y Cajal fellowship (RYC2020-028756-I) and by the grant PID2022-136224NB-C22, funded by MCIN/AEI/10.13039/501100011033/FEDER, UE. PB and GdV were supported by Proyecto de Consolidación Investigadora (CNS 2023-143822) from Spain’s Ministry of Science, Innovation and Universities. The work of JM is supported by ANID FONDECYT Postdoctorado Grant No. 3230626.

\bibliographystyle{JHEP-2}
\bibliography{Gravities.bib}

\end{document}